\documentclass[twocolumn,trackchanges]{aastex701} %linenumbers,
\usepackage{amssymb}
\usepackage{lipsum}
\usepackage{url}
\usepackage{bm}
\usepackage{amsmath, mathtools, xfrac}  % Advanced math commands

\begin{document}

\title{Dynamical and Photometric Analysis of NGC 146 and King 14: Evidence for a Co-Moving, Unbound Cluster Pair}

\correspondingauthor{D. Bisht}
\email{devendrabisht297@gmail.com}  

\author[orcid=0000-0002-8988-8434, sname=Bisht, gname=D.]{D. Bisht}\thanks{E-mail: devendrabisht297@gmail.com}
\affiliation{Indian Centre For Space Physics
466, Barakhola, Singabari road, Netai Nagar, Kolkata, West Bengal, 700099}
\email{devendrabisht297@gmail.com}

\author[orcid=0000-0001-7359-3300, sname=Jiang, gname=D.]{Ing-Guey Jiang}\thanks{E-mail: jiang@phys.nthu.edu.tw}
\affiliation{Department of Physics and Institute of Astronomy, National Tsing-Hua University, Hsinchu 30013, Taiwan}
\email{jiang@phys.nthu.edu.tw}

\author[orcid=0000-0002-2298-4026, sname=Elsanhoury, gname=W. H.]{W. H. Elsanhoury}
\affiliation{Department of Physics, College of Science, Arar, Northern Border Region, Saudi Arabia}
\email{elsanhoury@nbu.edu.sa}

\author[sname=Belwal, gname=K.]{K. Belwal}\thanks{E-mail: kuldeepbelwal1997@gmail.com}
\affiliation{Indian Centre For Space Physics
466, Barakhola, Singabari road, Netai Nagar, Kolkata, West Bengal, 700099}
\email{kuldeepbelwal1997@gmail.com}

\author[orcid=0000-0001-7940-3731, sname=Çınar, gname=Deniz Cennet]{D. C. Çınar}
\affiliation{Programme of Astronomy and Space Sciences, Institute of Graduate Studies in Science, Istanbul University, Istanbul, 34116, Turkey}
\email{denizcdursun@gmail.com}

\author[sname=Raj, gname=A.]{A. Raj}
\affiliation{Indian Centre For Space Physics
466, Barakhola, Singabari road, Netai Nagar, Kolkata, West Bengal, 700099}
\affiliation{Uttar Pradesh State Institute of Forensic Science (UPSIFS) Aurawan, P.O. Banthra, Lucknow 226401, U.P, India}
\email{raj@example.com}

\author[orcid=0009-0003-8446-4557, sname=Biswas, gname=S.]{Shraddha Biswas}
\affiliation{Indian Centre For Space Physics
466, Barakhola, Singabari road, Netai Nagar, Kolkata, West Bengal, 700099}
\email{sbiswas@example.com}

\author[orcid=0000-0002-4729-9316, sname=Dattatrey, gname=Arvind]{Arvind K. Dattatrey}
\affiliation{Indian Institute of Astrophysics, 560034 Bangalore, India}
\email{arvind@aries.res.in}

\author[orcid=0000-0002-6373-770X, sname=Rangwal, gname=Geeta]{Geeta Rangwal}
\affiliation{Aryabhatta Research Institute of Observational Sciences, Manora Peak, Nainital 263129, India}
\email{geetarangwal91@gmail.com}

\author[sname=Sariya, gname=Devesh]{Devesh P. Sariya}
\affiliation{Department of Physics and Institute of Astronomy, National Tsing-Hua University, Hsinchu 30013, Taiwan}
\email{deveshpath@gmail.com}

\author[sname= Singh Bisht, gname=Mohit]{Mohit Singh Bisht}
\affiliation{Indian Centre For Space Physics
466, Barakhola, Singabari road, Netai Nagar, Kolkata, West Bengal, 700099}
\email{mohitsinghbisht742@gmail.com}

\author[sname=Durgapal, gname=Alok]{Alok Durgapal}
\affiliation{Center of Advanced Study, Department of Physics, D. S. B. Campus, Kumaun University, Nainital 263002, India}
\email{alokdurgapal@gmail.como}


\begin{abstract}

To understand the nature of the NGC~146--King~14 cluster pair, we conducted a detailed photometric, astrometric, and dynamical study using multiwavelength data from \textit{Gaia}~DR3, Pan-STARRS1, WISE, and \textit{TESS}. Using a probabilistic approach, we identified 770 and 690 high-probability members of NGC~146 and King~14, respectively. Both clusters exhibit well-defined radial density profiles consistent with King models. We estimate the cluster ages as $20\pm5$~Myr and $50\pm10$~Myr from isochrone fitting, and distances of $2.98\pm0.33$~kpc and $2.51\pm0.23$~kpc from parallaxes after applying the Bailer-Jones criteria. The clusters show consistent mean proper motions. The mass function slopes ($1.51\pm0.18$ and $1.50\pm0.15$) are close to the Salpeter value, and the extinction follows a normal Galactic reddening law ($R_V \approx 3.1$). Three-dimensional mapping gives a projected separation of $\sim9$~pc. Orbit integration using the \texttt{galpy} \texttt{MWPotential2014} model shows that NGC\,146 and King\,14 move in nearly circular, disk-like orbits with similar mean orbital radii ($R_{\rm m} \sim 9$~kpc) and orbital periods of roughly 255~Myr. A dynamical separation of $\sim32$~pc indicates that both clusters share a common spatial and kinematic association, consistent with a co-moving pair. However, their relative velocity exceeds the escape velocity set by their combined mass, indicating they are not gravitationally bound. \textit{TESS} light curves reveal seven variable stars, including $\gamma$ Doradus, SPB stars, and eclipsing binaries, though only one is a likely member. Overall, the clusters likely formed within the same giant molecular cloud and now exist as an unbound co-moving pair.
\end{abstract}




%\keywords{open clusters and associations, mass function, stellar kinematics, stellar orbits}

%\tableofcontents



\section{Introduction}\label{introduction}

Open clusters (OCs) typically contain a few hundred to a few thousand stars formed from a single massive molecular cloud \citep{corsaro2017spin}. These stars are loosely bound by weak gravitational forces \citep{leroy2018forming}. OCs are valuable for studying the history of star evolution as they originate from the collapse and fragmentation of massive molecular clouds \citep{bate2003formation, harris1994supergiant}. Binary clusters are especially significant among OCs, as they help in understanding the formation, evolution, and characteristics of stars in the Galactic disk \citep{piecka2021aggregates}. Binary clusters can form when several massive and dense gas clumps within a giant molecular cloud collapse due to gravity. These clumps may give rise to clusters that are nearly the same age and located very close to each other \citep{piatti2010evidence, arnold2017binary, mora2019collision, darma2021formation}. Most multiple and embedded clusters quickly disperse or merge due to early disruption \citep{lada2003embedded}. In some cases, two clusters can become gravitationally bound during a close encounter, forming a binary system with significantly different ages \citep{van1996formation, de2009double}. The existence of binary and multiple OCs has been explored for decades. Many studies based on observational data suggested that around 8–10$\%$ of OCs could be genuine binaries \citep{subramaniam1995probable, loktin1997selection, de2009double}. In addition, studies focusing on binary and multiple OCs in the $Gaia$ era are still ongoing, and further investigations are needed to fully understand their nature and frequency \citep{Song_2023, Li2024, Haroon2024, Haroon2025, Palma_2024, TasdemirCinar2025}. 

The catalog of known OCs continues to expand, and membership determination has become more accurate thanks to the $Gaia$ mission. As a result, the identification of binary clusters has attracted growing interest in recent years. \citet{soubiran2019open} identified 21 cluster pairs with small separations; \citet{liu2019catalog} reported 56 candidate cluster groups based on 3D spatial positions. \citet{piecka2021aggregates} also identified 60 cluster groupings by examining shared members in low phase-space volumes using the \citet{Cantat-Gaudin-Anders_2020} catalog. These studies highlight the growing potential for discovering new binary cluster systems.

Binary and co-moving clusters are essential for understanding star and cluster formation and evolution \citep{casado2021list}.  They also offer key insights into the mechanisms driving cluster formation and development \citep{dalessandro2018unexpected}. They reveal how molecular clouds fragment and how Galactic forces impact cluster survival. Examining unbound or dissolving clusters shows how stars disperse into the field population. These systems also shed light on the Milky Way’s past cluster-formation and stellar-migration history \citet{angelo2022investigating, darma2021formation}. NGC 146 and King 14 are nearby OCs in the Perseus arm. Their similar distances and proximity have led to discussions of a possible physical connection or a binary nature. The available information for both clusters from the literature is summarized below-
\\\\

{\bf NGC 146}

NGC 146 (C 0030+630) is an open star cluster located in the constellation of Cassiopeia, approximately 1,800 light-years away from Earth. It is a relatively young cluster, estimated to be around 40 million years old. NGC 146 contains a mix of late B-type and early A-type stars, which contribute to its brightness. Observational studies, such as those conducted by \citet{kharchenko2005astrophysical}, have focused on the stellar population and overall dynamics of the cluster, uncovering insights into its structure and the impact of the surrounding interstellar medium on its evolution. \citet{rojas2010metal} also studied the relationships between stellar populations and the interstellar environment, highlighting NGC 146 as a target for further astrophysical studies due to its relatively rich stellar environment.
\\

{\bf King 14}

King 14 (C 0029+628) is an open cluster situated in the constellation of Cassiopeia, approximately 2,000 light-years from Earth. Characterized as a faint cluster, King 14 has a somewhat dispersed distribution of stars and is estimated to be about 100 Myr old, making it relatively older than others. The cluster consists primarily of F-type stars, with studies emphasizing its membership and stellar content. Research by \citet{king1962structure} provided early observations of the cluster's structure, while \citet{dias2002new} investigated its stellar population, shedding light on the characteristics and evolution of older OCs. These studies underscore the challenges posed by King 14's dispersed nature for observational research, yet they remain crucial for understanding the dynamics of such clusters.

The NGC 146–King 14 pair presents a unique case for investigating how young, co-moving OCs in the Galactic disk are physically connected and how they evolve dynamically. Studying such pairs is critical, as they serve as laboratories to examine the transition from bound binary clusters to unbound yet kinematically associated systems influenced by Galactic tidal forces. By comparing them, researchers gain insight into the formation of cluster pairs within giant molecular clouds and how their interactions shape the early dynamical evolution. Identifying variable stars within and around these clusters adds further value, since these variables trace recent star formation, stellar pulsations, and possible cluster membership. Analyzing these variables with high-precision TESS photometry deepens understanding of each cluster's stellar population. It provides context for the time-domain evolution and dynamical history of young OCs in the Milky Way.

The primary objective of this study is to investigate whether the OCs NGC 146 and King 14 constitute a physically related or co-moving binary cluster system. To address this, we determine their fundamental parameters, examine the mass functions, and analyze their internal kinematics using high-precision Gaia DR3 data. An orbital analysis is conducted to trace their past trajectories and evaluate a possible common origin. Additionally, we employ TESS photometric data to identify and characterize variable stars within and around the clusters. The presence of such variable stars yields insights into the stellar population, ongoing dynamical evolution, and potential field contamination. This comprehensive approach enables us to assess the physical connection, dynamical state, and evolutionary history of the NGC 146–King 14 system within the Galactic disk.

The paper is organized as follows. Section~\ref{sec: dataset} provides a comprehensive description of the datasets employed in this study. Section~\ref{sec: Membership} outlines the methodology used to determine high-probability stellar membership within the clusters. In Section~\ref{sec: structral_parameter}, we derive the structural parameters of the clusters, including core and tidal radii. Section~\ref{sec: Age} presents the estimation of cluster distances from parallax measurements and the determination of their ages through theoretical isochrone fitting. Section~\ref{sec: dynamical_analysis} examines the dynamical evolution and orbital properties of the clusters. Section~\ref{sec: binarity} investigates the nature of the NGC 146–King 14 pair through their spatial separations and orbital analysis. In Section~\ref{sec: sed_variable}, we perform spectral energy distribution analysis and systematically identify and classify variable stars within the clusters. The key results and implications of our study are discussed in  Section~\ref{sec: Conclusion}.

\section{Data} {\label{sec: dataset}}

In this study, we investigate the neighboring OCs NGC~146 and King~14, which exhibit indications of a possible binary or co-moving nature. We combine astrometric, photometric, and time-series data from major surveys, including $Gaia$~DR3, WISE, Pan-STARRS1, 2MASS, and TESS, to conduct a comprehensive multi-wavelength analysis.  This integrated approach allows us to examine the clusters’ morphology, spatial distribution, and internal kinematics in detail. Additionally, high-precision TESS photometry enables the detection and characterization of variable stars in the vicinity of these clusters, offering further insights into their stellar content and dynamical evolution.

\begin{figure}[h]
\centering
\includegraphics[width=0.75\linewidth]{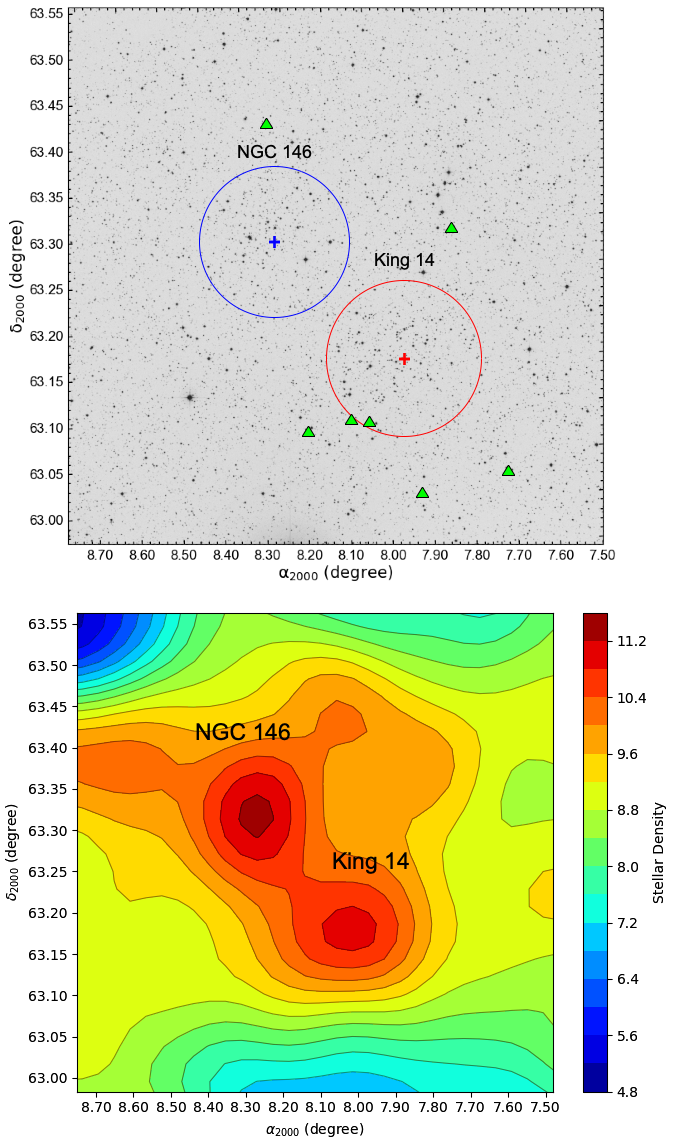}\\
\caption{{\textbf{Top:}} { Identification charts of the clusters NGC\,146 (blue circle) and King\,14 (red circle), based on Digitized Sky Survey (DSS) images. The circles indicate the approximate angular extent of each cluster, and the denser central regions correspond to their probable cores. The identified variable stars towards the cluster regions are marked as triangles. \textbf{Bottom:} Stellar surface density maps of the cluster pair, where the color scale represents the stellar density in stars per arcmin$^{2}$.}}
\label{id}
\end{figure}

Figure~\ref{id} displays the identification chart and the corresponding stellar surface density map for the region containing the OCs NGC~146 and King~14. 
The upper panel presents an image from the Digitized Sky Survey (DSS), a publicly available digital collection of astronomical photographs, covering a field of 30 arcminutes and highlighting the projected spatial configuration of the two clusters.  The circles mark the approximate boundaries of the cluster. The symbols represent the identified variable stars. This chart effectively shows the relative positions and orientations of NGC~146 and King~14 within the observed field. The lower panel illustrates the stellar surface density distribution derived from $Gaia$~DR3 data. The data reveal two distinct and prominent density enhancements corresponding to the centres of NGC~146 and King~14. 
The smooth transition of the density contours between the clusters suggests a possible spatial proximity or partial overlap in their projected distributions. This indicates that they may share a similar location in the Galactic plane. The DSS image was obtained from the STScI Digitized Sky Survey archive\footnote{\url{https://stdatu.stsci.edu/}}.

\subsection{Gaia DR3}

The Gaia mission, launched by the European Space Agency in 2013, operated until its observational phase ended in 2025. After spacecraft operations concluded, Gaia has continued to yield new data releases as the processing and calibration of the full mission dataset progress. Its goal is to construct a precise, extensive three-dimensional map of the Milky Way Galaxy. This map enables investigation of the intrinsic properties and kinematic behaviour of its stellar populations \citep{prusti2016gaia}. The third data release of Gaia provides astrometric measurements for about 1.46 billion celestial sources (\citet{GaiaDR3}, \citet{vallenari2023gaia}). These include positions ($\alpha, \delta$), trigonometric parallaxes, and proper motions ($\mu_{\alpha} \cos\delta, \mu_{\delta}$). Additionally, \textit{Gaia}~DR3 \citet{vallenari2023gaia} offers photometric observations in three broad passbands: the unfiltered $G$ band (white light), the $BP$ band (blue photometer), and the $RP$ band (red photometer) \citep{prusti2016gaia, brown2021gaia}. We applied several selection criteria to the collected data, requiring a Renormalised Unit Weight Error (RUWE) of $\leq$ 1.4 \citep{lindegren2021gaia}. In all cases, the RUWE values were $\leq$ 1.4, confirming the reliability of the Gaia astrometric solutions for the stars included in our analysis. The selected stars with trigonometric parallaxes of approximately $\varpi \approx 0.50$ mas, proper motions within $\pm0.5$ mas yr$^{-1}$, and $G$-band magnitudes of $\leq 20.1$, based on the mean values reported in the literature.

\begin{figure*}
\centering 
\vspace{-1cm}
\includegraphics[width=0.4\linewidth]{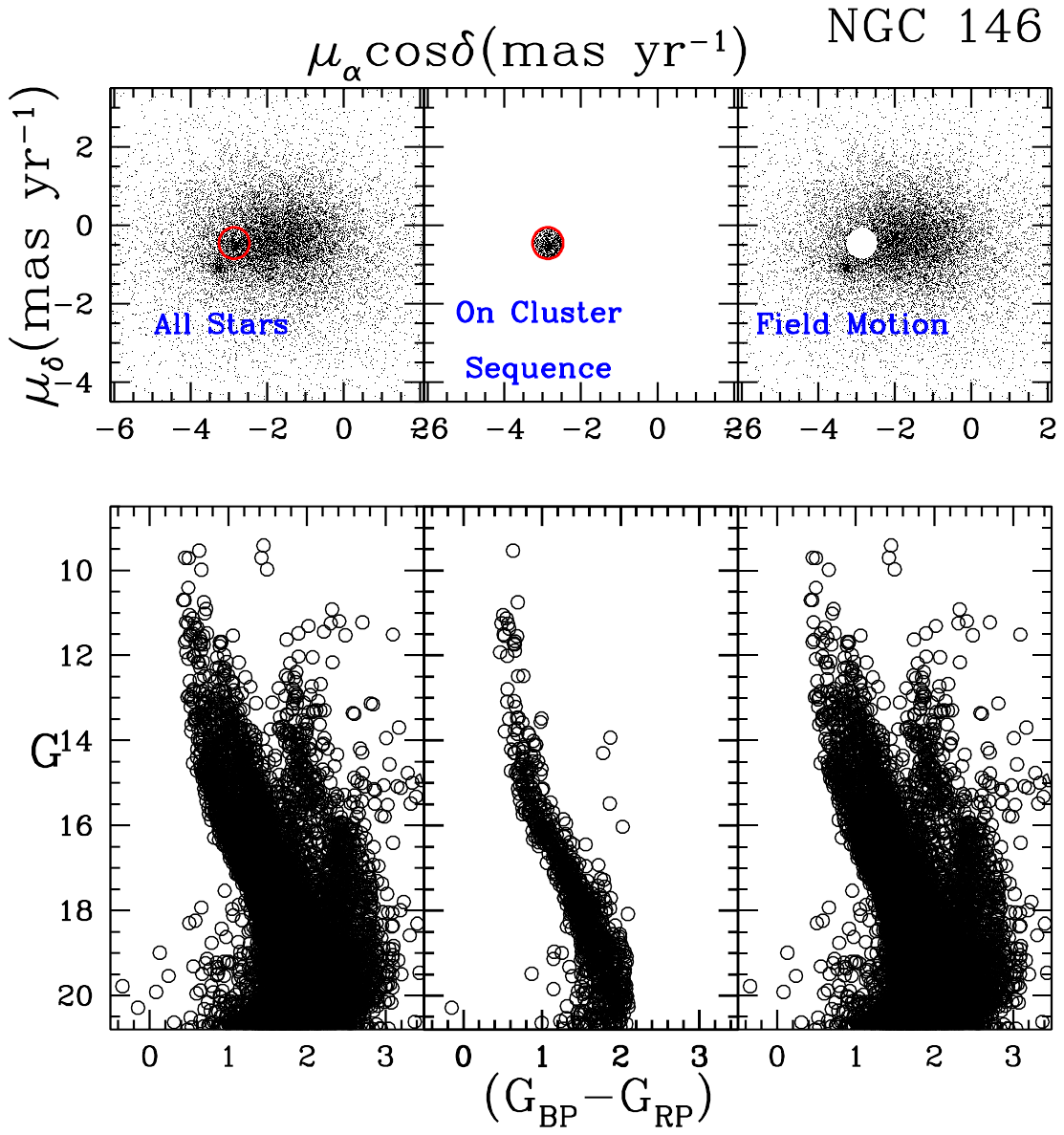}	
\includegraphics[width=0.4\linewidth]{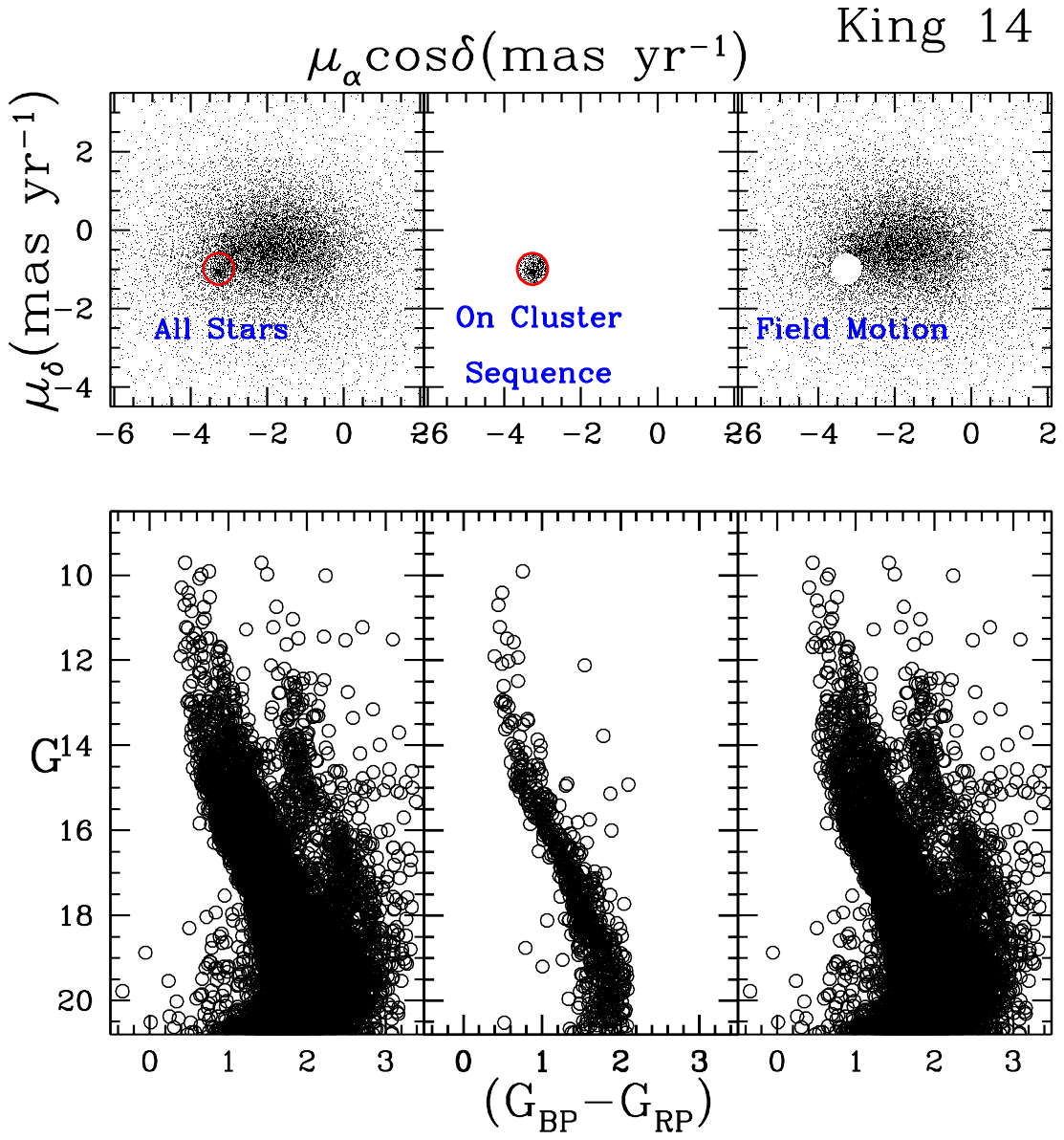}	
\vspace{-2cm}
\caption{\textbf{Top:} { Vector point diagrams (VPDs) for NGC\,146 (left group) and King\,14 (right group), showing the distribution of proper motions in right ascension ($\mu_{\alpha}\cos\delta$) and declination ($\mu_{\delta}$). For each cluster, the three panels correspond to all stars (left), stars lying on the cluster sequence (middle), and field stars (right). The red circles with radii of 0.5 and 0.6 mas\,yr$^{-1}$ for NGC\,146 and King\,14, respectively, indicate the proper-motion selection used to define probable cluster members. \textbf{Bottom:} Corresponding Gaia DR3 colour--magnitude diagrams, $G$ versus $(G_{\rm BP}-G_{\rm RP})$, for the same samples (all stars, probable members, and field stars) in the direction of NGC\,146 (left) and King\,14 (right).}}
	\label{fig: vpd}
\end{figure*}

\subsection{ALLWISE}

The Wide-field Infrared Survey Explorer data release (\textit{ALLWISE}) mission provides mid-infrared photometric measurements of celestial sources in four bands, with effective wavelengths centered at 3.35~$\mu$m ($W1$), 4.60~$\mu$m ($W2$), 11.56~$\mu$m ($W3$), and 22.09~$\mu$m ($W4$) \citep{wright2010wise}. In the present study, we extracted photometric data from the \textit{ALLWISE} source catalog for the regions surrounding the clusters NGC 146 and King 14. These mid-infrared observations are particularly valuable for probing the presence of circumstellar dust and detecting potential infrared excesses among cluster members, which may indicate ongoing or past disk activity.

\subsection{2MASS}
This study used 2MASS (Two Micron All-Sky Survey) data for these clusters. This data set has been collected via the two highly automated 1.3-m telescopes, one at Mt Hopkins, Arizona (AZ), USA, and the other at CTIO, Chile, with 3-channel cameras (256 $\times$ 256 array of HgCdTe detectors). The 2MASS database comprises photometric data in the near-infrared $J$, $H$, and $K$ bands, reaching limiting magnitudes of 15.8, 15.1, and 14.3, respectively. This data has a signal-to-noise ratio (S/N) greater than 10. We performed a cross-match of our dataset with 2MASS data using the Topcat\footnote[2]{\url{https://www.star.bris.ac.uk/~mbt/topcat/}} software.

\subsection{PANSTARS1}

The Panoramic Survey Telescope and Rapid Response System 1 \citep{hodapp2004design} provides photometric data in five broadband filters: $g$, $r$, $i$, $z$, and $y$, covering a wavelength range from approximately 400~nm to 1~$\mu$m  \citep{stubbs2010precise}. These filters have effective wavelengths of 481, 617, 752, 866, and 962 nm, respectively \citep{schlafly2012photometric}\citep{tonry2012pan}. The typical 5$\sigma$ limiting magnitudes for point sources in the $g$, $r$, $i$, $z$, and $y$ bands are 23.3, 23.2, 23.1, 22.3, and 21.4~mag, respectively \citep{chambers2016pan}.

\subsection{TESS}

The \textit{Transiting Exoplanet Survey Satellite} (TESS) is equipped with four 2K$\times$2K charge-coupled devices (CCDs), offering a combined field of view (FOV) of 24$^{\circ}$~$\times$~96$^{\circ}$ and an angular resolution of approximately $\sim21^{\prime\prime}\,\mathrm{pixel}^{-1}$
\citep{ricker2015transiting}.
 TESS observes the sky in segments called \textit{sectors}, each spanning two orbits (27 days), with the FOV shifting by 27$^{\circ}$ along the ecliptic between sectors. It performs continuous-time series photometry in the 600--1000~nm range (roughly equivalent to the Cousins $I$-band), suitable for detecting stellar variability. During the primary mission (Sectors~1--26), TESS provided full-frame images (FFIs) at a 30-minute cadence. In the first extended mission (Sectors~27--55), the cadence improved to 10 minutes, and from Sector~56 onward, short-cadence (200~s) target pixel files (TPFs) are available for selected sources. All data products are processed using the TESS Science Processing Operations Center (SPOC) pipeline\footnote{\url{https://heasarc.gsfc.nasa.gov/docs/tess/documentation.html}}. In this study, we utilized TESS photometric data from multiple sectors (17, 18, 54, 58, 78, and 85) to analyze stellar variability towards the clusters King 14 and NGC 146. All the {\it TESS} data used in this paper can be found in the Mikulski Archive for Space Telescopes (MAST) \citet{tess_mast_2021}.

\section{Kinematic Membership Analysis of stars} {\label{sec: Membership}}

NGC 146 and King 14 are young OCs situated near the Galactic plane, with Galactic latitudes of $0.504^{\circ}$ and $0.379^{\circ}$, respectively. Their positions in these low-latitude regions result in the observed stellar fields being heavily contaminated by foreground and background field stars. This contamination can significantly bias the determination of cluster parameters such as distance, reddening, age, and mass function. Therefore, obtaining a clean sample of probable cluster members is a critical step in our analysis. To determine cluster membership, we selected stars based on their locations in both the vector point diagram (VPD) and the color-magnitude diagrams (CMDs). Probable cluster members were identified as those lying within a circular region of radius 0.5 and 0.6~mas~yr$^{-1}$ for clusters NGC 146 and King 14, centred on the mean proper motion in the VPD. We initially adopted mean proper‐motion values for NGC 146 and King 14 from the literature and defined circular regions of radius 0.50 and 0.66 mas~yr$^{-1}$ around these values to extract preliminary candidate members. These radii reflect the observed spread in proper motions and represent the typical combination of intrinsic dispersion and Gaia DR3 uncertainties for stars in this magnitude range. Using this preliminary sample, we computed refined proper-motion centers, shown in Fig. \ref{fig: vpd}. Subsequently, we used them as input parameters for the maximum-likelihood membership analysis described below. Stars located outside this region were considered field contaminants. The top panels of Fig. \ref{fig: vpd} display the proper motion distribution: all stars (left), selected probable members (middle), and field stars (right). The bottom panels show the corresponding CMDs. The CMDs (middle) constructed from the selected members reveal a well-defined main sequence for both clusters. The VPDs are plotted using proper motions in right ascension ($\mu_{\alpha*} = \mu_{\alpha} \cos \delta$) and declination ($\mu_{\delta}$).

 We derived membership probabilities using the method of \citet{balaguer1998determination}. This approach modifies and generalizes the classical maximum-likelihood method from \citet{sanders1971improved}. The Sanders method was based on the two-component Gaussian model of  \citet{vasilevskis1958relative}. This method minimizes field contamination and has been successfully applied in several recent Gaia-based studies \citep{ bisht2020investigation, sariya2021astrometric, sariya2023gaia, panwar2024low, 2024AJ....167..188B,chand2025long}. For the cluster and field star distributions, two distinct frequency distribution functions, $\phi_c^{\nu}$ and $\phi_f^{\nu}$, are constructed for each $i^{\text{th}}$ star. These functions are defined as follows:

\begin{figure}
	\centering 
	\hbox{
 \includegraphics[width=0.5\linewidth]{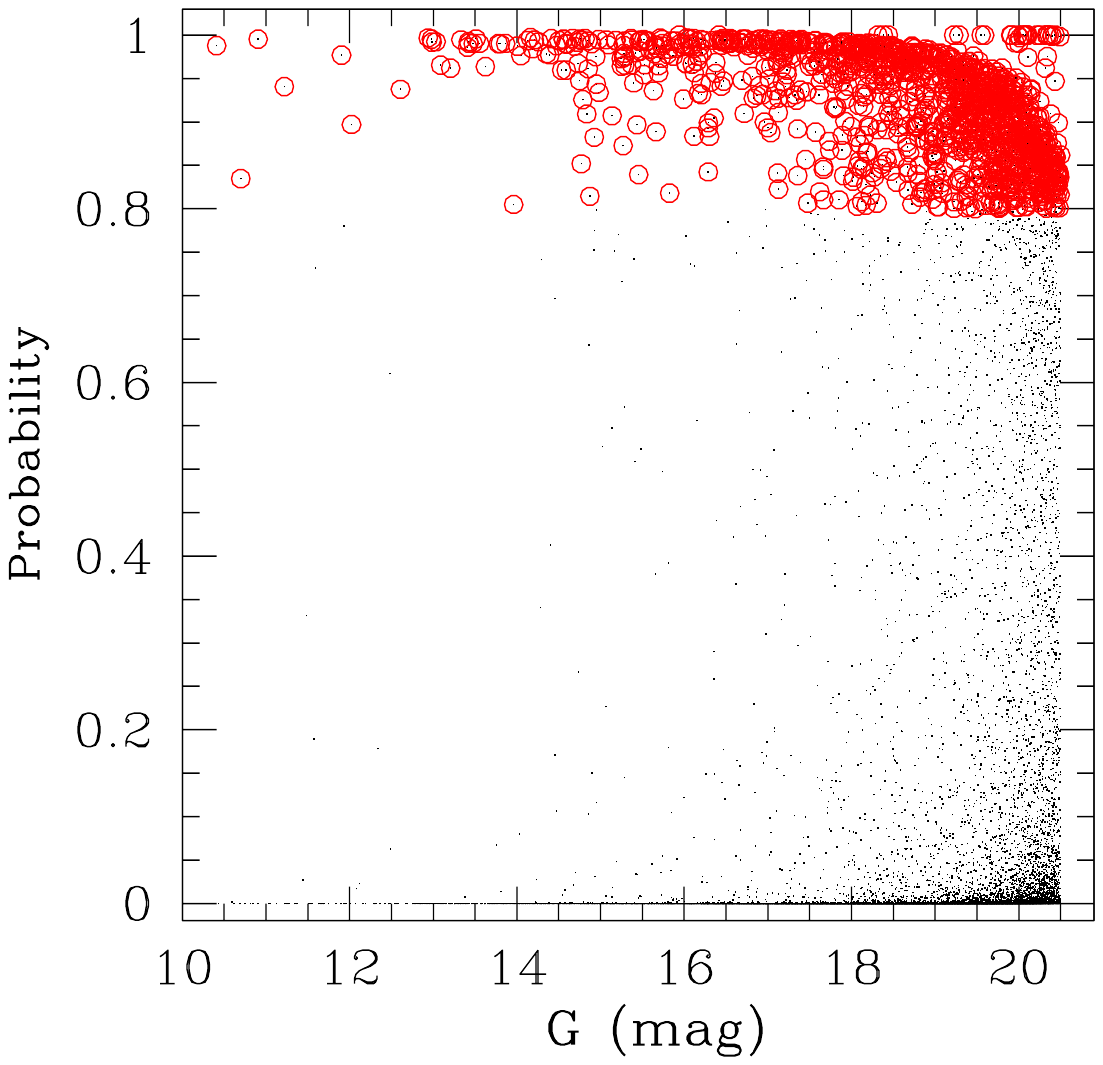}	
     \includegraphics[width=0.5\linewidth]{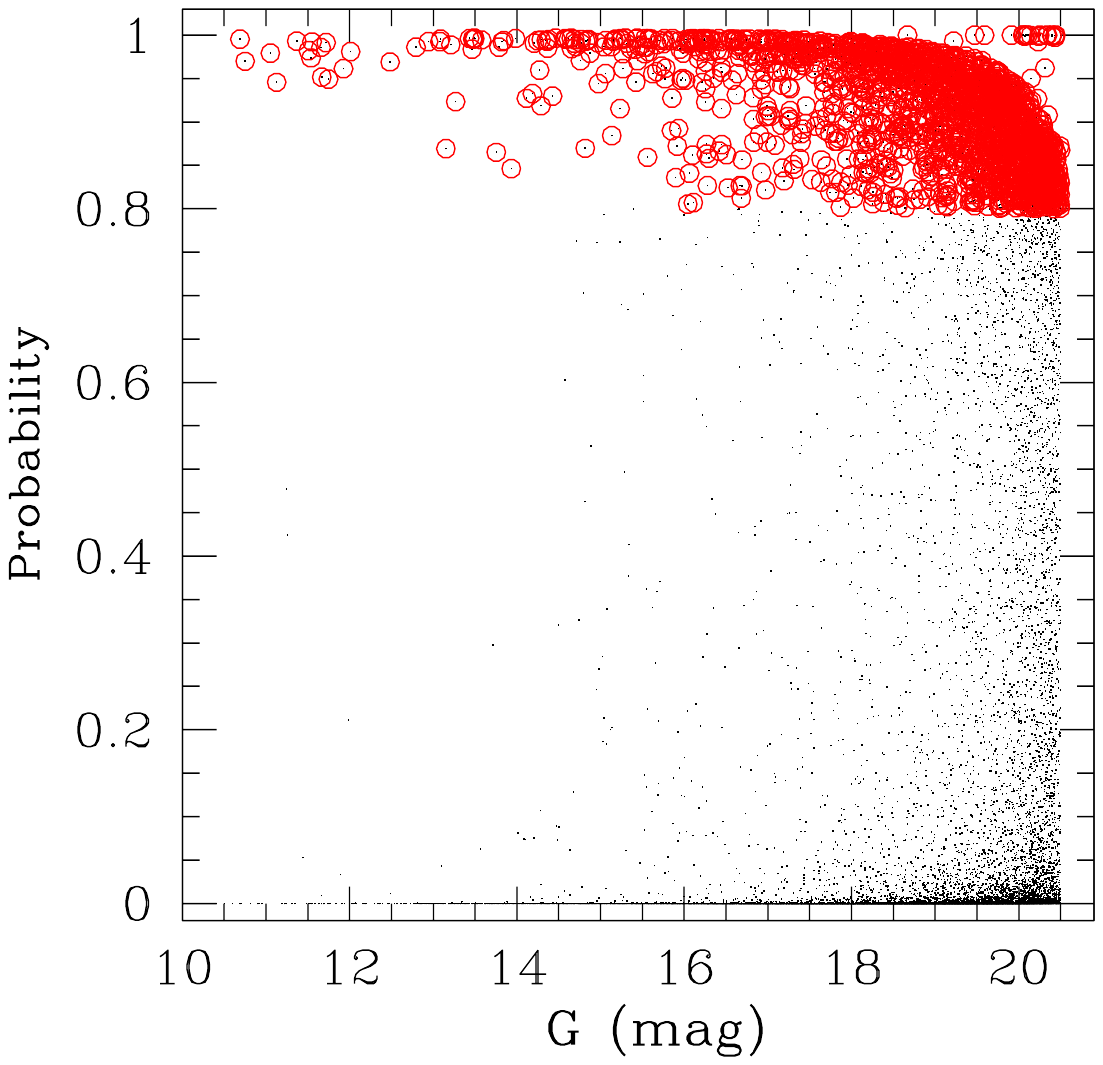}	
	}
\vspace{-1cm}
 \caption{Membership probability as a function of $Gaia$ $G$-band magnitude for stars in the regions of NGC 146 (left) and King 14 (right), derived using the method of \citet{balaguer1998determination}. Red circles indicate stars with membership probability greater than 80$\%$, considered as probable cluster members.}
	\label{fig: probability_magnitude}
\end{figure}

\begin{center}
   $\phi_c^{\nu} =\frac{1}{2\pi\sqrt{{(\sigma_c^2 + \epsilon_{xi}^2 )} {(\sigma_c^2 + \epsilon_{yi}^2 )}}}$

$\times$ exp$\{{ -\frac{1}{2}[\frac{(\mu_{xi} - \mu_{xc})^2}{\sigma_c^2 + \epsilon_{xi}^2 } + \frac{(\mu_{yi} - \mu_{yc})^2}{\sigma_c^2 + \epsilon_{yi}^2}] }\}$ \\
\end{center}
%\end{equation}
\begin{center}
and\\
\end{center}
%\begin{equation}
\begin{center}
$\phi_f^{\nu} =\frac{1}{2\pi\sqrt{(1-\gamma^2)}\sqrt{{ (\sigma_{xf}^2 + \epsilon_{xi}^2 )} {(\sigma_{yf}^2 + \epsilon_{yi}^2 )}}}$

$\times$ exp$\{{ -\frac{1}{2(1-\gamma^2)}[\frac{(\mu_{xi} - \mu_{xf})^2}{\sigma_{xf}^2 + \epsilon_{xi}^2}}
-\frac{2\gamma(\mu_{xi} - \mu_{xf})(\mu_{yi} - \mu_{yf})} {\sqrt{(\sigma_{xf}^2 + \epsilon_{xi}^2 ) (\sigma_{yf}^2 + \epsilon_{yi}^2 )}} + \frac{(\mu_{yi} - \mu_{yf})^2}{\sigma_{yf}^2 + \epsilon_{yi}^2}]\}$\\
\end{center}
where ($\mu_{xi}$, $\mu_{yi}$) are the PMs of the star $i^{th}$.  PM errors are represented by ($\epsilon_{xi}$, $\epsilon_{yi}$). The PM center of the cluster is given by ($\mu_{xc}$, $\mu_{yc}$), while ($\mu_{xf}$, $\mu_{yf}$) represent the center of the field PM values. The intrinsic PM dispersion for cluster stars is denoted by $\sigma_c$, and $\sigma_{xf}$ and $\sigma_{yf}$ provide the intrinsic PM dispersions for the field populations. The correlation coefficient $\gamma$ is calculated as:

%\begin{equation}
\begin{center}
$\gamma = \frac{(\mu_{xi} - \mu_{xf})(\mu_{yi} - \mu_{yf})}{\sigma_{xf}\sigma_{yf}}$.
\end{center}

Stars with proper motion (PM) errors of $\le$0.5 mas yr$^{-1}$ have been used to determine $\phi_c^{\nu}$ and $\phi_f^{\nu}$. A group of stars is found at $\mu_{xc}$ = $-$2.86 mas yr$^{-1}$, $\mu_{yc}$ = $-$0.45 mas yr$^{-1}$ for NGC 146 and $\mu_{xc}$ = $-$3.27 mas yr$^{-1}$, $\mu_{yc}$ = $-$0.98 mas yr$^{-1}$ for King 14. Assuming distances of 2.76 and 2.36 kpc  from \citet{Hunt2024} for clusters and radial velocity dispersion of 1 km s$^{-1}$ for open star clusters \citep{girard1989relative}, the expected dispersion ($\sigma_c$) in PMs would be approximately 0.08 mas yr$^{-1}$ for both the clusters. Observations indicate that OCs typically exhibit internal velocity dispersions of 0.5–2~km~s$^{-1}$ \citep{geller2010wiyn}, which motivates the adopted value in this analysis. The membership probabilities remain relatively stable for velocity dispersions within this range; however, adopting higher values tends to increase the level of field-star contamination \citep{belwal2025unveiling}. All cluster and field proper-motion distribution parameters were derived using the maximum-likelihood method of \citet{balaguer1998determination}, which optimizes the two-component Gaussian model to the Gaia DR3 data. For field region stars, we have estimated ($\mu_{xf}$, $\mu_{yf}$) = ($-$1.8, 0.4) mas yr$^{-1}$ for NGC 146 and ($\mu_{xf}$, $\mu_{yf}$) = ($-$3.2, $-$1.8) mas yr$^{-1}$ for King 14, with ($\sigma_{xf}$, $\sigma_{yf}$) = (3.5, 3.9) and (4.2, 3.4) mas yr$^{-1}$ for both the clusters respectively.

Considering the normalized numbers of cluster stars and field stars as $n_{c}$ and $n_{f}$ respectively (i.e., $n_c + n_f = 1$), the total distribution function can be calculated as:
\begin{center}
$\phi = (n_{c}~\times~\phi_c^{\nu}) + (n_f~\times~\phi_f^{\nu})$,  \\
\end{center}
As a result, the membership probability for the $i^{th}$ star is given by:
\begin{center}
$P_{\mu}(i) = \frac{\phi_{c}(i)}{\phi(i)}$. 
\end{center}

Using this method, we identified 770 and 690 stars as cluster members for NGC 146 and King 14, respectively, with membership probabilities higher than $80\%$ and $G \leq 20$ mag. In Figure \ref{fig: probability_magnitude}, we plot the likelihood of membership versus magnitude $G$ for both clusters. The \citet{cantat2018gaia} catalog provides membership probabilities for all clusters studied. We matched our likely members with this catalog and identified 138 stars for NGC 146 and 153 stars for King 14, which are common. In the CMDs (see Figure \ref{fig:cmd}), blue dots represent the matched stars, while black dots indicate those with membership probabilities higher than $80\%$. We adopted a membership probability threshold of $80\%$ to minimize contamination, particularly at the fainter end of the cluster sequence. Several threshold values were tested, and we found that adopting lower probabilities led to increased contamination from field stars in the faint region of the CMD. Thus, a threshold of $80\%$ was adopted to achieve an optimal compromise between member-star and field-star contamination. The mean proper motions derived for these clusters are $(\mu_{\alpha}\cos\delta, \mu_{\delta}) = (-2.81 \pm 0.14, -0.47 \pm 0.21)$ mas yr$^{-1}$ for NGC 146 and $(-3.23 \pm 0.18, -0.97 \pm 0.21)$ mas yr$^{-1}$ for King 14. These values are in good agreement with the measurements reported by \citet{dias2021updated}, validating our membership determination and supporting the kinematic coherence of the NGC 146–King 14 pair.
 
The close agreement in proper motions, parallaxes, and spatial positions of both clusters provides strong evidence for a common bulk motion through the Galaxy. Their kinematic coherence, along with similar ages and distances, strongly indicates a co-moving origin. These results suggest that both clusters may represent a physically associated pair formed within the same giant molecular cloud complex. The detailed assessment of their relative separation, orbital parameters, and dynamical binding is discussed in the following sections to examine whether they constitute a primordial binary cluster system or a dynamically unbound co-moving pair within the Galactic disk.

\section{Physical Characterization of OCs}
{\label{sec: structral_parameter}}
\subsection{Radial Density Profile}

The center coordinates of each cluster, NGC~146 (RA = 8.262, Dec = 63.293) degrees and King~14 (RA = 7.965, Dec = 63.162) degrees, were determined as the positions corresponding to the maximum stellar surface density. These locations were identified from a two-dimensional stellar density map, constructed by performing uniform binning in right ascension and declination. The derived central coordinates are in good agreement with those reported by \citet{dias2021updated} and are consistent with the positional uncertainties provided in the catalogue of \citet{cantat2018gaia}.

To analyze the stellar distribution, we constructed radial density profiles (RDPs) for NGC 146 and King 14 by dividing the cluster region into concentric annular rings of width 1.0 arcmin, centered on the estimated cluster centers. The stellar surface density in the $i^{\text{th}}$ ring, $\rho_i$, was calculated using the standard relation:
\[
\rho_i = \frac{N_i}{A_i},
\]
where $N_i$ is the number of stars within the $i^{\text{th}}$ annulus, and $A_i = \pi \left(r_{i+1}^2 - r_i^2\right)$ is its area.

\begin{figure}
\centering 
\includegraphics[width=0.49\linewidth]{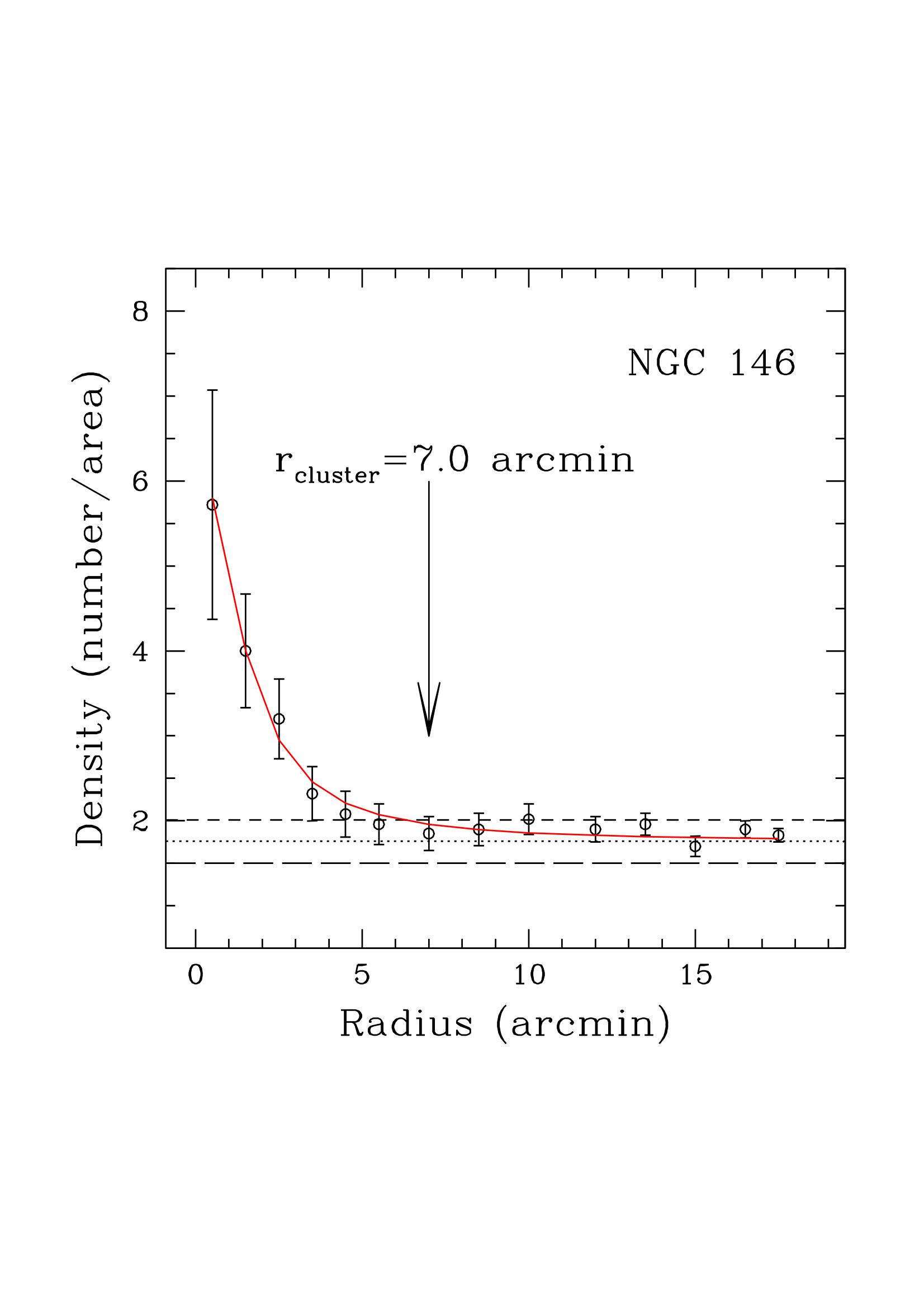}	\includegraphics[width=0.49\linewidth]{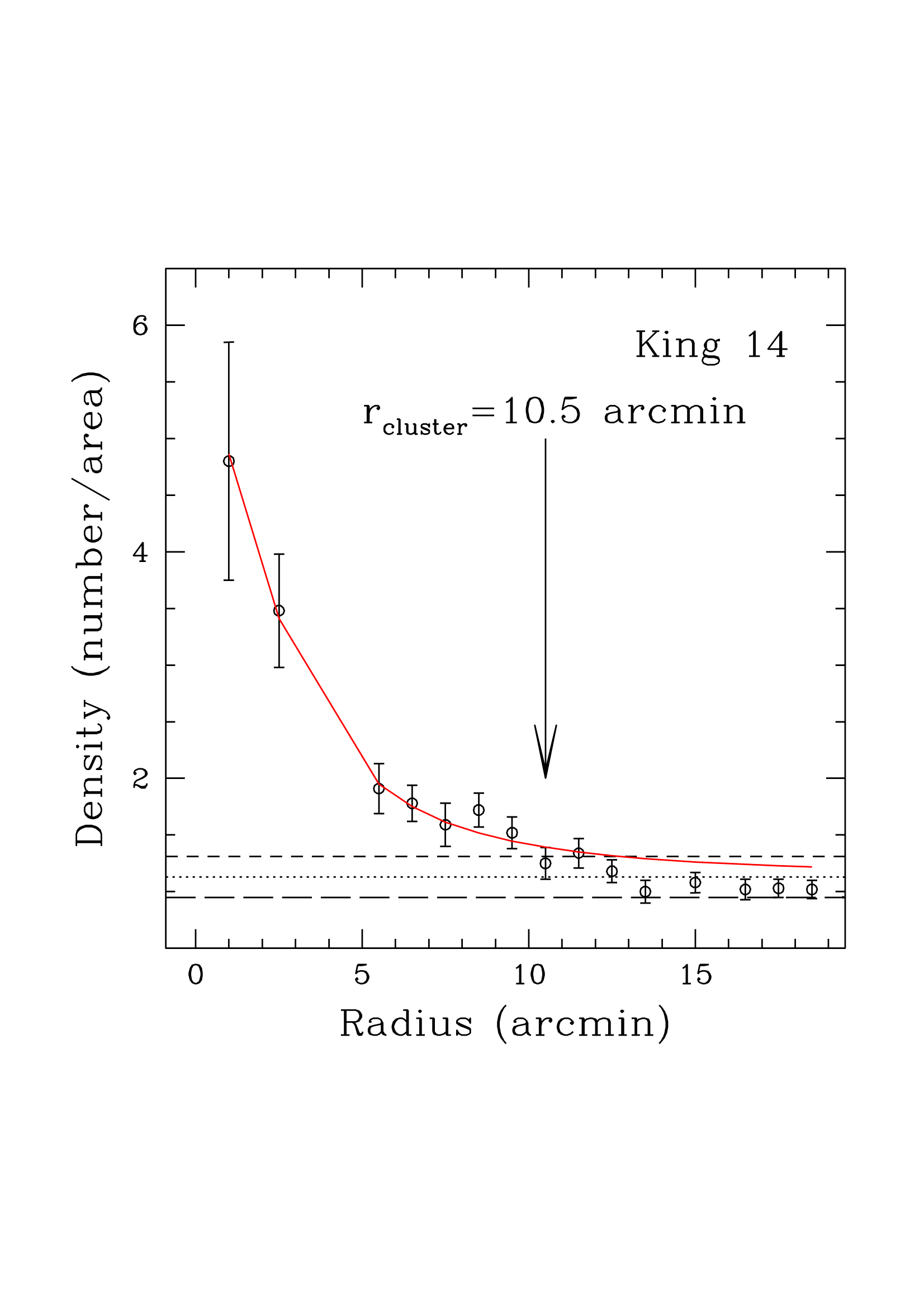}	
\vspace{-1.2cm}
 \caption{RDPs of NGC~146 (left) and King~14 (right), constructed using the surface stellar density in concentric annuli centered on each cluster. The observed data points (black dots) represent the stellar density, with vertical error bars indicating Poisson uncertainties. The solid red curves show the best-fit \citet{king1962structure} model profiles. The horizontal dashed lines represent the estimated background field star density, while the arrows mark the adopted cluster radii ($r_{\mathrm{cluster}}$) of 7.0 arcmin for NGC~146 and 10.5 arcmin for King~14. These fits enable the determination of structural parameters such as core radius, background density, and limiting radius.}
	\label{fig: rdp}
\end{figure}

To further characterize the structural properties of the clusters, we fitted the observed RDPs using the modified empirical King profile \citep{king1962structure}:
\begin{equation}
    f(r) = f_{0} \left[\frac{1}{\sqrt{1 + (r/r_{c})^{2}}} - \frac{1}{\sqrt{1 + (r_{t}/r_{c})^{2}}}\right]^2 + f_{b},
\end{equation}
where $f(r)$ is the stellar surface density at radius $r$, $f_0$ is the central density above the background level, $f_b$ is the background field star density, $r_c$ is the core radius, and $r_t$ is the tidal radius of the cluster. All estimated structural parameters are listed in Table \ref{tab: strutral_parameters}. The radius of the cluster was chosen as the radial distance at which the stellar surface density flattens and merges with the background field star density. Based on this criterion, we found $r_{\text{lim}} \sim 7.0^{\prime}$ for NGC 146 and $r_{\text{lim}} \sim 10.5^{\prime}$ for King~14, as shown in Fig.~\ref{fig: rdp}. Beyond these radii, the stellar density remains nearly constant, demonstrating the change from cluster to field-dominated regions. In addition to the cluster radius, this model provided a good fit to the observed profiles, yielding estimates of the core radius, tidal radius, and central density that offer insights into the dynamical state and spatial extent of the clusters under study.

\begin{table}
\centering
\caption{Structural parameters of the clusters under study. Background and central densities are in units of stars per arcmin$^2$. Core radius ($r_c$) and limiting radius ($r_{\text{lim}}$) are given in arcminutes and parsecs. $r_t$ is tidal radius in pc.}
\small
\begin{tabular}{lcc}
\hline\hline
\textbf{Parameter} & \textbf{NGC 146} & \textbf{King 14} \\
\hline
$f_0$ (stars/arcmin$^2$)      & 19.28 & 12.17 \\
$f_b$ (stars/arcmin$^2$)      & 6.20  & 3.20  \\
$r_c$ (arcmin)                & 0.93  & 2.20  \\
$r_c$ (pc)                    & 0.66  & 1.70  \\
$r_t$ (pc)                    & 11.0   & 14.0
\\
$\delta_c$                    & 4.1   & 4.8   \\
$r_{\text{lim}}$ (arcmin)     & 7.0   & 10.5   \\
$c$                           & 0.58  & 0.47  \\
\hline
\end{tabular}
\label{tab: strutral_parameters}
\end{table}

According to \citet{maciejewski2007ccd}, the limiting radius ($r_{\text{lim}}$) of an open cluster typically lies in the range of approximately $2r_c$ to $7r_c$, depending on the dynamical state and external tidal effects acting on the cluster. For NGC~146 and King~14, our derived values of $r_{\text{lim}}$ and $r_c$ yield $r_{\text{lim}}/r_c$ ratios of $\sim 7.53$ and $\sim 4.78 $, respectively, which fall well within this expected range. This consistency indicates that the structural parameters obtained from our King model fit well with the general properties of OCs as reported in the literature.

To further assess the structural concentration of the clusters, we computed the density contrast parameter $\delta_c$, defined as:
\[
\delta_c = 1 + \frac{f_0}{f_b},
\]
The parameter $\delta_c$ provides a measure of how prominently the cluster stands out against the field. The resulting $\delta_c$ values are found to be $\sim$4.1 for NGC~146 and $\sim$4.8 for King 14. These values are somewhat lower than the contrast range ($7 \leq \delta_c \leq 23$) reported for compact, rich clusters by \citet{bonatto2007open}, suggesting that both clusters are relatively loose or moderately concentrated. Such low contrast values are typical of young or dynamically evolving clusters in dense Galactic fields.

The derived structural parameters provide valuable insights into the internal morphology and dynamical state of the clusters. NGC 146 has a relatively small core radius ($r_{\mathrm{c}} = 0.66$ pc), with most stars concentrated toward the center, indicating a higher degree of central concentration than King 14. King 14 exhibits a larger core radius ($r_{\mathrm{c}}$ = 1.70 pc) compared to NGC 146. This difference may indicate a more dynamically evolved structure, but it may also arise from differences in the initial size of the parent molecular cloud. A larger core radius alone does not demonstrate physical expansion. Detecting true dynamical expansion would require observing systematic radial gradients in the internal proper-motion field; however, the Gaia DR3 proper-motion uncertainties at the distance of these clusters are too large to reveal such subtle signatures. Thus, while we cannot assert greater dynamical evolution with certainty, the larger core radius provides a strong indication consistent with that scenario. Both clusters exhibit modest density-contrast parameters ($\delta_c \sim 4$–5), typical of young OCs in dense Galactic environments, where tidal interactions and external perturbations play significant roles in shaping their evolution. The RDPs of both clusters display extended outer regions and gradual declines in stellar density, suggesting possible tidal distortion or halo overlap due to interactions with the Galactic gravitational field or with each other. Based on the King model fits, the estimated tidal radii are $\sim$11 pc for NGC 146 and $\sim$14 pc for King 14, indicating that both clusters are well confined within their tidal boundaries. The larger tidal radius of King 14, consistent with its older age and higher mass, implies it has undergone a longer period of dynamical evolution under Galactic tidal influence. The $r_t/r_c$ ratios of $\sim17$ and $\sim8$ for NGC 146 and King 14, respectively, are typical of moderately concentrated young clusters. The comparable structural morphology and overlapping spatial extents of the two systems support the view that NGC 146 and King 14 likely originated within the same molecular complex and are presently evolving as a co-moving, dynamically unbound pair within the Galactic disk.

\subsection{Extinction Law from Optical to Mid-Infrared Photometry}

\begin{figure*}
\centering 
\includegraphics[width=0.40\textwidth,height = 14cm]{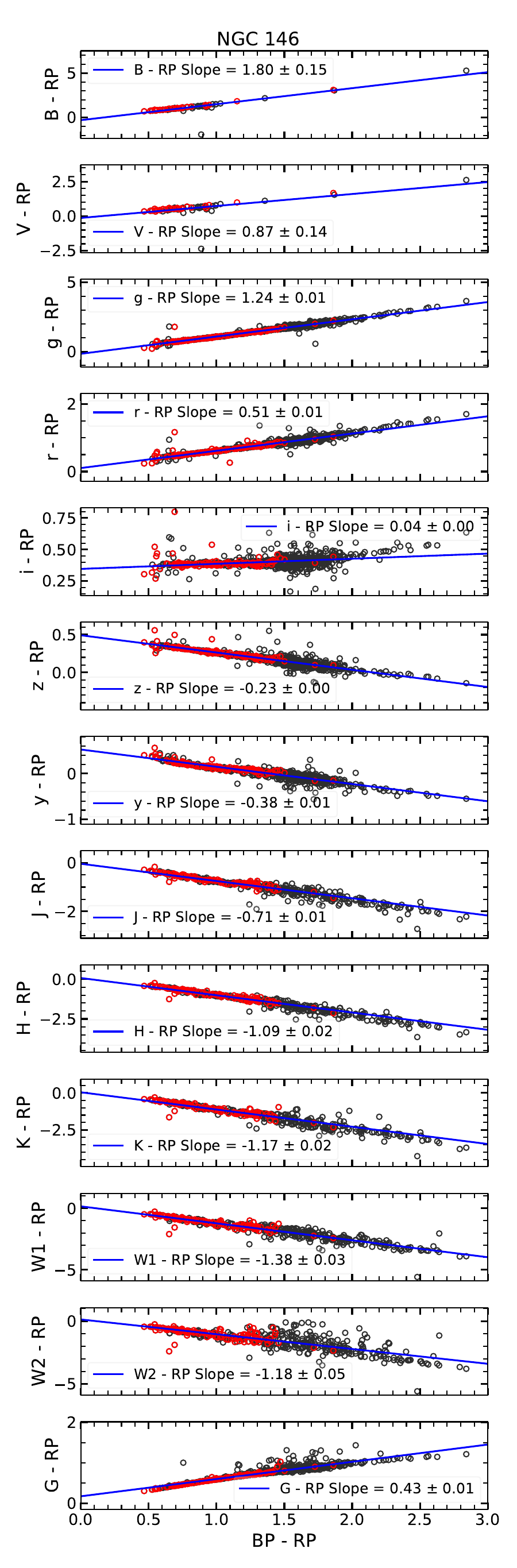}	
\includegraphics[width=0.40\textwidth,,height = 14cm]{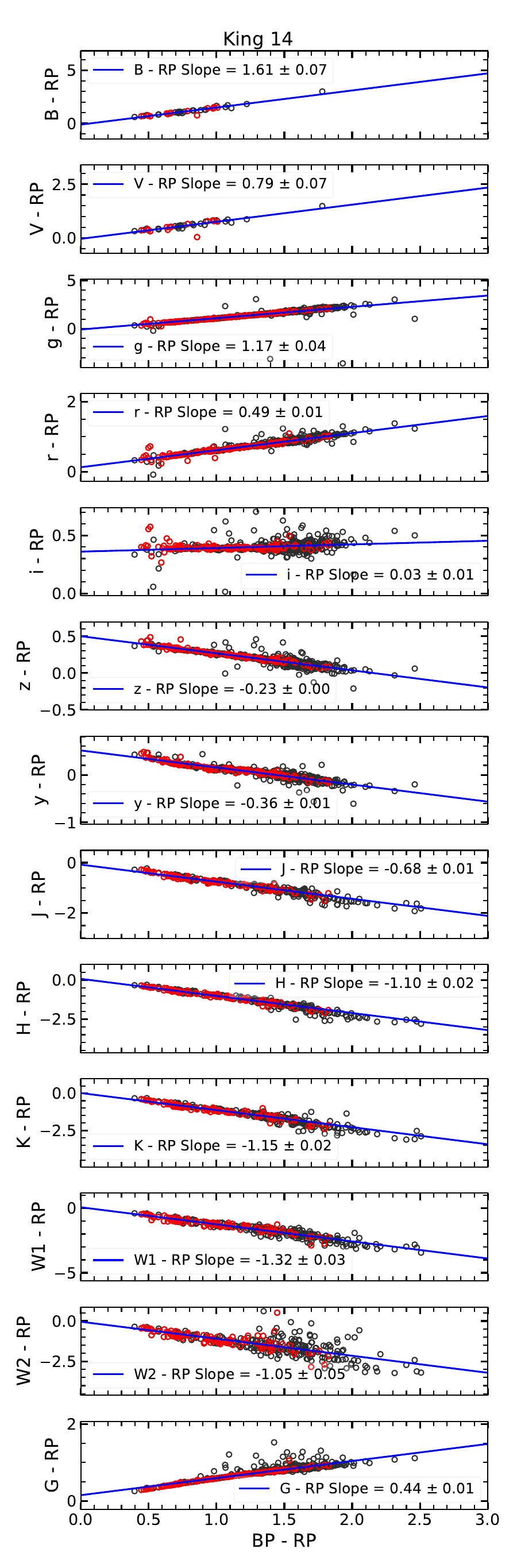}	
\caption{Two-color diagrams (TCDs) for the clusters NGC 146 (left panels) and King 14 (right panels), constructed using combinations of optical to near-infrared photometric bands (e.g., $B$, $V$, $g$, $r$, $i$, $z$, $y$, $J$, $H$, $K$, $W1$, $W2$, $G$) against $Gaia$ $G_{\rm RP}$ magnitudes. The diagrams illustrate the photometric transformations and inter-band correlations, with linear fits overplotted in each panel. These TCDs are employed to probe the interstellar extinction law across different wavelengths and help in constraining the total-to-selective extinction ratio ($R_{\rm V}$), offering insights into the dust properties along the line of sight toward the clusters.}
	\label{fig:tcd_plot}
\end{figure*}

In this section, we use all the datasets to examine the extinction law from the optical to the mid-infrared region towards the clusters under study. The resulting two-color diagrams (TCDs), plotted as $(\lambda - G_{\rm RP})/( G_{\rm BP} - G_{\rm RP})$, are presented in Fig.~\ref{fig:tcd_plot} for all clusters. Here, $\lambda$ represents the filters other than $G_{\rm RP}$. All stars shown in Fig.~\ref{fig:tcd_plot} plotted in black represent the probable cluster members, while the red open circles indicate the matched stars from \citet{cantat2018gaia}. A linear fit was applied to the data points, and the resulting slopes are listed in Table \ref{tab: color_color_plot}. The estimated slope values are in good agreement with those reported by \citet{wang2019optical} and \citet{bisht2020comprehensive}. We calculated the ratio $\frac{A_V}{E(B-V)}$ as 3.02 for NGC 146 and 3.03 for King 14. These values are very close to the standard value of 3.1, indicating that the reddening law towards the regions of both clusters is consistent with the normal interstellar extinction law. The close agreement in the extinction law and reddening values shows that both clusters are affected by a similar interstellar medium. This similarity suggests that they are at comparable distances along the same line of sight and likely share the same intervening dust layer. The following sections examine their spatial separation and kinematics for further insight into their possible association.

\begin{table}
\centering
\footnotesize
\caption{Derived color-excess ratios $(\lambda - G_{\mathrm{RP}}) / (G_{\mathrm{BP}} - G_{\mathrm{RP}})$ for various photometric filters along the line of sight toward the OCs NGC~146 and King~14. These ratios were obtained using multiwavelength photometric data from \textit{Johnson}, \textit{Pan-STARRS}, \textit{2MASS}, \textit{ALLWISE}, and \textit{Gaia} passbands to characterize the wavelength dependence of interstellar reddening in the direction of the clusters.}
\begin{tabular}{cccc}
\hline\hline
Filter ($\lambda$)  & Wavelength   &  \multicolumn{2}{c}{$\frac{\lambda-G_{\rm RP}}{G_{\rm BP}-G_{\rm RP}}$} \\
                         &     [nm]               &  NGC 146   & King 14   \\
    \hline
    Johnson B      &   445     &      1.796 $\pm$ 0.154      &    1.610  $\pm$  0.068   \\  
    Johnson V      &   551     &      0.867 $\pm$ 0.144      &    0.795  $\pm$  0.067   \\  
    Pan-STARRS g   &   481     &     1.243  $\pm$  0.009     &    1.170  $\pm$  0.041  \\   
    Pan-STARRS r   &   617     &     0.511  $\pm$  0.006     &    0.490  $\pm$  0.007  \\ 
    Pan-STARRS i   &   752     &     0.045  $\pm$  0.004     &    0.035  $\pm$  0.005  \\ 
    Pan-STARRS z   &   866     &     $-$0.228 $\pm$  0.004     &   $-$0.235  $\pm$  0.004   \\ 
    Pan-STARRS y   &   962     &     $-$0.378 $\pm$  0.006     &   $-$0.360  $\pm$  0.006   \\ 
    2MASS J        &   1234.5  &     $-0.708$ $\pm$  0.013     &   $-$0.679  $\pm$  0.017   \\ 
    2MASS H        &   1639.3  &     $-1.085$ $\pm$  0.016     &   $-$1.095  $\pm$  0.018   \\ 
    2MASS K        &   2175.7  &     $-1.167$ $\pm$  0.022     &   $-$1.149  $\pm$  0.025   \\ 
    ALLWISE W1        &   3317.2  &     $-1.378$ $\pm$  0.025     &   $-$1.322  $\pm$  0.025   \\ 
    ALLWISE W2        &   4550.1  &     $-$1.176 $\pm$  0.046     &   $-$1.053  $\pm$  0.054   \\ 
    $Gaia$ $G$         &   641.9   &      0.428 $\pm$  0.006     &   0.443   $\pm$  0.009    \\ 
    \hline
    \end{tabular}
    \label{tab: color_color_plot}
\end{table}

\section{Fundamental Parameter Estimation of the Cluster Pair} {\label{sec: Age}}

\subsection{Distance of clusters from trigonometric parallax}
To calculate the distance between the clusters under study, we first selected the most probable members with a probability greater than 80$\%$. Next, we fitted a Gaussian curve to the parallax histogram to determine the mean trigonometric parallax, excluding stars with negative parallax.
We obtained mean parallax values of 0.320 mas for NGC 146 and 0.380 mas for King 14. Following this, we applied a zero-point offset of $-0.021$ mas to the mean trigonometric parallax, as recommended by \citet{groenewegen2021parallax}. We estimated the cluster distances by directly inverting the trigonometric parallaxes, which may be inaccurate. This approach yields values of 3.125 $\pm$ 0.450 kpc for NGC 146 and 2.632 $\pm$ 0.310 kpc for King 14, respectively.
We followed the method of \citet{bailer2015estimating}, which shows that a probabilistic approach yields more precise estimates, particularly when accounting for trigonometric parallax errors. Building on this insight, they proposed a Bayesian approach that incorporates both the measured trigonometric parallax and its uncertainty. We therefore adopted the methodology from \citet{bailer2018estimating} to measure the distance. A comprehensive description of the methods we used is provided in \citet{belwal2025unveiling}. Consequently, the estimated distances, $2.98 \pm 0.33$ kpc and $2.51 \pm 0.23$ kpc, are in good agreement with those reported by \citet{cantat2018gaia, Hunt2024}. We used these distance values for further analysis of the cluster.

Determining the age and distance of clusters is vital to understanding their evolution, dynamics, and role in the Milky Way’s structure. In this study, we utilized multi-band photometric data from $Gaia$, $WISE$, and $Pan-STARRS$ to construct precise color-magnitude diagrams (CMDs) for NGC 146 and King 14. We include the near-infrared (J, J–H) CMD to provide an independent consistency check using 2MASS photometry. However, the isochrones reproduce only the bright part of the NIR main sequence. For J $ \geq$ 13–14 mag, 2MASS photometry becomes increasingly incomplete, and photometric uncertainties multiply, leading to a broadened, poorly defined lower main sequence. In addition, for young clusters (20–50 Myr), the low-mass main sequence is nearly vertical in (J–H), limiting the diagnostic power of faint NIR data. Therefore, we primarily rely on $Gaia$ CMDs to determine the cluster parameters and use the 2MASS CMD only as a complementary check at the bright end. To minimize field star contamination and ensure reliable cluster membership, we applied selection criteria based on proper motion and trigonometric parallax. This astrometric filtering allowed us to identify probable members with high confidence. We performed isochrone fitting using PARSEC models by \citet{marigo2017new} at solar metallicity ($Z=0.02$), employing photometric sets: $(G, G_{BP} - G_{RP})$ from $Gaia$, $(J, J-H)$ from 2MASS, and $(g, g-i)$ from Pan-STARRS. Since the colour-excess ratios (Table \ref{tab: color_color_plot}) indicate that the clusters follow a normal extinction law (Rv $\approx$ 3.1), we adopt standard extinction coefficients for all passbands and derive the absolute reddening directly from CMD isochrone fitting. Parameters from isochrone fitting for each cluster are detailed below:

\subsection*{NGC 146}

The CMDs for NGC 146 (see Fig.~\ref{fig:cmd}) exhibit a well-defined and tight main sequence extending down to fainter magnitudes, indicative of a relatively young stellar population. Isochrone fitting was performed for $\log(\text{age})$ values between 7.2 and 7.4. The best global fit was achieved for $\log(\text{age}) = 7.3$, corresponding to an age of $20 \pm 5$ Myr. The cluster parameters were estimated via isochrone fitting using the chi-square minimization technique, yielding a reduced chi-square value of 0.85. The apparent distance modulus was estimated to be $(m-M) = 13.90 \pm 0.24$ mag, yielding a heliocentric distance of $2.87 \pm 0.46$ kpc.  This distance is consistent with the value reported by \citet{cantat2018gaia} and agrees well with the parallax-based distance listed in Table~\ref{tab:fundamental_params}. The young age and compact structure of NGC 146 suggest that the cluster is dynamically young and possibly still undergoing early evolutionary processes.

\subsection*{King 14}

\begin{figure*}
	\centering 
    \vspace{-0.6cm}
    \includegraphics[width=13.5cm,height=6.0cm]{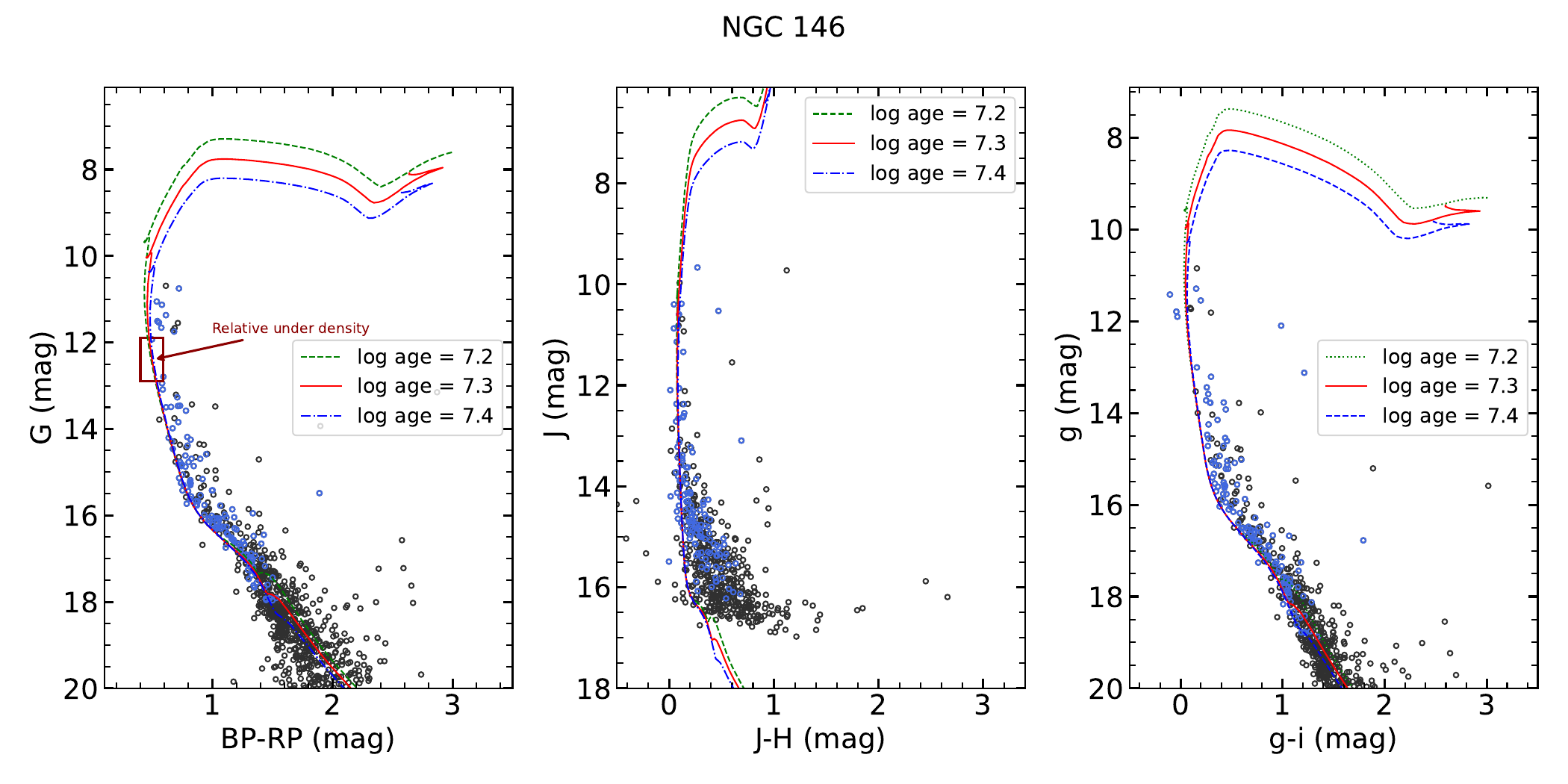}	
    \includegraphics[width=13.5cm,height=6.0cm]{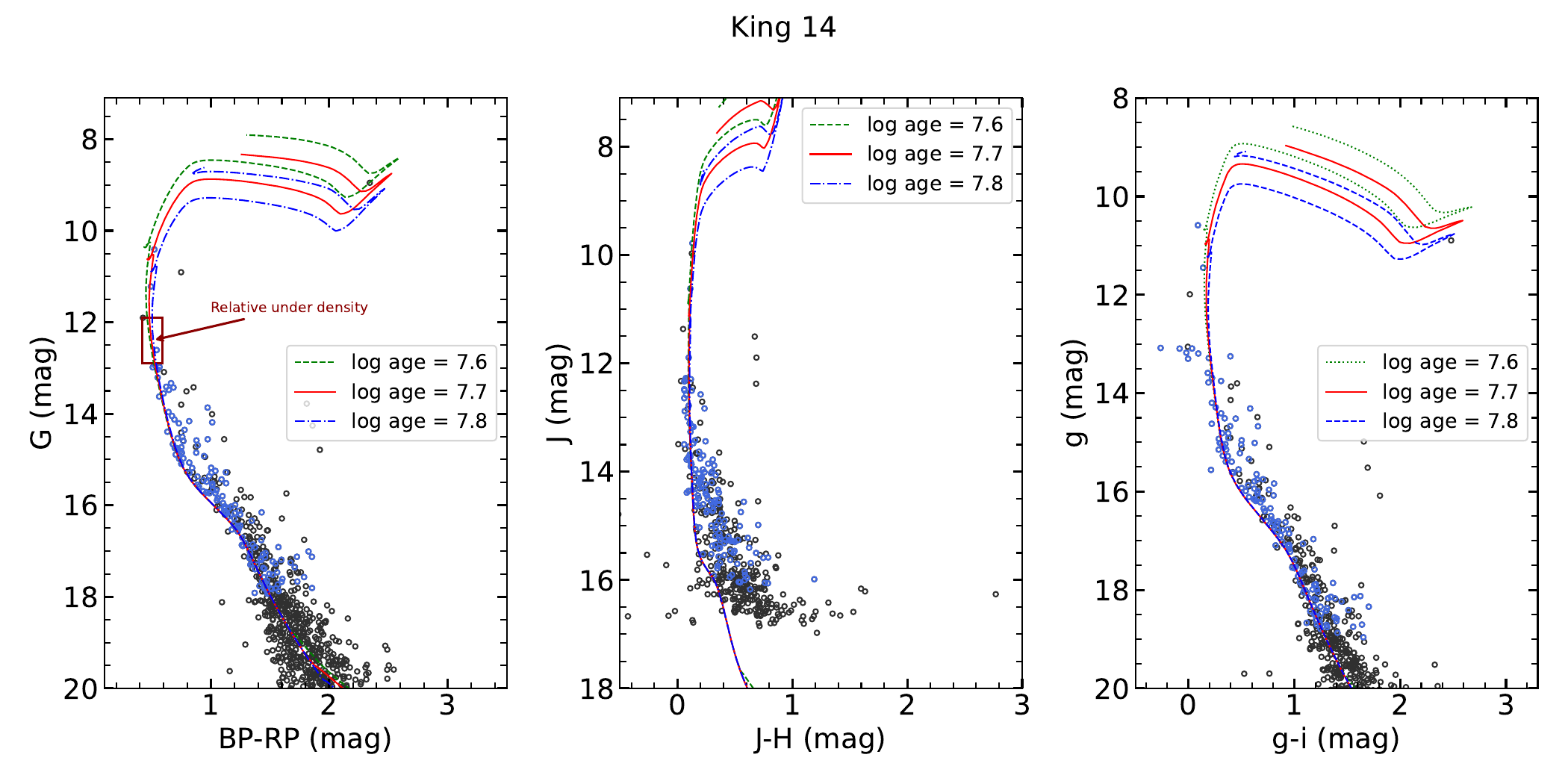}
    \caption{Color-Magnitude Diagrams (CMDs) for the OCs NGC 146 (top panels) and King 14 (bottom panels) in three different photometric systems: $Gaia$ ($G$ vs. $G_{\rm BP} - G_{\rm RP}$), 2MASS ($J$ vs. $J-H$), and Pan-STARRS ($g$ vs. $g-i$). Black points represent probable cluster members identified based on our membership criteria, while blue points indicate stars that are cross-matched with the \citet {cantat2020painting} catalog. Overplotted are PARSEC theoretical isochrones log(age) = 7.2, 7.3, and 7.4 for NGC 146 and  7.6, 7.7, and 7.8 for King 14, used to estimate the cluster's ages and ensure consistency across photometric bands. The alignment of cluster sequences with the isochrones across all photometric bands supports the reliability of the adopted membership selection and provides consistent age estimates for both clusters.}
	\label{fig:cmd}
\end{figure*}

King 14 exhibits a relatively more dispersed main sequence in the CMD (see Fig.~\ref{fig:cmd}), indicative of a slightly older age and possibly some degree of dynamical evolution. For this cluster, isochrones with $\log(\text{age})$ between 7.6 and 7.8 were overlaid, and the best fit was found at $\log(\text{age}) = 7.7$, corresponding to an age of $50 \pm 10$ Myr. The cluster parameters were estimated via isochrone fitting using the chi-square minimization technique, yielding a reduced chi-square value of 0.91. The apparent distance modulus was derived as $(m-M) = 13.50 \pm 0.20$ mag, which corresponds to a heliocentric distance of $2.47 \pm 0.34$ kpc. This value is in good agreement with the parallax and consistent with previous studies. The cluster's more evolved stellar content and wider main sequence support the age estimate.

\medskip

In both NGC~146 and King~14, the CMDs reveal a relative underdensity of stars near $G\simeq12$--$13$~mag (Fig.~\ref{fig:cmd}). Since these clusters form a physical pair likely originating from the same molecular cloud, the presence of a similar feature in both CMDs is notable. Such underdensities, or main-sequence ``gaps''
, can arise from unresolved binaries, which shift stars upward in brightness, as well as from the brief evolutionary timescale near the main-sequence turnoff, which naturally reduces the number of stars in this region. The similarity of this feature in both clusters, therefore, suggests that they share comparable stellar populations, binary fractions, and initial conditions, consistent with a coeval origin. The detailed information about the Main-sequence gaps in CMDs has also been reported in previous studies \citep{bohm1974gap, sagar1978gap, rachford2000relationship, hasan2023gaps}. Overall, the derived parameters summarized in Table~\ref{tab:fundamental_params} indicate that both clusters are relatively young OCs located at similar distances within the Perseus arm of the Milky Way. Their close spatial proximity and comparable ages suggest a possible physical connection, which will be further investigated through orbital and dynamical analyses.

For the radial velocity estimation, we used Gaia~DR3 measurements for three stars in NGC~146 and six stars in King~14. Given the small sample sizes, we adopted the weighted mean method, which accounts for the individual uncertainties of each measurement. For NGC~146, this yields a mean radial velocity of $-88.457 \pm 4.613~\mathrm{km\,s^{-1}}$. Our value is consistent, within uncertainties, with that reported by \citet{Hunt2024}, which was based on a single star. In the case of King~14, radial velocity values were available for six stars; however, two stars were excluded as outliers because their velocities deviated by more than 3$\sigma$ from the mean of the sample. Using the remaining four stars, we obtain a mean radial velocity of $-79.075 \pm 5.348~\mathrm{km\,s^{-1}}$.

%%%%%%%%%%%%%######################################%%%%%%%%%%%%%%%%%%%%%%%%
\begin{table*}
\centering
\vspace{-0.4cm}
\caption{Comparison of the cluster parameters derived in the present study with those reported in previous works for clusters NGC~146 and King~14. The listed parameters include positional coordinates (R.A., Dec.), kinematic properties (proper motions and radial velocity), parallaxes, distances, ages, and Galactic positions ($R_{\mathrm{GC}}$, $X_{\mathrm{GAL}}$, $Y_{\mathrm{GAL}}$, $Z_{\mathrm{GAL}}$).}
\small
\resizebox{\textwidth}{!}{%
\begin{tabular}{c c c c c c c c c c c c c}

\hline

Author & R.A & Dec. & $\mu_{\alpha}cos(\delta)$ & $\mu_{\delta}$ & Parallax  & $d_{plx}$ &  Age & R$_{GC}$ & X$_{GAL}$ & Y$_{GAL}$ &  Z$_{GAL}$ & Radial Velocity  \\
    &  Deg.  & Deg.  & mas yr$^{-1}$ & mas yr$^{-1}$ & mas  &  (kpc)  & (Myr) & (kpc) & (kpc) & (kpc) & (kpc) & (Km s$^{-1}$)\\

\hline

&  &  &  & & & \textbf{NGC 146} &  &  & & &  \\
\hline
\textbf{Present study} & 8.262 $\pm$ 0.345 & 63.293 $\pm$ 0.154 &    $-2.812 \pm 0.145$ & $-0.475 \pm 0.210$ & 0.320 $\pm$ 0.105 & 2.98$^{+0.33}_{-0.26}$ & 20 $\pm$ 5 &  10.059 $\pm$ 0.186 & $-$1.529 $\pm$ 0.169  & 2.558 $\pm$ 0.283 & 0.026 $\pm$ 0.010  & $ -88.457 \pm 4.613$ \\ 
\citet{kharchenko2013vizier}& 08.226 & +63.305  & $-$4.170 &	$-$2.230 & -- & 2.800 & 50.2 & -- &  --  &  --  & -- \\ 
\citet{Cantat-Gaudin-Anders_2020} & 08.2624 &  +63.316 & $-$2.864 & $-$0.451 &  0.307  &   3141 &  35.5   & 10.310  & $-$1.611  &     2.695   &    0.028   \\
\citet{dias2021updated} & 008.2590 & 63.3302 & $-$2.873 $\pm$ 0.087 &	$-$0.458 $\pm$ 0.107 &	0.311 $\pm$ 0.036 &	 2.685 $\pm$ 0.146 & 	19.1 $\pm$ 5.2 & -- &  --  &  --  & -- \\
\citet{Hunt2024}   & 008.2632  &  63.3047   & $-$2.841 &   $-$0.508 &  0.3342 &  2.761 &  17.5 & -- &  --  &  --  & -- & $-$99.85 \\  
\citet{subramaniam2005ngc}   & 8.271  & 63.301   & -- & -- & -- & 3.470$^{.335}_{-0.305}$ & 10-16 & -- & -- & -- & -- \\
\hline

&  &  &  &  &  & \textbf{King 14}  &  &  &  &  & \\
\hline
\textbf{Present study} & 7.965 $\pm$ 0.321  & 63.162 $\pm$ 0.136   &$-3.230 \pm 0.185$ & $-0.975 \pm 0.213$ & 0.380 $\pm$ 0.128 & 2.51$^{+0.30}_{-0.23}$ & 50 $\pm$ 10 &  9.725 $\pm$ 0.146 & $-$1.282 $\pm$ 0.118  & 2.158 $\pm$ 0.200 & 0.016 $\pm$ 0.010    & $ -79.075 \pm 5.348$ \\ 
\citet{kharchenko2013vizier} & 008.014 & 63.181 &  $-$3.75  &	$-$1.55 & -- & 2.600 & 50.1 & -- &  --  &  --  & -- & -- \\ 
\citet{Cantat-Gaudin-Anders_2020} & 07.987	  & 63.163 & $-$3.267 	& $-$0.985 	& 0.402 &  2.420 & 144.6 & 9.800 &   $-$1.236  &   2.080    &   15 & -- \\
\citet{dias2021updated}& 08.004 & +63.164& $-$3.268 $\pm$ 0.158 &	$-$0.995 $\pm$ 0.165  & 	0.400 $\pm$	0.064 &	 2.213 $\pm$ 0.064 & 45.9 $\pm$ 23.3 & -- &  --  &  --  & -- \\  	  	
\citet{Hunt2024} & 007.98607 & 63.17061 & $-$3.257 &  $-$1.084 &   0.3927  &   2.358 &   48.2  & -- &  --  &  --  & -- & --\\ 
\citet{netopil2006photometric}   & 8.013    &  63.156  & -- & -- & -- & 2.96 $\pm$ 0.42 & 79.4 & 10.330 & -- & -- & 0.020 \\
\hline
\end{tabular}%
}
\label{tab:fundamental_params}
\end{table*}

\section{Dynamical Study of the OCs}{\label{sec: dynamical_analysis}}

\subsection{Luminosity Function and Mass Function}

\begin{figure}
	\centering 
 \hbox{
 \includegraphics[width=0.5\linewidth]{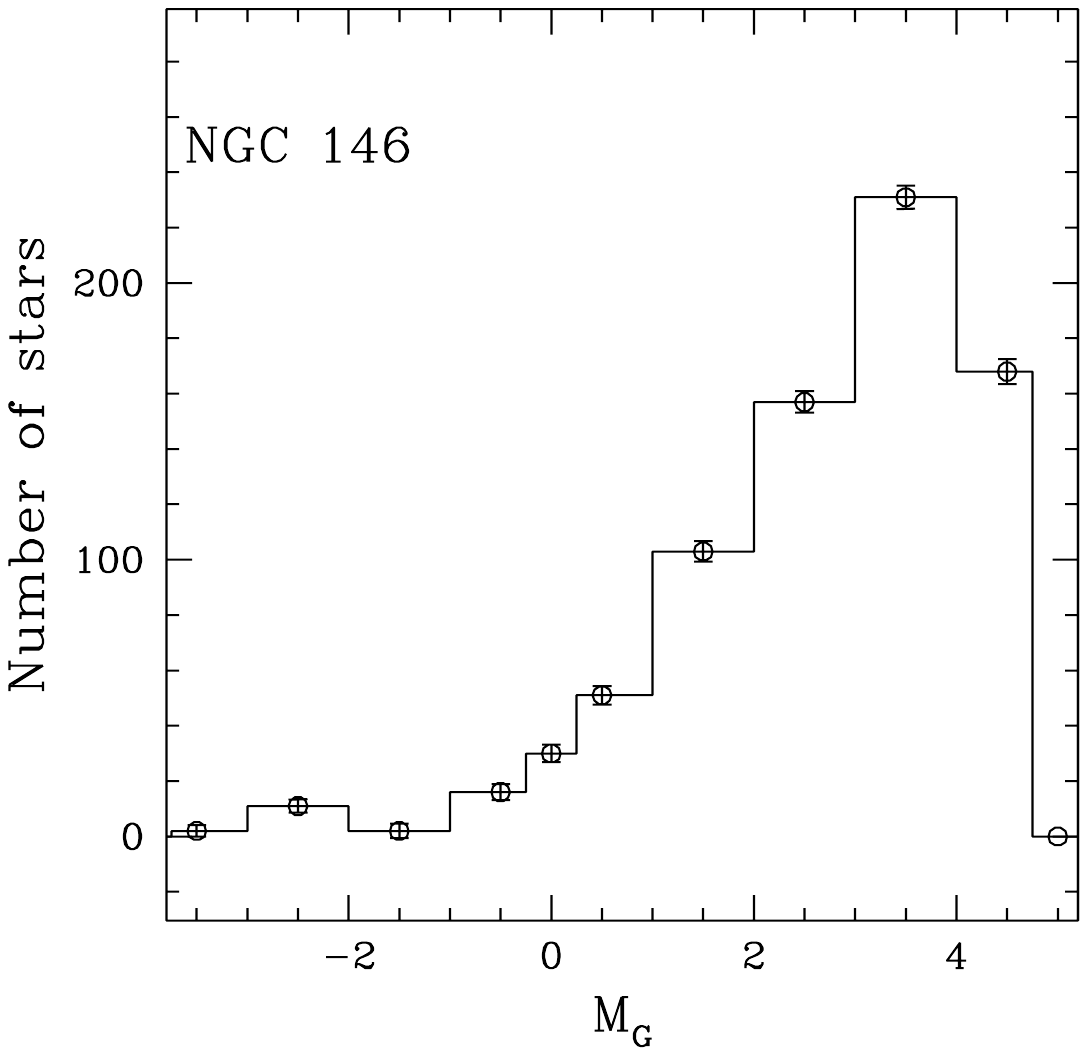}	
 \includegraphics[width=0.5\linewidth]{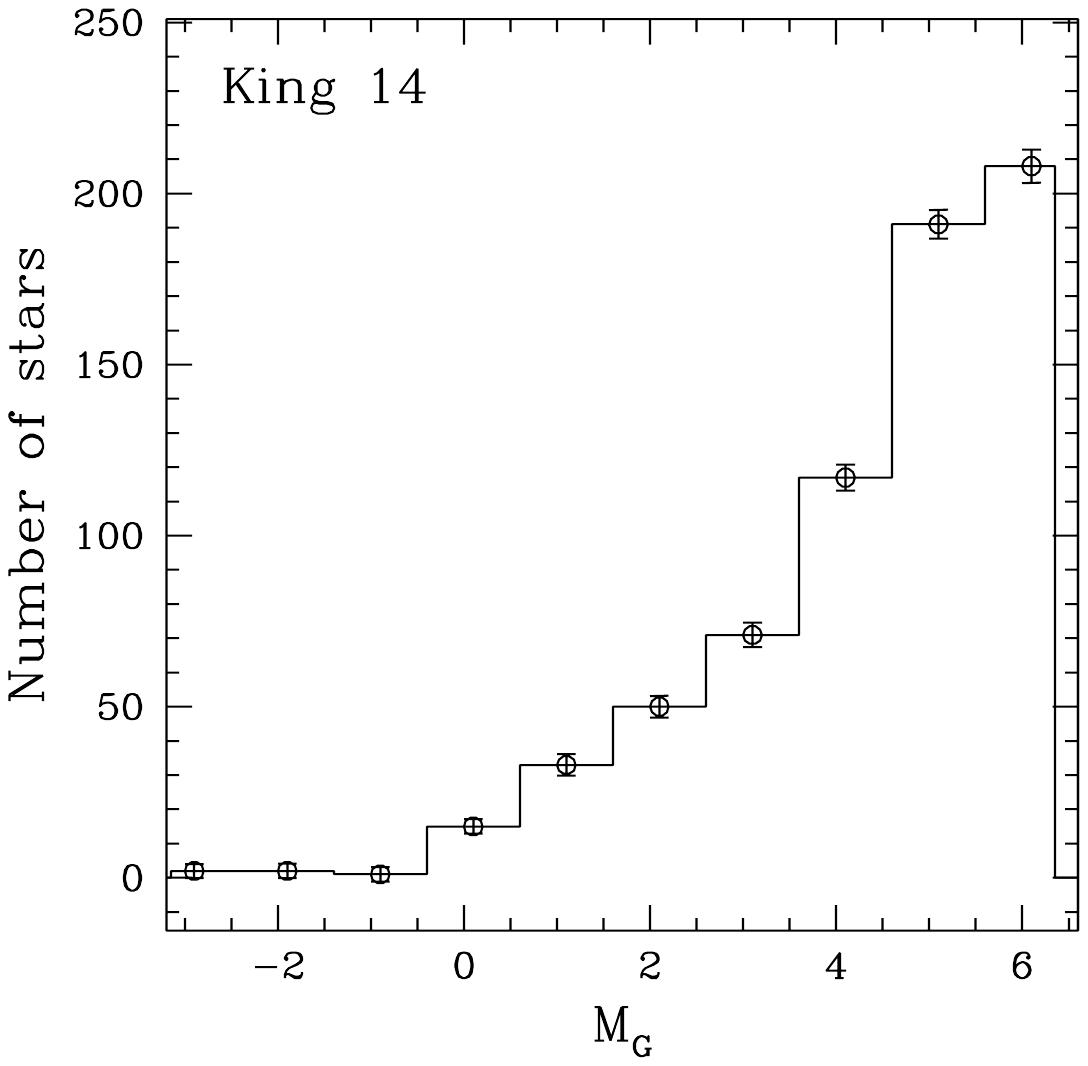}
}
\vspace{-1.8cm}
\hbox{
\includegraphics[width=0.5\linewidth]{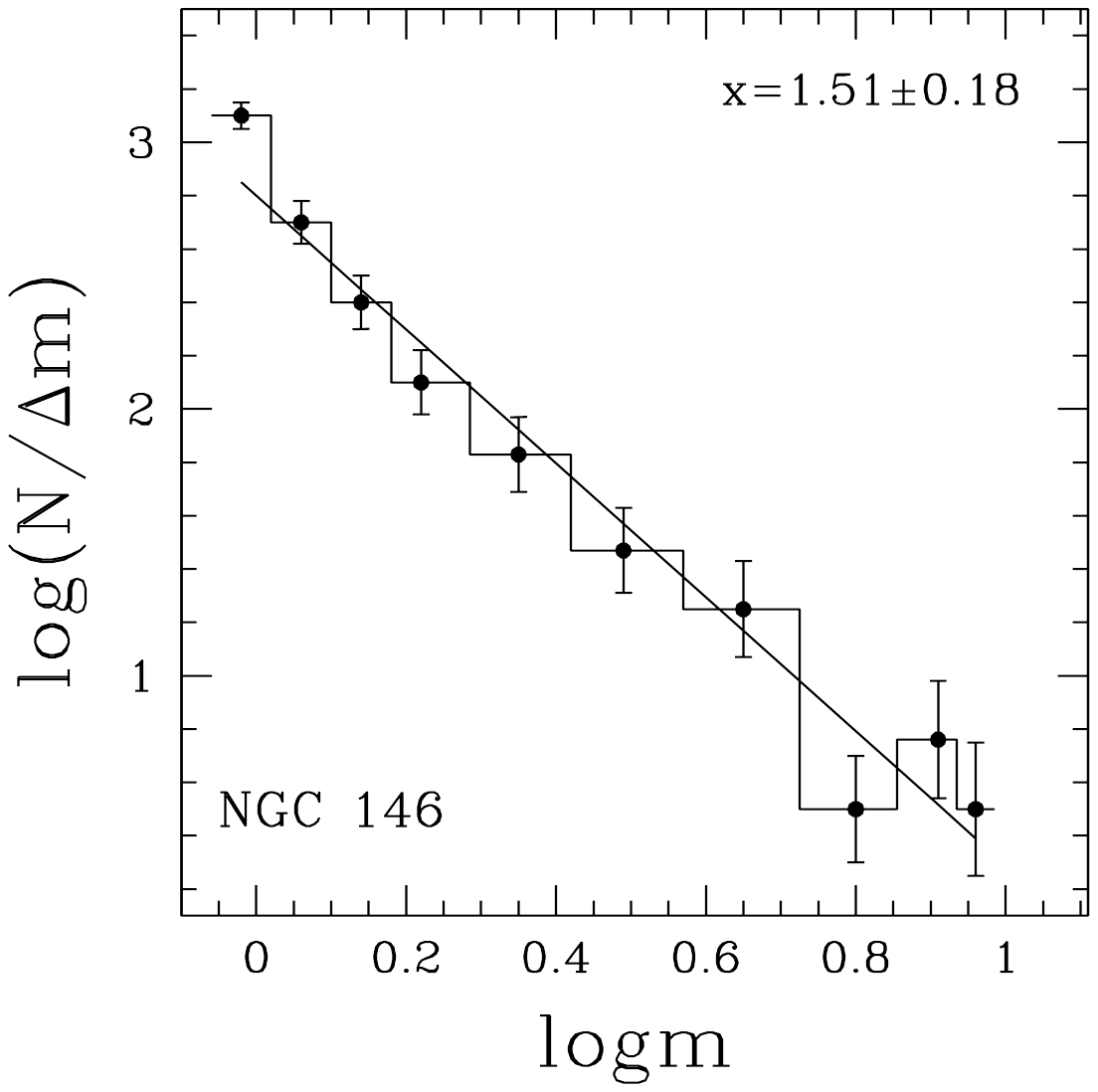}	
\includegraphics[width=0.5\linewidth]{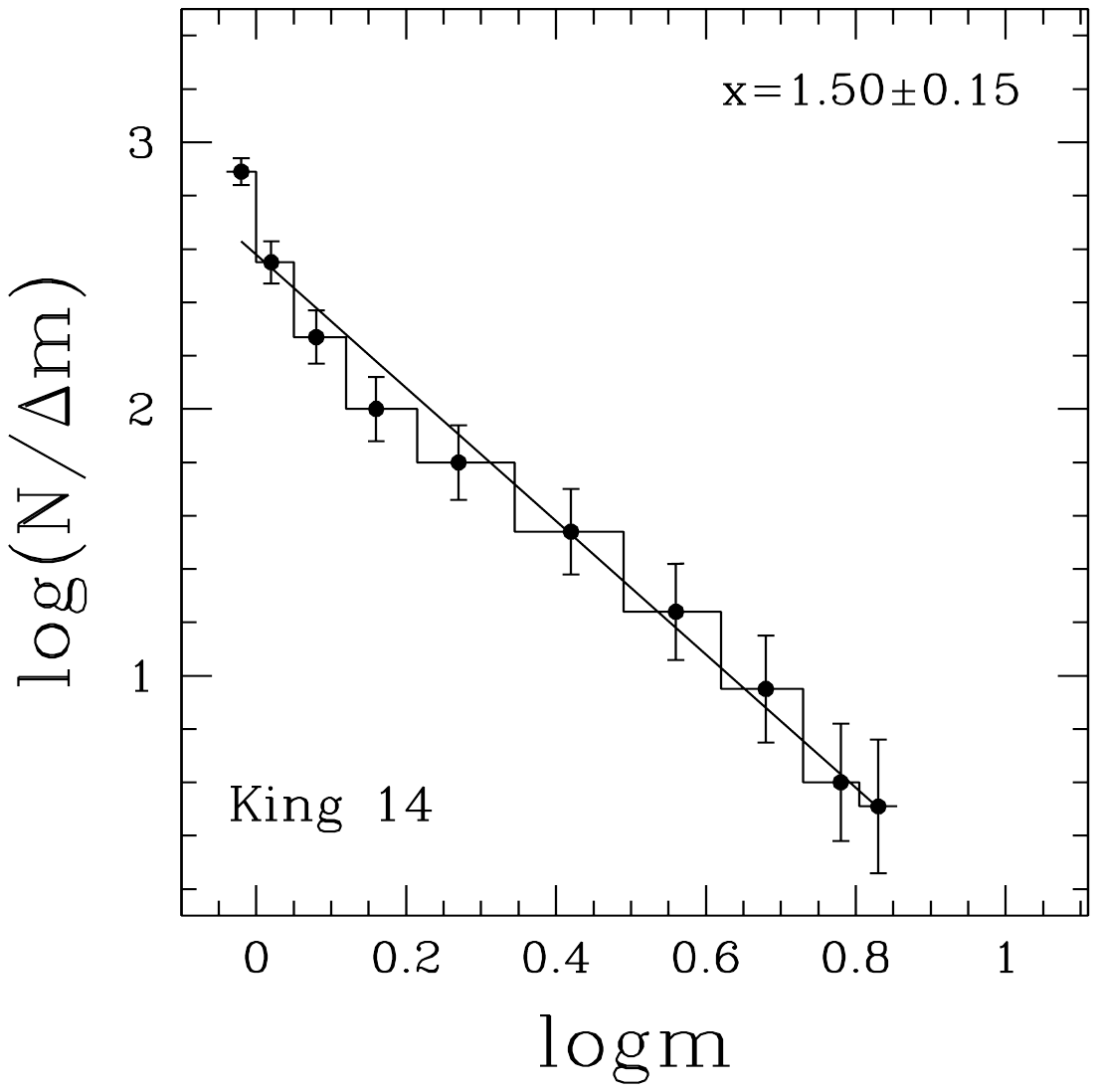}
}
\vspace{-1cm}
 \caption{Luminosity and mass functions for the star clusters NGC 146 and King 14. The functions are plotted with fitted Salpeter power-law distributions, shown as solid lines in each panel. The derived slope (x) of the mass function is indicated within the respective panels, with uncertainties estimated using Poisson statistics ($1/\sqrt{N}$), where $N$ is the number of stars in each bin.}

	\label{lf_mf}%
\end{figure}

The luminosity function (LF) and Mass function (MF) depend primarily on cluster membership and are also connected to the well-known mass-luminosity relationship. We have used $G$ versus $(G_{\rm BP}-G_{\rm RP})$ CMD to construct LF. We converted the $G$ magnitudes of main-sequence stars into the absolute magnitudes using the distance modulus and reddening. A histogram is constructed with 1.0 mag intervals as shown in the top panels of Fig. \ref{lf_mf}. Inspection of the LF plots for both clusters reveals a steady increase in stellar density toward fainter magnitudes. There is a prominent increase in the last few bins. This behaviour likely reflects the majority of low-mass stars in our sample. However, the rise could also be partially affected by residual observational biases, such as photometric incompleteness or stellar crowding at the faint end, even though our analysis is restricted to highly probable stars brighter than 20 mag and with membership probabilities above 80 percent.

We have used theoretical isochrones from \citet{marigo2017new} to convert absolute magnitudes into stellar masses, enabling a transformation of the LF into the MF. This transformation was carried out by mapping each magnitude bin to its corresponding mass bin using isochrones appropriate to each cluster’s age, metallicity, and reddening. The resulting MF is expressed as the logarithm of the number of stars per unit mass interval versus the logarithm of mass, and is plotted in the lower panels of Fig.~\ref{lf_mf}.

The MF follows a power-law distribution of the form:
\begin{equation}
\log \left(\frac{dN}{dM}\right) = -(1 + x) \log(M) + \mathrm{constant},
\end{equation}
where $dN$ is the number of stars in a mass bin $dM$ centered at mass $M$, and $x$ is the MF slope. A linear fit to the plotted data in logarithmic space provides the slope $x$.

\begin{table}[h]
\centering
\caption{Summary of the derived MF slopes and stellar mass estimates for the clusters NGC~146 and King~14. The table lists the stellar mass range used in the analysis, the corresponding MF slope, total cluster mass, and mean stellar mass. 
}
\begin{center}
\begin{tabular}{lcccc}
\hline\hline
\textbf{Object} & \textbf{ Mass} & \textbf{MF} & \textbf{Total} & \textbf{Mean} \\
&       Range &            Slope & mass & mass \\
&      \textbf{$M_{\odot}$} &  & \textbf{$M_{\odot}$} & $M_{\odot}$\\
\hline
NGC 146  & $0.9-9.7$&$1.51\pm0.18$&$1492$&$1.93$ \\
King 14  & $0.9-6.8$&$1.50\pm0.15$&$1250$&$1.81$ \\
\hline
\end{tabular}
\label{tab: massf_tab}
\end{center}
\end{table}

For NGC~146 and King~14, the estimated MF slopes are $x = 1.51 \pm 0.18$ and $x = 1.50 \pm 0.15$, respectively. The corresponding total and mean stellar masses for both clusters are listed in Table~\ref{tab: massf_tab}. These values are slightly steeper than the canonical Salpeter \citep{salpeter1955luminosity} slope of $x = 1.35$. This suggests a mild excess of low-mass stars in the observed mass range. Such slopes are typical for young OCs in the Galactic disk. They support the hypothesis of a nearly universal initial mass function (IMF), with minor deviations that may arise from local star-formation conditions or early dynamical evolution. The close similarity in MF slopes between the two clusters further implies that they have experienced comparable star-formation histories and subsequent dynamical evolution. Their consistent MF slopes, along with similar ages, distances, and kinematics, strengthen the argument that NGC 146 and King 14 may have originated within the same parental molecular cloud complex. A slightly steeper slope may also indicate ongoing mass segregation or the preferential loss of higher-mass members. This can occur due to internal relaxation or external tidal interactions, both of which are expected during the early stages of cluster evolution.

\subsection{Orbital Study of OCs}

We conducted a detailed dynamical analysis of NGC\,146 and King\,14, spatially and kinematically associated OCs, to explore their orbital properties within the Milky Way. The adopted methodology was specifically tailored for these two clusters \citep{Yontan2022, Tasdemir2025, Cinar2025}. Such studies are crucial for understanding the formation and long-term survival of binary cluster systems in the Galactic disk.

Orbital computations were performed using the axisymmetric {\sc MWPotential2014} Galactic potential model implemented in the {\sc galpy} software package \citep{Bovy2015}. This model includes critical Galactic constants, such as the Galactocentric distance of the Sun ($R_{\rm gc} = 8.20 \pm 0.10$ kpc), the circular velocity at the solar radius ($V_{\rm rot} = 220$ km\,s$^{-1}$), and the vertical displacement of the Sun from the Galactic mid-plane ($Z_0 = 25 \pm 5$ pc) \citep{Bovy2012}.

The input parameters used in the orbit integration comprise equatorial coordinates ($\alpha$, $\delta$), heliocentric distance ($d$), mean proper motion components ($\mu_\alpha\cos\delta$, $\mu_\delta$), and systemic radial velocities. These quantities were derived from high-precision astrometric and spectroscopic measurements compiled in Table~\ref{tab:fundamental_params}. With these six-dimensional phase-space coordinates, we integrated the orbits of both clusters forward in time over intervals consistent with their estimated ages, using a 1 Myr timestep \citep{Bostanci2015, Elsanhoury2025}.

\begin{figure}
\centering
\includegraphics[width=1\linewidth]{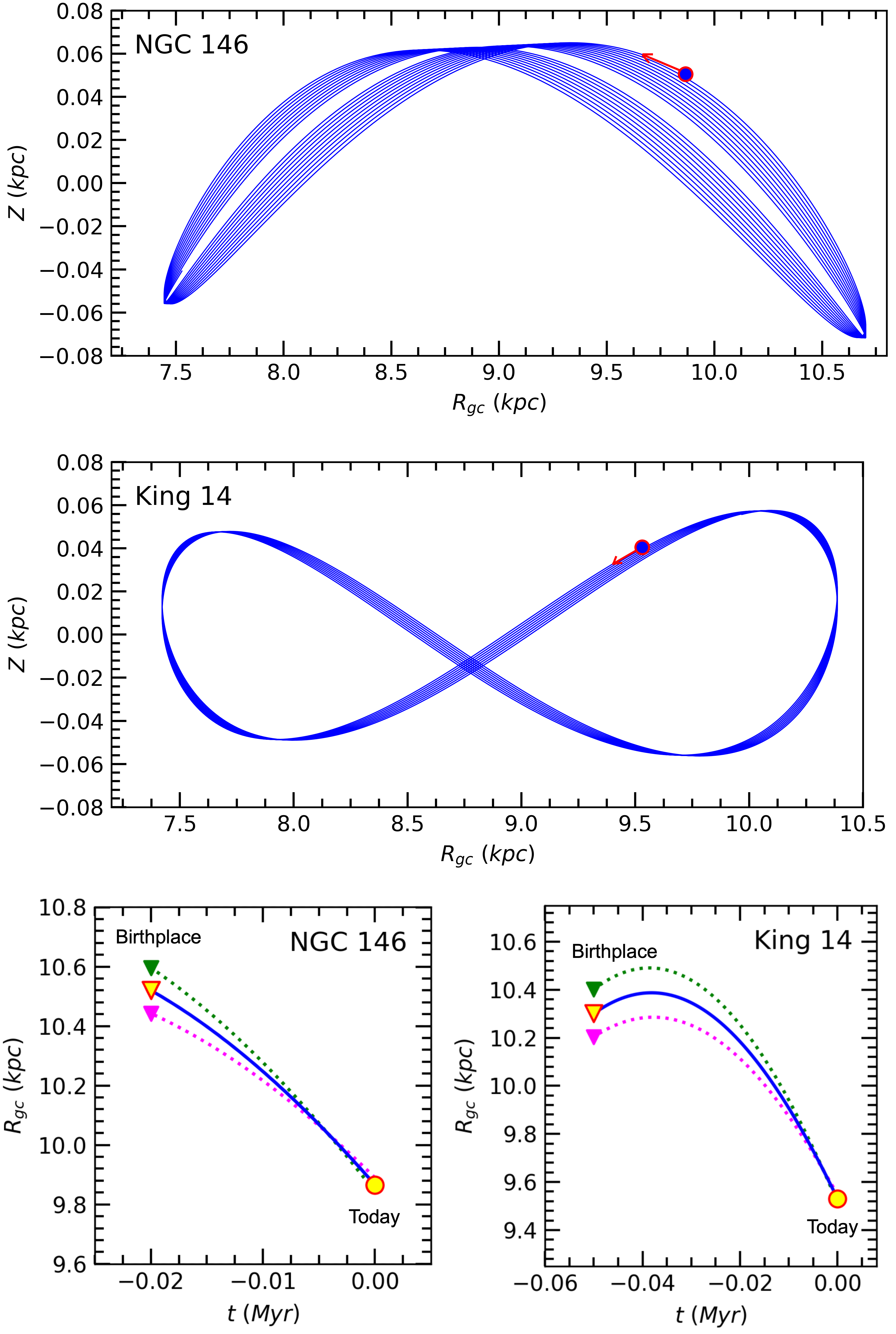}
\caption{The Galactic orbits and birth radii of NGC\,146 and King\,14 are illustrated in two projections: $Z~\times~R_{\rm gc}$ (top panels) and $R_{\rm gc}~\times~t$ (bottom panels). In the bottom panel, present-day positions are indicated by filled yellow circles, while the estimated birth locations are shown as filled triangles. The dotted curves represent orbital paths computed by propagating uncertainties in the input parameters.}
\label{fig:orbits}
\end{figure}

From the resulting orbital trajectories, we derived the apogalactic distance ($R_{\rm a}$), perigalactic distance ($R_{\rm p}$), mean orbital radius ($R_{\rm m} = (R_{\rm a} + R_{\rm p}) / 2$), orbital eccentricity ($e$), maximum vertical excursion from the Galactic plane ($Z_{\rm max}$), and orbital period ($P_{\rm orb}$). These results are listed in Table~\ref{Tab:dyn_params}.

\begin{table}[h]
%\tiny
\centering
\renewcommand{\arraystretch}{1.4}
\setlength{\tabcolsep}{2.8pt}
\caption{Derived orbital parameters for clusters NGC~146 and King~14. The listed quantities include the maximum distance from the Galactic plane ($Z_{\mathrm{max}}$), apogalactic distance ($R_{\mathrm{a}}$), perigalactic distance ($R_{\mathrm{p}}$), mean orbital radius ($R_{\mathrm{m}}$), orbital eccentricity ($e$), orbital period ($ P_{\rm orb}$), and present Galactocentric distance ($R_{\mathrm{teo}}$).}
%\footnotesize
\begin{tabular}{l|cc}
\hline\hline
\textbf{Parameter} & \textbf{NGC 146} & \textbf{King 14} \\
\hline
$Z_{\rm max}$ (kpc) & 0.072 $\pm$ 0.031 & 0.058 $\pm$ 0.025 \\
$R_{\rm a}$ (kpc)   & 10.703 $\pm$ 0.133 & 10.387 $\pm$ 0.103 \\
$R_{\rm p}$ (kpc)   & 7.446 $\pm$ 0.208  & 7.422 $\pm$ 0.270 \\
$R_{\rm m}$ (kpc)   & 9.075 $\pm$ 0.171 & 8.905 $\pm$ 0.187 \\
$e$                 & 0.179 $\pm$ 0.020  & 0.167 $\pm$ 0.023 \\
$P_{\rm orb}$ (Myr)   & 258 $\pm$ 1 & 253 $\pm$ 2 \\
$R_{\rm teo}$ (kpc) & 10.52 $\pm$ 0.15 & 10.30 $\pm$ 0.19 \\ 
\hline
\end{tabular}
\label{Tab:dyn_params}
\end{table}

NGC\,146 follows a nearly circular Galactic orbit with an eccentricity of ($e = 0.179 \pm 0.020$), oscillating between $R_{\rm p} = 7.446 \pm 0.208$ kpc and $R_{\rm a} = 10.703 \pm 0.133$ kpc. Its mean Galactocentric radius is $R_{\rm m} = 9.075 \pm 0.171$ kpc, and the cluster reaches a maximum vertical height of only $Z_{\rm max} = 0.072 \pm 0.031$ kpc above the Galactic plane. The orbital period of NGC\,146 is calculated as $258 \pm 1$ Myr. King\,14 exhibits similarly low orbital eccentricity ($e = 0.167 \pm 0.023$), with radial excursions ranging from $R_{\rm p} = 7.422 \pm 0.270$ kpc to $R_{\rm a} = 10.387 \pm 0.103$ kpc. Its mean orbital radius is $R_{\rm m} = 8.905 \pm 0.187$ kpc, and its maximum vertical amplitude is $Z_{\rm max} = 0.058 \pm 0.025$ kpc. The orbital period is computed as $253 \pm 2$ Myr.

The very small $Z_{\rm max}$ values for both systems indicate that their orbits remain tightly confined to the Galactic mid-plane, consistent with expectations for young thin-disk clusters. Their nearly circular orbits and comparable orbital periods suggest they have experienced similar dynamical conditions throughout their lifetimes. Using the traceback early orbital radius method ($R_{\rm teo}$), the inferred formation radii are now very close: NGC\,146 at $R_{\rm teo} = 10.53 \pm 0.15$ kpc and King\,14 at $R_{\rm teo} = 10.30 \pm 0.19$ kpc, implying that both clusters likely formed at similar Galactocentric distances.

\begin{figure}
\centering
\includegraphics[width=0.7\linewidth]{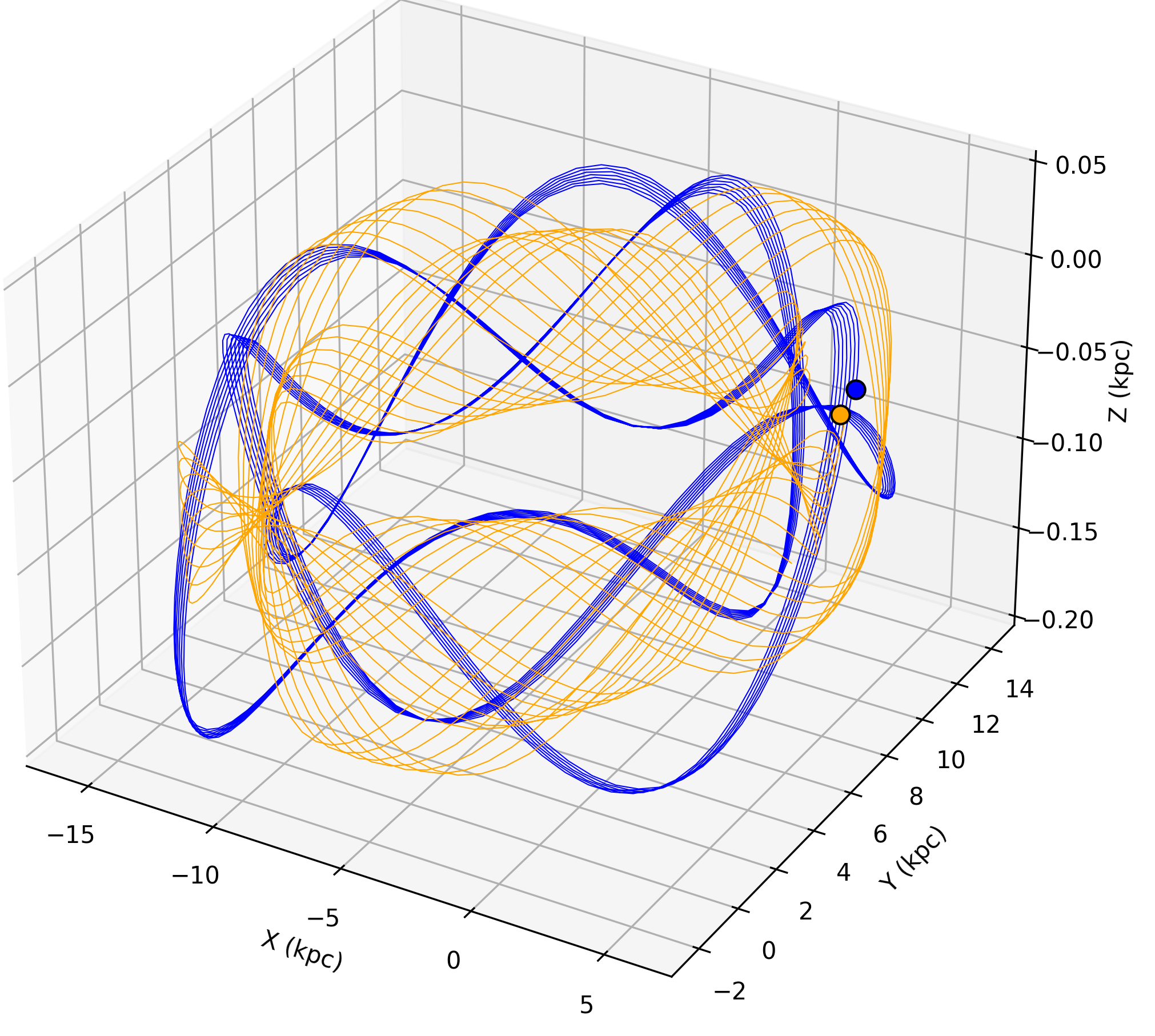}\\
\includegraphics[width=0.75\linewidth]{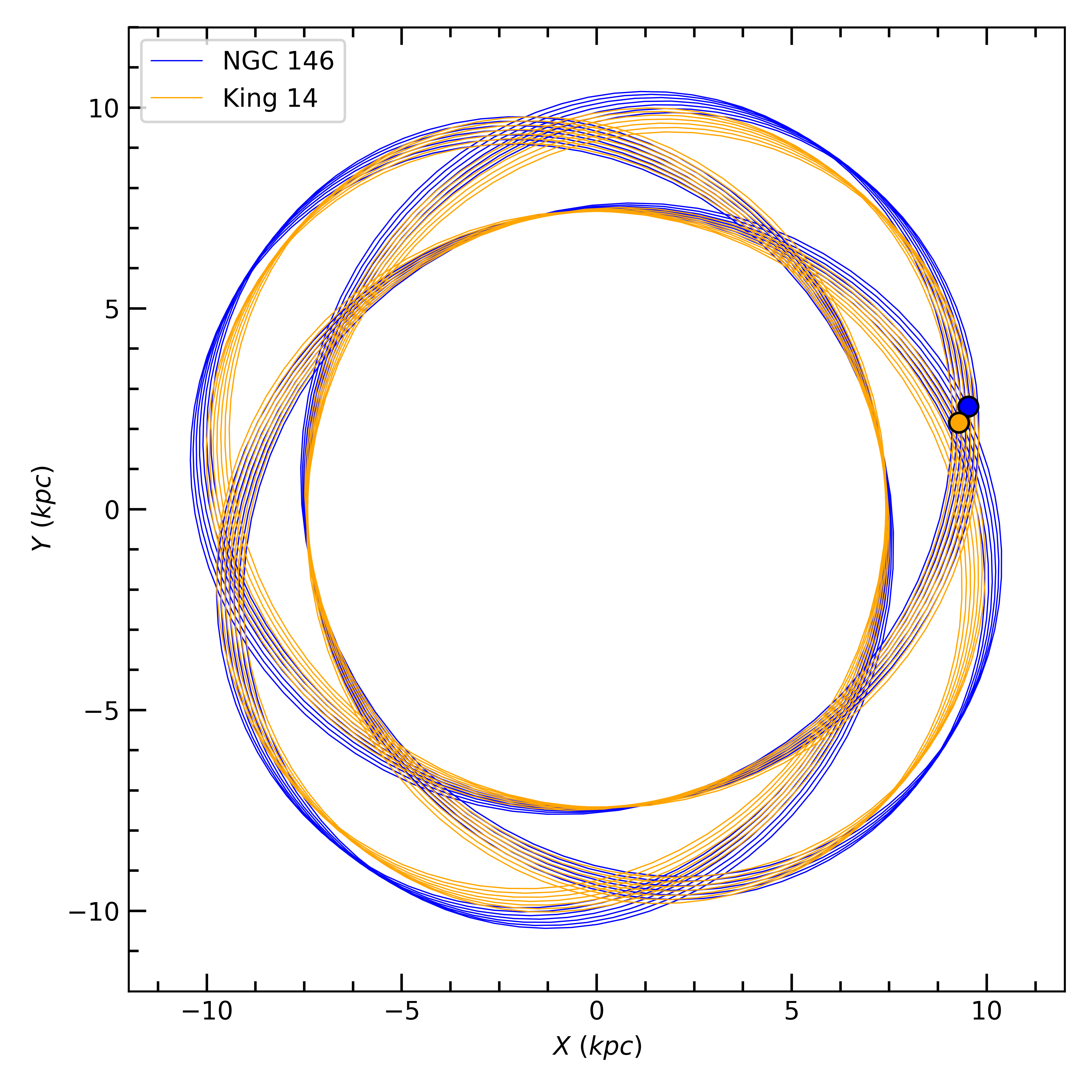}
\caption{The orbits of NGC 146 and King 14 in the Milky Way. The top panel shows a three-dimensional representation of the orbits of NGC 146 (blue) and King 14 (orange), while the bottom panel displays the same orbits projected onto the XY plane.}
\label{fig:orbits2}
\end{figure}

Figure~\ref{fig:orbits2} shows the positions of NGC\,146 and King\,14 in both the Galactic $x$--$y$ plane and in three-dimensional space. Both clusters move in disk-like, low-eccentricity orbits with small vertical excursions, staying close to the Galactic plane. Their current velocities and orbital paths are similar, but they are not gravitationally bound. Table~\ref{Tab:dyn_params} confirms this: the clusters have similar orbital radii and periods, but their separation is too large for them to form a bound system. Overall, NGC\,146 and King\,14 form a co-moving pair that travels together through the Galactic disk without being gravitationally connected.

\section{Dynamical Relationship of the NGC 146-King 14 Cluster Pair} {\label{sec: binarity}}

In this study, we investigate the possible physical association between the young OCs NGC 146 and King 14, which are located in close angular proximity on the sky. The fundamental parameters derived in this work indicate that both clusters are broadly consistent with having formed in the same Galactic environment. NGC 146 and King 14 exhibit comparable ages, $\log(t) \approx 7.3$ and $\log(t) \approx 7.7$, and lie at similar heliocentric distances of $2.98 \pm 0.33$~kpc and $2.51 \pm 0.23$~kpc, respectively. Their mean proper motions derived from \textit{Gaia}~DR3, $(\mu_{\alpha}\cos\delta, \mu_{\delta})$, are also very close, indicating minimal relative tangential motion within the measurement uncertainties. The similarity in age, distance, and motion suggests a likely common origin, possibly from the same parental molecular cloud complex.

To assess their spatial association, we estimated the three-dimensional separation between the two clusters using the method of \citet{subramaniam1995probable}, which employs the clusters' equatorial coordinates and distances. This yields a positional separation of $8.97$~pc, consistent with the previously reported value of $8.64$~pc. Such a slight separation places the pair among close cluster pairs ($R < 20$~pc), which are typically considered candidate binaries or primordial clusters. Building on this two-dimensional analysis, we next explored their actual three-dimensional separation.

Adopting cluster masses of $M_1 = 1492\,M_\odot$ for NGC~146 and $M_2 = 1250\,M_\odot$ for King~14, the total mass of the system is $M_{\rm tot} = 2742\,M_\odot$, giving $\sqrt{M_{\rm tot}} \approx 52.4$. Using our updated orbital solutions obtained with \texttt{galpy}, the clusters exhibit similar orbital periods and eccentricities, with $(P_{orb}, e) = (258~{\rm Myr}, 0.18)$ for NGC~146 and $(253~{\rm Myr}, 0.17)$ for King~14. We have used the relationship given by \citet{de2010evolution} to calculate the separation between these two clusters:
\begin{equation}
P_{\rm orb}~({\rm Myr}) = 94 \left( \frac{S_0}{1+e_0} \right)^{3/2} \frac{1}{\sqrt{M_1 + M_2}},
\end{equation}

 where $P_{\rm orb}$ is the orbital period, $S_0$ is the present-day three-dimensional separation between the clusters, $e_0$ is the orbital eccentricity of the pair, and $M_1$ and $M_2$ are their total masses. We have estimated $S_0 \approx 32.3$~pc for NGC~146 and $S_0 \approx 31.7$~pc for King~14. The close consistency of these values supports a characteristic dynamical separation of $S_0 \approx 32$~pc for the pair. This quantity represents the effective orbital semi-major axis the clusters would possess if they formed a gravitationally bound binary system and should not be interpreted as the present-day geometric distance. Instead, it is a theoretical separation scale derived from their total mass, orbital period, and orbital eccentricity, corresponding to the radius required to maintain a bound configuration. Although this relatively small separation indicates a spatially coherent system, dynamical considerations show that the clusters are not gravitationally bound.

The escape velocity for a binary cluster system is given by \citep{2008gady.book.....B}:

\begin{equation}
v_{\rm esc} = \sqrt{\frac{2GM_{\rm tot}}{R}},
\end{equation}

where $G = 4.302\times10^{-3}$~pc~(km~s$^{-1}$)$^2~M_{\odot}^{-1}$. At a separation of $\sim33$~pc, the escape speed is $v_{\rm esc} \simeq 1$~km~s$^{-1}$. Since the relative velocity between the clusters is of the order of a few km~s$^{-1}$, the system is dynamically unbound.

The Galactic orbital analysis shows that both clusters follow nearly circular, thin-disk orbits with similar guiding radii and orbital periods of $\sim250$--$260$~Myr. Their low eccentricities, comparable orbital parameters, and similarity in age, proper motion, and distance indicate that NGC~146 and King~14 are a co-moving pair that likely originated in the same giant molecular cloud. At present, however, the low escape velocity and non-zero relative motion demonstrate that the pair is dispersing rather than gravitationally bound. The NGC~146–King~14 system thus represents a co-moving, co-born but dynamically unbound cluster pair, providing a valuable case study for the formation and subsequent dissolution of primordial binary clusters in the Galactic disk.

\section{Characterization of Variable Stars in the Vicinity of NGC 146 and King 14} {\label{sec: sed_variable}}

 Variable stars are powerful tracers of stellar content and the dynamical evolution of star clusters. Their pulsation or eclipsing behavior provides strong constraints on fundamental parameters. These include mass, luminosity, radius, and evolutionary state. This information allows precise calibration of cluster distance, age, and composition. Several classes of variable stars, such as eclipsing binaries, contact systems, and massive pulsators, are strongly influenced by the cluster's internal dynamics. Processes like two-body relaxation, mass segregation, binary interactions, and tidal stripping affect their distribution, survival, and frequency within a cluster. As a result, the presence and location of variable stars clearly reveal signs of dynamical ageing and cluster evolution. Identifying and characterizing variable stars gives insight into both the cluster’s stellar population and the dynamical processes shaping its long-term evolution. In the NGC~146–King~14 system, identifying and cataloguing variable stars helps establish a more complete picture of the stellar population along the line of sight. While only one of the detected variables appears to be a likely cluster member, the remaining field variables still provide helpful information for future time-domain studies and for distinguishing cluster-associated variability from foreground or background sources. Motivated by this context, we conducted a focused variability analysis based on \textit{TESS} photometry. Physical parameters were estimated only for the likely cluster-member star using spectral energy distribution (SED) fitting, which allowed us to place this object on the H–R diagram and assess its pulsational nature. The remaining variable stars are treated as field objects; their light curves are presented for completeness, but no further physical interpretation is attempted in this work due to insufficient membership probability and limited observational coverage.

\subsection{Derivation of Stellar Parameters from SED Fitting}

To investigate the nature of the variable star TIC~444457513, identified as a likely member of the King~14 cluster, we performed a detailed SED fitting analysis. As spectroscopic parameters such as effective temperature and luminosity are unavailable for this source, SED modeling provides a reliable means to estimate its fundamental physical properties, including effective temperature ($T_{\rm eff}$), surface gravity ($\log g$), metallicity ([Fe/H]), interstellar extinction ($A_{\rm V}$), and distance ($d$) \citep{Cinar2024, Yousef2025b}.

The identification of TIC~444457513 as a cluster member was based on its spatial location, photometric consistency with the cluster sequence, and distance concordance. We utilized a Bayesian SED fitting methodology via the {\sc ARIADNE} \citep{Vines2022} framework, which incorporates a diverse suite of model atmospheres and accounts for interstellar extinction. The nested sampling algorithm implemented by {\sc dynesty} allows rigorous propagation of uncertainties in the derived parameters.

TIC~444457513 exhibits an effective temperature of $T_{\rm eff} = 11,826$ K and $\log g = 3.49$ cgs. Its estimated distance (2388 pc) and extinction ($A_{\rm V} = 1.22$ mag) are in excellent agreement with previous estimates for King~14, further strengthening its membership probability. The derived metallicity of [Fe/H] = $-$0.21 dex is consistent with the metal-poor nature of young open clusters located in the outer Galactic disk.

\begin{figure*}
\centering
\hbox{
\includegraphics[width=0.60\linewidth]{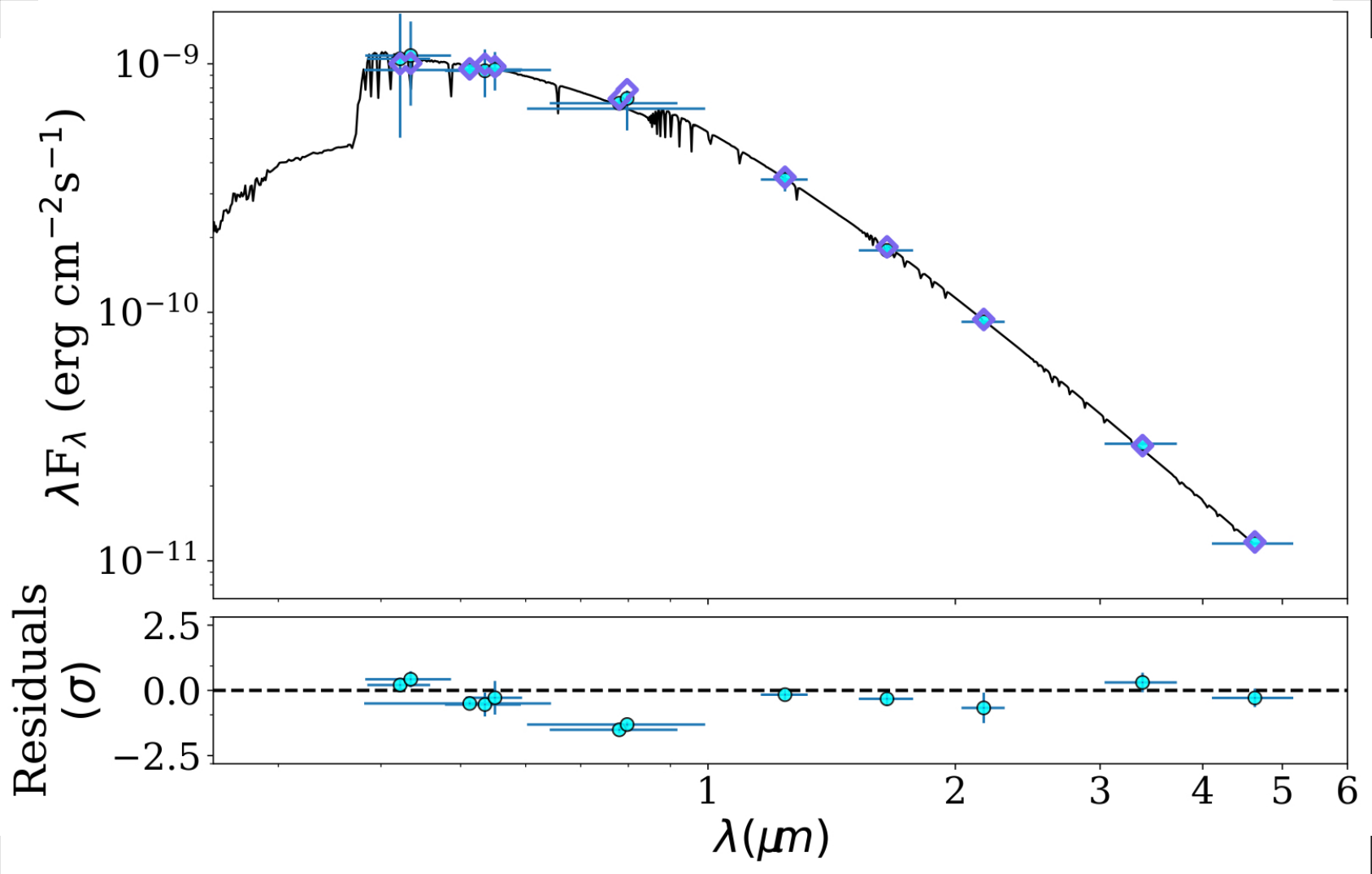}
\includegraphics[width=0.40\linewidth]{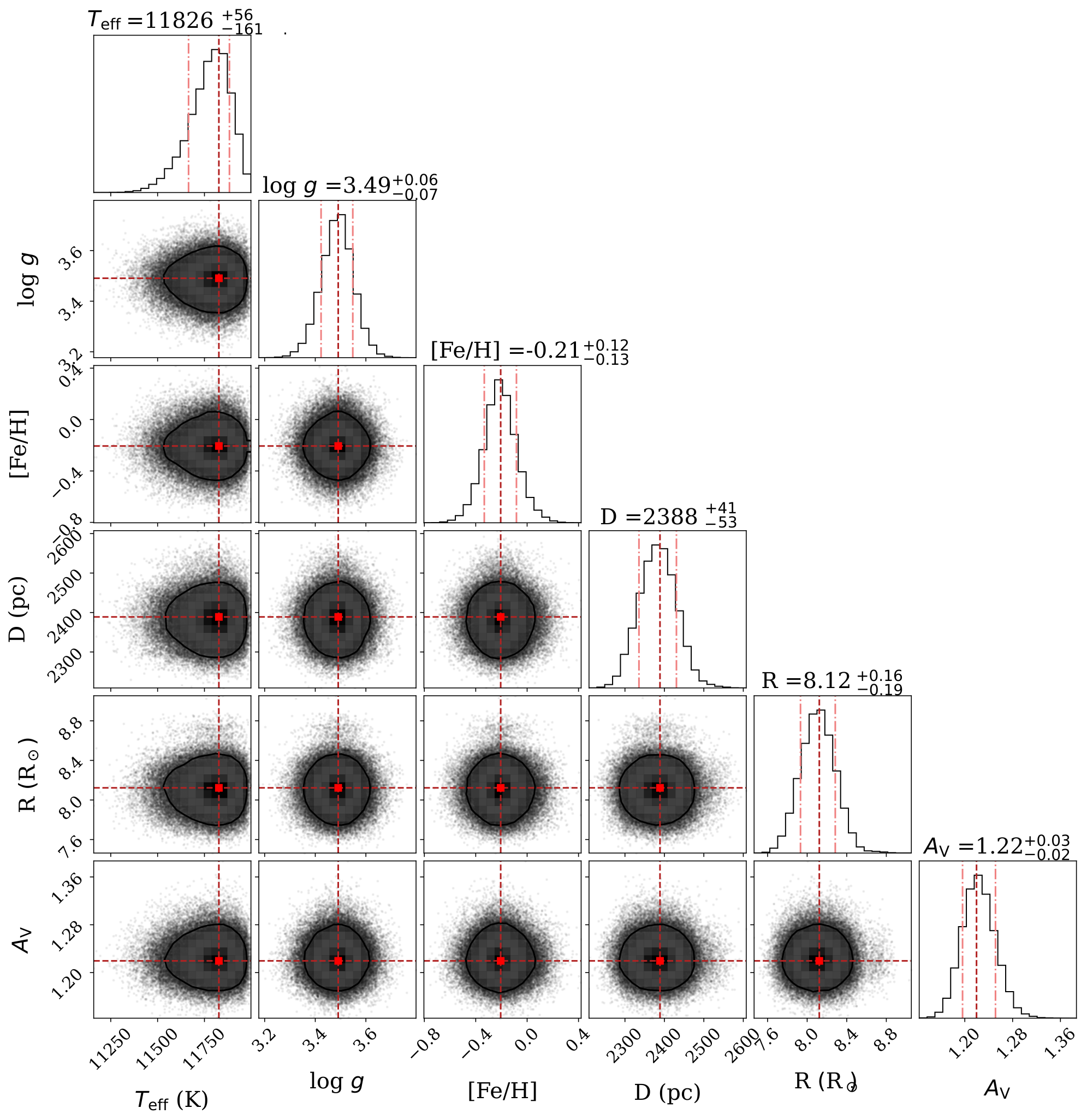}
}
\caption{Observed and modelled spectral energy distribution (SED) of the King~14 member star TIC~444457513 (left), showing the best-fit model (solid line) and observed fluxes with uncertainties (blue points). The lower panel displays the residuals between the observed and modelled fluxes. The right panel shows the posterior probability distributions of the fitted stellar parameters from the Bayesian SED-fitting analysis.}
\label{fig:SED}
\end{figure*}

It is important to note that the metallicity inferred from SED fitting for TIC~444457513 may be affected by methodological limitations. Early-type B stars possess weak or absent metallic absorption features, making accurate metallicity determination challenging. SED-based estimates for hot stars are inherently uncertain due to their smooth, continuum-dominated spectra. Additionally, photometric variability and potential circumstellar material may introduce systematic biases. Thus, the subsolar metallicity derived for TIC~444457513 may not reflect its intrinsic chemical abundance, but rather the limitations of SED fitting for hot variable stars.

Despite these caveats, SED fitting provides a robust and self-consistent framework for deriving the astrophysical parameters of TIC~444457513 in the absence of high-resolution spectroscopy. The combination of SED fitting and astrometric cross-validation allows us to characterise this variable star reliably within the King~14 cluster context.

\subsection{H-R Diagram}
\begin{figure}
    \centering
    \includegraphics[width=\linewidth]{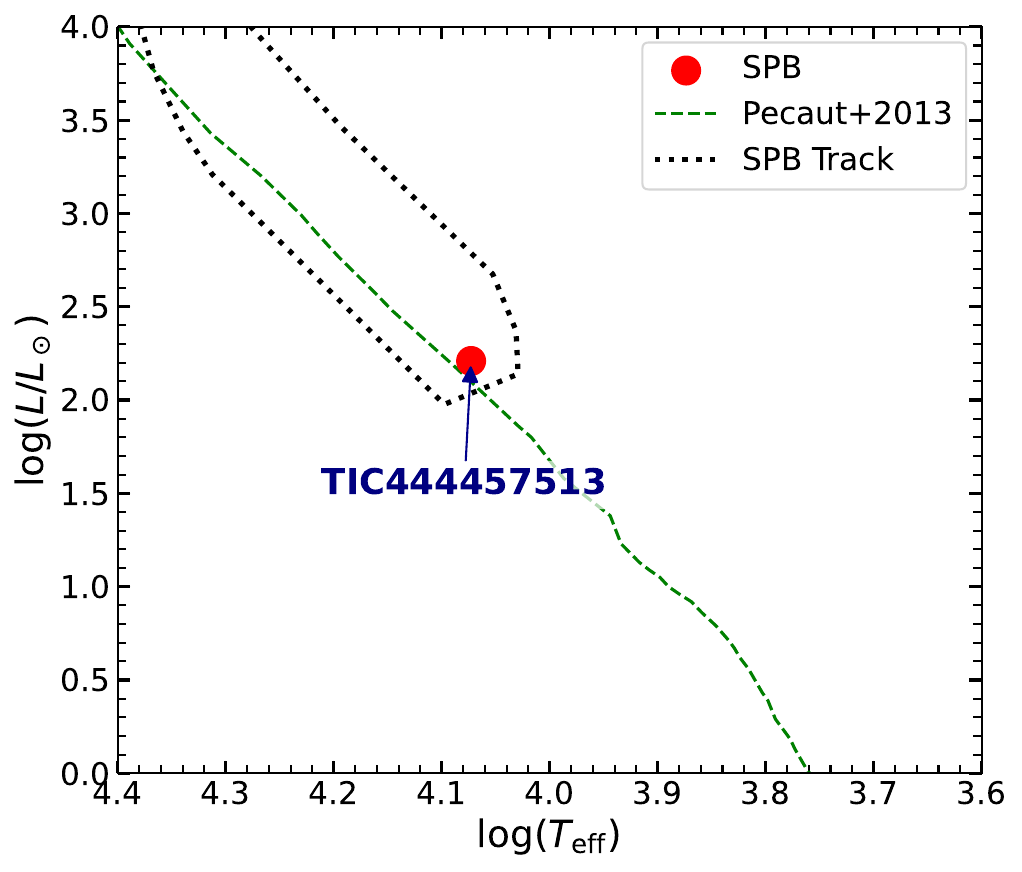}
    \caption{H-R Diagram showing the position of selected TESS-observed star. The plot shows $\log(L/L_\odot)$ versus $\log(T_{\rm eff})$, with a red circle indicating a member of the King 14 cluster. The green dashed line corresponds to a theoretical isochrone \citet{pecaut2013intrinsic}, and the black dotted region marks the classical instability strip. This diagram helps to identify the evolutionary status of the observed star.}
    \label{fig: HR_diagram}
\end{figure}

We constructed a Hertzsprung–Russell (H–R) diagram using SED-derived luminosities and effective temperatures. Since most detected variable stars are field sources and not relevant to the cluster analysis, we plot only the variable star that is likely a member of King~14, TIC~444457513 (Fig.~\ref{fig: HR_diagram}). The star lies within the theoretical instability region of slowly pulsating B-type (SPB) stars, and we overplot the SPB instability strip and the \citet{pecaut2013intrinsic} main-sequence track for reference. The remaining variable stars are not included in the diagram, as they appear to be field stars and their H–R positions do not provide additional constraints on cluster properties. Instead, we present their light curves separately without further physical interpretation. Periods were estimated using Lomb–Scargle analysis, and key variability parameters are summarized in Table~\ref{tab: Variable_Stars_detials}.

\begin{table*}
\centering
\vspace{-0.3cm}

\caption{The table presents the basic parameters of the variable stars identified in the observed clusters. The columns include the TIC ID, right ascension (RA) and declination (DEC), period (in hours), effective temperature (T$_{eff}$ in K), stellar luminosity, membership status, and type of variability. Membership is indicated as 'M' for probable cluster members and 'F' for field stars. The identified variability types include $\gamma$ Doradus stars, Slowly Pulsating B-type (SPB) stars, and an eclipsing binary. The stars are grouped by their respective clusters: NGC 146 and King 14.}
\scriptsize  % or \small for slightly larger text
\setlength{\tabcolsep}{3pt}  % default is 6pt; reduce to make columns narrower
\renewcommand{\arraystretch}{1.1}  % adjust row height if needed
\begin{tabular}{c c c c c c c c c}
\hline
ID & RA (J2000)  & DEC (J2000) & Period (h) &  log (T$_{eff}$) K   & log(L/L$_\odot$)   & Membership  & Type  \\
\hline
&  &   &   \bf{ King 14} &  &  &  &  \\
\hline
 TIC 444457577 &	8.09467 &	63.13414  &  0.78 & 3.8155  & 1.1139 &  F  & $\gamma$ Dor \\
 TIC 428480797 &	7.84173 &	63.33072  &  1.64 & 3.8468  & 1.1461 &  F  & $\gamma$ Dor \\
 TIC 444457513 & 8.08334 &   63.12024  &  45.20 & 4.0728 & 2.4329 &  M  &   SPB   \\
 TIC 419531798 & 8.20497 &   63.09827  &  197.1 & 3.5548 & 2.4579 &  F  &   EB     \\
 TIC 428479573 & 7.91712 &   63.02618  &  --    & 3.5414 & 2.7372 &  F  &  Misc   \\
 TIC 428479652 & 7.71905 &   63.04594  &  --    & 3.5990 & 3.3234 &  F  &  Misc   \\
\hline

&  &  &  \bf{NGC 146} & &  &  & \\
\hline
TIC 419524657 &  8.29757  &  63.42534  &  179.52  & 3.7865  & 1.2553 &  F  & Misc \\

\hline
  \end{tabular}
\label{tab: Variable_Stars_detials}
\end{table*}

\subsection{Classification of Variable Stars}

\begin{figure*}
    \centering

    % First Row (3 columns)
    \begin{minipage}{0.32\textwidth}
        \centering
        \includegraphics[width=\linewidth,height=5.7cm]{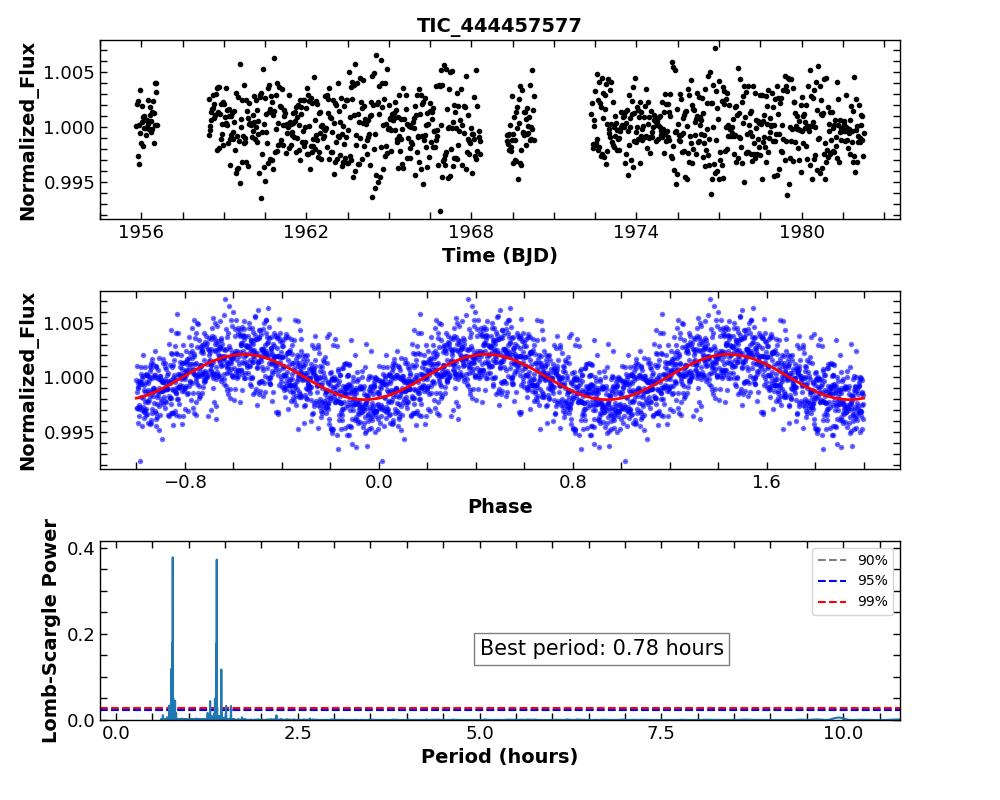}
    \end{minipage}
    \hfill
    \begin{minipage}{0.32\textwidth}
        \centering
        \includegraphics[width=\linewidth,height=5.7cm]{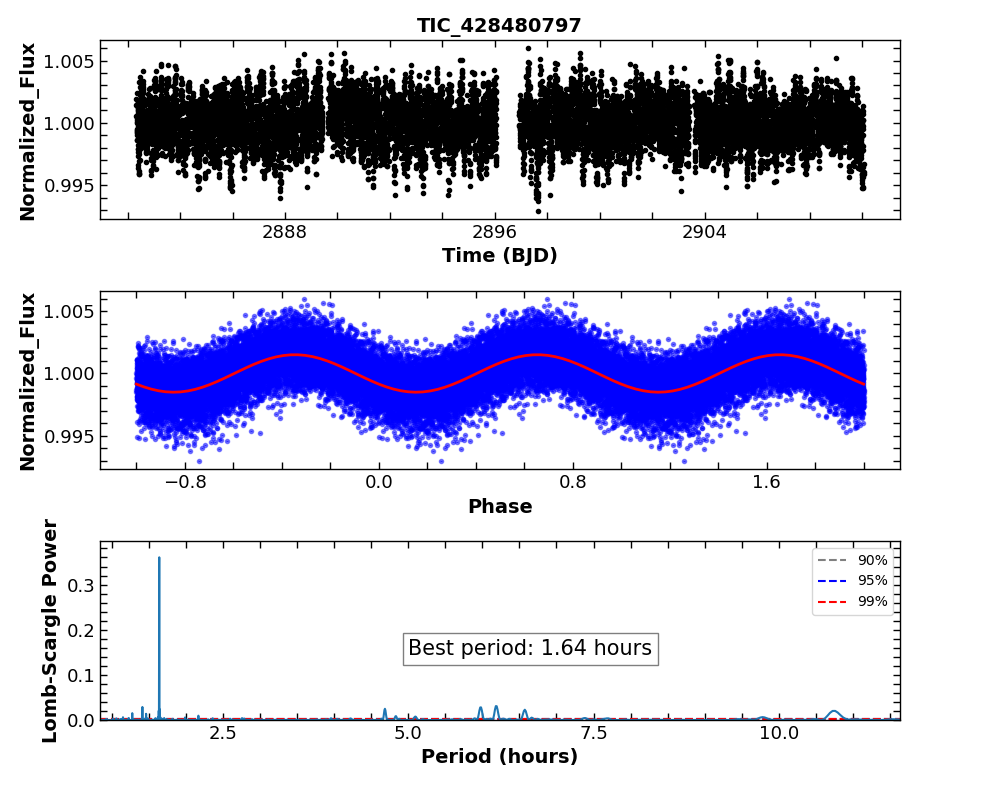}
    \end{minipage}
    \hfill
    \begin{minipage}{0.32\textwidth}
        \centering
        \includegraphics[width=\linewidth,height=5.7cm]{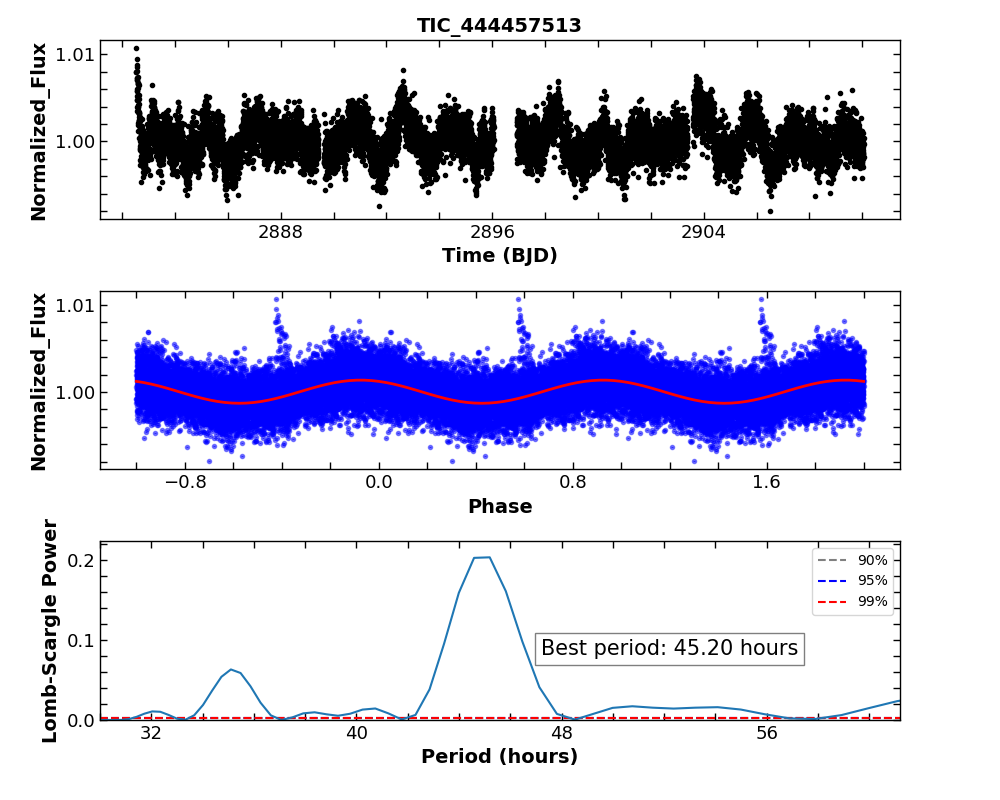}
    \end{minipage}

    \vspace{0.5em}

    % Second Row (3 columns again, last one has two vertical plots)
    \begin{minipage}{0.32\textwidth}
        \centering
        \includegraphics[width=\linewidth,height=5.7cm]{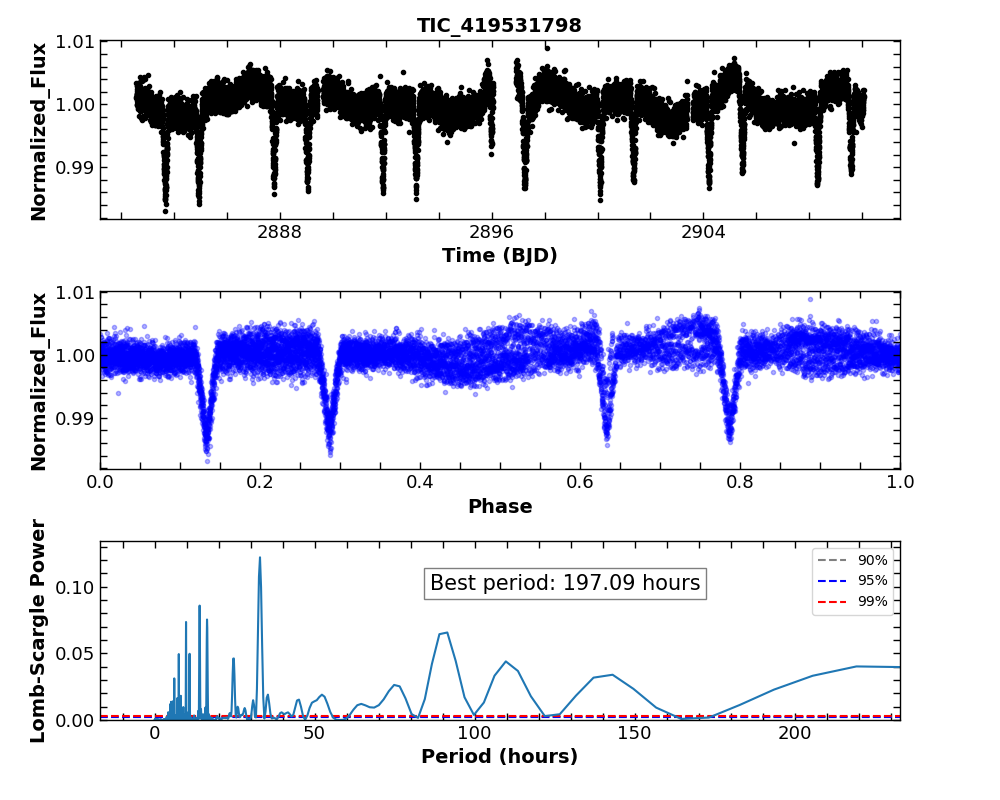}
    \end{minipage}
    \hfill
    \begin{minipage}{0.32\textwidth}
        \centering
        \includegraphics[width=\linewidth,height=5.7cm]{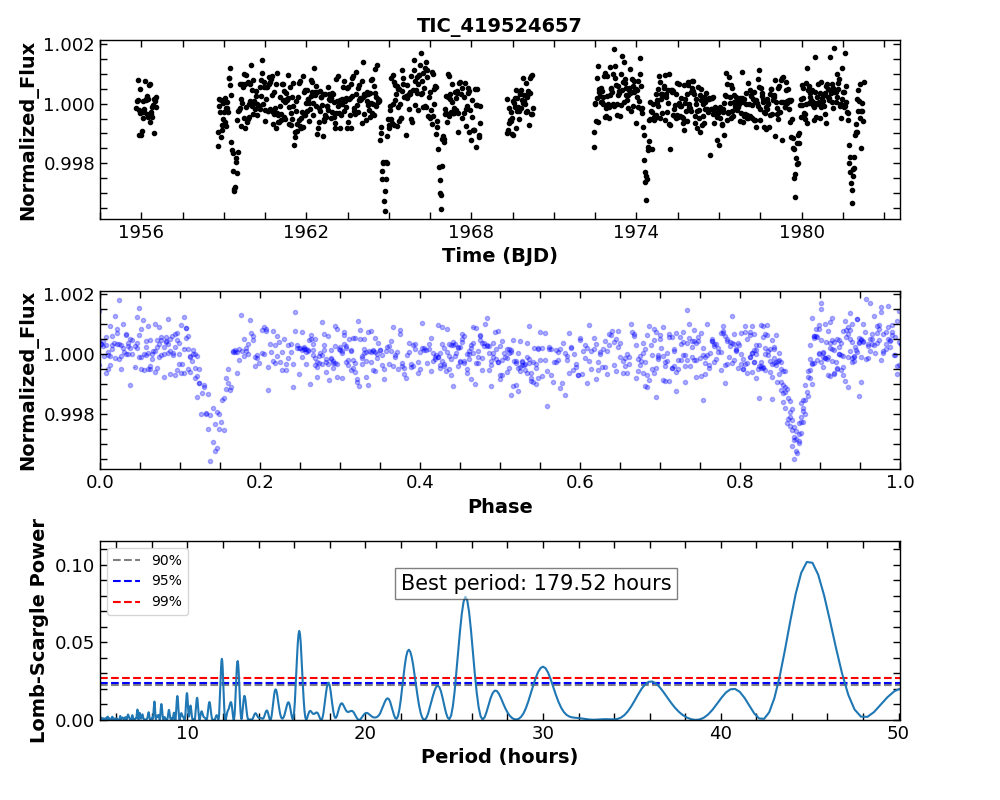}
    \end{minipage}
    \hfill
    \begin{minipage}{0.32\textwidth}
        \centering
        \includegraphics[width=\linewidth,height=2.1cm]{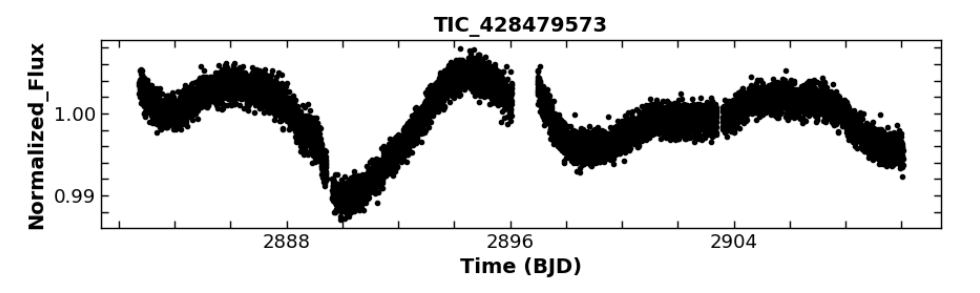} \\
        \vspace{0.2cm}
        \includegraphics[width=\linewidth,height=2.1cm]{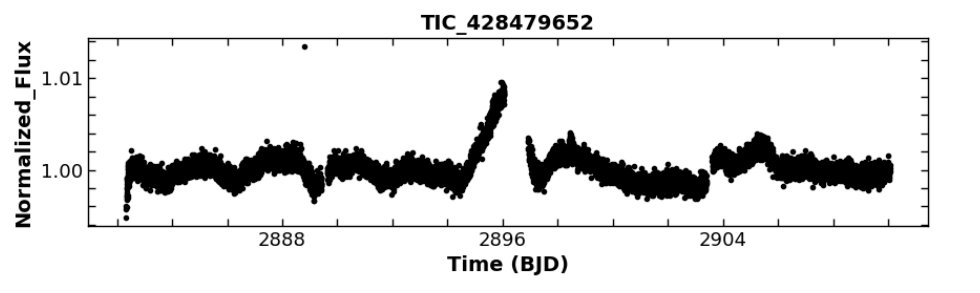}
    \end{minipage}

    \caption{Observed light curves and periodogram analyses of selected variable stars identified in the clusters NGC~146 and King~14 using \textit{TESS} data. For most stars, three vertically arranged panels are shown: the \textit{TESS} light curve (top), the phase-folded light curve with the fitted model overlaid in blue (middle), and the corresponding Lomb–Scargle periodogram (bottom). For TIC~426879573 and TIC~426176052, only the \textit{TESS} light curves are displayed, as reliable periodicities could not be determined due to irregular or long-term variability.}
    \label{fig:light_curve}
\end{figure*}

\subsubsection{Slowly pulsating B-type stars}
The upper main sequence of the H–R diagram hosts SPB stars, which are main-sequence stars of spectral types B2 to B9, with effective temperatures ranging from 22,000 K to 11,000 K. These stars exhibit non-radial, high-order gravity-mode (g-mode) oscillations that are primarily excited by the $\kappa$-mechanism at the iron opacity bump \citep{gautschy1993non}. SPB stars typically display variability on timescales between 0.3 and 5 days \citep{de2007observational}, although more extended periods have also been observed, such as the 188-day rotational modulation of KIC 10526294 \citep{papics2014kic}. In the present analysis of the King 14 cluster, one SPB star was identified: TIC 444457513, exhibiting a pulsation period of 45.20 hours (1.88 days) and an effective temperature of 11,826 K. The astrometric and photometric characteristics of this object strongly support its membership in the cluster. Its position in the H–R diagram is consistent with the theoretical SPB instability strip, as illustrated in Fig.~\ref{fig: HR_diagram}. The observed light curve and phase-folded counterpart, shown in Fig.~\ref{fig:light_curve}, clearly demonstrate periodic g-mode variability. The identification of a bona fide SPB star within the King 14 cluster is of considerable scientific significance. SPB stars situated in OCs benefit from accurately determined ages and metallicities, which are essential for advanced stellar modelling and asteroseismic investigations. This detection contributes to the limited catalog of confirmed SPB stars in intermediate-age clusters, deepening our understanding of pulsational phenomena in B-type stars within such environments.

\subsection{Field and Miscellaneous Variable Stars}

Miscellaneous (Misc) variable stars exhibit variability that does not fit clearly into established classes due to limited or inconclusive data. As these stars are not confirmed cluster members, we refrain from detailed physical interpretation. Continued photometric monitoring may help determine their nature in future studies. The presence of Misc and field variables in the vicinity of NGC~146 and King~14 is relevant primarily from a contamination perspective. Identifying such variables helps differentiate cluster-related variability from unrelated foreground or background stars, improving membership assessments and ensuring cleaner samples for cluster-variable analyses. In this study, we identified three miscellaneous variables: TIC~428479573, TIC~428479652, and TIC~419524657. Owing to the limited number of photometric epochs ($\approx$30), long-period behaviour could not be reliably recovered, and we therefore classify all three as Misc variables without further interpretation. TIC~428479573 and TIC~428479652 have been previously flagged as long-period or semi-regular variables in the literature \citep{chen2020zwicky, watson2006international}, but our dataset does not provide sufficient phase coverage to confirm these periods. TIC~419524657 displays structured flux variations suggestive of eclipsing behavior; however, incomplete phase coverage prevents a conclusive classification, and it is therefore also retained in the Misc category. Although these variables are not used directly to infer cluster properties in the present study, documenting them remains valuable for future time-domain investigations. Their inclusion provides a more complete census of variability in the surrounding region, which can benefit large-scale variability surveys and help distinguish true cluster members from contaminants in subsequent membership analyses. These stars may also serve as useful comparison objects for calibration or benchmarking when more extensive photometric or spectroscopic datasets become available.

% The star ID TIC 419524657 was situated near the $\gamma$ Dor instability strip and their period 3.0 days, it may be a  $\gamma$ Dor variable star but we classified as  Misc variable sue to the postion on the H-R Diagram.

\section{Conclusions} {\label{sec: Conclusion}}

In this study, we have accomplished a comprehensive photometric, kinematic, and orbital investigation of the young OCs NGC~146 and King~14, aiming to understand their stellar content, dynamical state, and the nature of this co-moving, dynamically unbound cluster pair. By combining multiwavelength data from \textit{Gaia}~DR3, \textit{ALLWISE}, Pan-STARRS1, and \textit{TESS}, we have characterized cluster membership, internal structure, stellar populations, orbital dynamics, and variable star content. The main findings of this analysis are summarized below:

\begin{itemize}
  
    \item Membership analysis using the probabilistic method of \citet{balaguer1998determination} and \textit{Gaia}~DR3 astrometric data identified 770 and 690 probable members for NGC~146 and King~14, respectively, each with membership probabilities above 80$\%$. The clear separation of clusters from field stars in the vector-point diagram supports the reliability of this selection. The resulting clean color–magnitude diagrams show defined main sequences, validating membership determination and providing a foundation for photometric and kinematic analyses.

       \item  Structural analysis using King profile fitting yields compact cores ($r_{\mathrm{c}} = 0.66$~pc for NGC~146 and $1.70$~pc for King~14) and limiting radii ($r_{\mathrm{lim}} = 7.0'$ and $10.5'$). The derived tidal radii are $r_{\mathrm{t}} = 11.0$~pc for NGC~146 and $r_{\mathrm{t}} = 14.0$~pc for King~14, corresponding to tidal-to-core radius ratios of $r_{\mathrm{t}}/r_{\mathrm{c}} \approx 16.67$ and $8.23$, respectively. These relatively moderate ratios and concentration parameters ($\delta_{c}\sim4$–5) suggest that both systems are only loosely bound and dynamically evolving. Such structural characteristics, together with their young ages, are consistent with clusters undergoing early-stage mass loss and tidal interaction with the Galactic potential.
    
     \item The clusters exhibit comparable heliocentric distances ($2.98 \pm 0.33$~kpc for NGC~146 and $2.51 \pm 0.23$~kpc), mean proper motions, and spatial locations, indicating a strong kinematic association. However, the derived orbital separation and escape velocities confirm that the pair is not gravitationally bound. Instead, NGC~146 and King~14 form a co-moving, dynamically unbound cluster pair that likely originated within the same giant molecular complex in the Perseus arm.
    
    \item Isochrone fitting shows that NGC~146 is young ($20 \pm 5$~Myr), while King~14 is slightly older ($50 \pm 10$~Myr). The modest age difference suggests formation during closely related star-formation episodes within a common environment, though not from a single collapse event. Both clusters are still in early evolutionary stages, making them valuable probes for studying the transition from bound binary clusters to unbound associations.
    
    \item The mass function slopes ($x = 1.51 \pm 0.18$ for NGC~146 and $x = 1.50 \pm 0.15$ for King~14) agree well with the classical \citet{salpeter1955luminosity} value, implying a normal stellar initial mass function. Evidence of mild mass segregation in both clusters indicates early internal dynamical evolution, possibly enhanced by their former proximity as a binary pair.
    
    \item Galactic orbit integration using the \texttt{galpy} \texttt{MWPotential2014} model shows that both clusters follow nearly circular, low-$Z_{\mathrm{max}}$ orbits confined to the Galactic thin disk, with orbital periods of $\sim$255~Myr and mean orbital radii of $R_{\rm m} \sim 9$~kpc. These results indicate that NGC\,146 and King\,14 form a co-moving but dynamically unbound pair.

   \item The SED-based analysis confirms that TIC~444457513 is a probable member of King~14. Its distance modulus and line-of-sight extinction are in excellent agreement with the cluster parameters. The subsolar metallicity derived for this early-type variable ([Fe/H]~$=-0.21 \pm 0.13$ dex) is lower than some literature estimates, but this likely reflects the intrinsic limitations of SED fitting for hot stars, where continuum-dominated spectra make chemical abundance determinations challenging, rather than a true deviation from the cluster composition. Overall, the stellar parameters obtained are consistent with predictions from single-star evolutionary models, demonstrating the reliability of the {\sc ARIADNE} framework for characterising stellar properties in the absence of high-resolution spectroscopy.
    
    \item Analysis of \textit{TESS} photometric time series identified several variable stars in the fields of NGC~146 and King~14, including $\gamma$~Doradus and SPB pulsators as well as eclipsing binaries. Only one of these variables appears to be a likely cluster member; the remaining sources are field stars and are not analyzed further for cluster properties. However, we provide their light curves for completeness, as cataloguing variable stars in the surrounding region is valuable for future time-domain studies and for distinguishing cluster-associated variability from unrelated field sources. The confirmed member may offer additional constraints on cluster parameters through future pulsation modeling with improved photometric coverage.

    \item Altogether, the NGC~146–King~14 system provides an excellent laboratory for studying the formation and early dynamical evolution of binary OCs in the Milky Way. Their similar kinematic, structural, and orbital properties, combined with a modest age difference, strongly support a common origin within the same giant molecular cloud. However, their present-day spatial configuration and dynamical separation of $\sim32$~pc, together with relative velocities exceeding the binding threshold for their combined mass, indicate that the clusters no longer constitute a gravitationally bound binary system. Instead, they form a co-moving but dynamically independent pair that is gradually dispersing under the influence of the Galactic tidal field. This system, therefore, offers a valuable test case for evaluating theoretical models of primordial cluster pairing, tidal disruption, and long-term orbital evolution in the Galactic disk. Future high-resolution spectroscopic observations and forthcoming \textit{Gaia} DR4 data will be crucial for refining radial velocities, chemical abundances, and orbital constraints, allowing a more precise assessment of the clusters' kinematic alignment and evolutionary link.
    
\end{itemize}

\section*{Acknowledgements}  We are grateful to the anonymous referee for the constructive and insightful comments that significantly improved the clarity and quality of this work. Ing-Guey Jiang acknowledges the support from the National Science and Technology Council (NSTC), Taiwan, under the grants  (NSTC 113-2112-M-007-030 and NSTC 114-2112-M-007-029). We acknowledge the use of archival data from the European Space Agency’s \textit{Gaia} \url{(https://www.cosmos.esa.int/gaia)} mission, processed by the \textit{Gaia} Data Processing and Analysis Consortium (DPAC, https:\url{//www.cosmos.esa.int/web/gaia/dpac/consortium)}, as well as other publicly available photometric data sets from the 2MASS, Pan-STARRS1, and AllWISE. This study also utilized data from the Transiting Exoplanet Survey Satellite (TESS) mission, funded by NASA’s Explorer Program. This research has used the VizieR catalogue access tool, operated by the CDS, Strasbourg, France, and the NASA Astrophysics Data System (ADS). We thank the teams responsible for these surveys for making their data publicly available, which has been essential for the success of this study. D. Bisht also acknowledges the valuable discussions with collaborators and colleagues, which significantly improved the quality of this work.

\software{
\texttt{ARIADNE} \citep{Vines2022},
\texttt{galpy} \citep{Bovy2015}
} 


\bibliography{example}{}

@misc{tess_mast_2021,
doi = {10.17909/T9-NMC8-F686},
url = {http://archive.stsci.edu/doi/resolve/resolve.html?doi=10.17909/t9-nmc8-f686},
author = {{TESS Team}},
title = {TESS Light Curves - All Sectors},
publisher = {STScI/MAST},
year = {2021}
}

@ARTICLE{GaiaDR3,
       author = {{Gaia Collaboration} and {Vallenari}, A. and {Brown}, A.~G.~A. and {Prusti}, T. and {de Bruijne}, J.~H.~J. and {Arenou}, F. and {Babusiaux}, C. and {Biermann}, M. and {Creevey}, O.~L. and {Ducourant}, C. and {Evans}, D.~W. and {Eyer}, L. and {Guerra}, R. and {Hutton}, A. and {Jordi}, C. and {Klioner}, S.~A. and {Lammers}, U.~L. and {Lindegren}, L. and {Luri}, X. and {Mignard}, F. and {Panem}, C. and {Pourbaix}, D. and {Randich}, S. and {Sartoretti}, P. and {Soubiran}, C. and {Tanga}, P. and {Walton}, N.~A. and {Bailer-Jones}, C.~A.~L. and {Bastian}, U. and {Drimmel}, R. and {Jansen}, F. and {Katz}, D. and {Lattanzi}, M.~G. and {van Leeuwen}, F. and {Bakker}, J. and {Cacciari}, C. and {Casta{\~n}eda}, J. and {De Angeli}, F. and {Fabricius}, C. and {Fouesneau}, M. and {Fr{\'e}mat}, Y. and {Galluccio}, L. and {Guerrier}, A. and {Heiter}, U. and {Masana}, E. and {Messineo}, R. and {Mowlavi}, N. and {Nicolas}, C. and {Nienartowicz}, K. and {Pailler}, F. and {Panuzzo}, P. and {Riclet}, F. and {Roux}, W. and {Seabroke}, G.~M. and {Sordo}, R. and {Th{\'e}venin}, F. and {Gracia-Abril}, G. and {Portell}, J. and {Teyssier}, D. and {Altmann}, M. and {Andrae}, R. and {Audard}, M. and {Bellas-Velidis}, I. and {Benson}, K. and {Berthier}, J. and {Blomme}, R. and {Burgess}, P.~W. and {Busonero}, D. and {Busso}, G. and {C{\'a}novas}, H. and {Carry}, B. and {Cellino}, A. and {Cheek}, N. and {Clementini}, G. and {Damerdji}, Y. and {Davidson}, M. and {de Teodoro}, P. and {Nu{\~n}ez Campos}, M. and {Delchambre}, L. and {Dell'Oro}, A. and {Esquej}, P. and {Fern{\'a}ndez-Hern{\'a}ndez}, J. and {Fraile}, E. and {Garabato}, D. and {Garc{\'\i}a-Lario}, P. and {Gosset}, E. and {Haigron}, R. and {Halbwachs}, J. -L. and {Hambly}, N.~C. and {Harrison}, D.~L. and {Hern{\'a}ndez}, J. and {Hestroffer}, D. and {Hodgkin}, S.~T. and {Holl}, B. and {Jan{\ss}en}, K. and {Jevardat de Fombelle}, G. and {Jordan}, S. and {Krone-Martins}, A. and {Lanzafame}, A.~C. and {L{\"o}ffler}, W. and {Marchal}, O. and {Marrese}, P.~M. and {Moitinho}, A. and {Muinonen}, K. and {Osborne}, P. and {Pancino}, E. and {Pauwels}, T. and {Recio-Blanco}, A. and {Reyl{\'e}}, C. and {Riello}, M. and {Rimoldini}, L. and {Roegiers}, T. and {Rybizki}, J. and {Sarro}, L.~M. and {Siopis}, C. and {Smith}, M. and {Sozzetti}, A. and {Utrilla}, E. and {van Leeuwen}, M. and {Abbas}, U. and {{\'A}brah{\'a}m}, P. and {Abreu Aramburu}, A. and {Aerts}, C. and {Aguado}, J.~J. and {Ajaj}, M. and {Aldea-Montero}, F. and {Altavilla}, G. and {{\'A}lvarez}, M.~A. and {Alves}, J. and {Anders}, F. and {Anderson}, R.~I. and {Anglada Varela}, E. and {Antoja}, T. and {Baines}, D. and {Baker}, S.~G. and {Balaguer-N{\'u}{\~n}ez}, L. and {Balbinot}, E. and {Balog}, Z. and {Barache}, C. and {Barbato}, D. and {Barros}, M. and {Barstow}, M.~A. and {Bartolom{\'e}}, S. and {Bassilana}, J. -L. and {Bauchet}, N. and {Becciani}, U. and {Bellazzini}, M. and {Berihuete}, A. and {Bernet}, M. and {Bertone}, S. and {Bianchi}, L. and {Binnenfeld}, A. and {Blanco-Cuaresma}, S. and {Blazere}, A. and {Boch}, T. and {Bombrun}, A. and {Bossini}, D. and {Bouquillon}, S. and {Bragaglia}, A. and {Bramante}, L. and {Breedt}, E. and {Bressan}, A. and {Brouillet}, N. and {Brugaletta}, E. and {Bucciarelli}, B. and {Burlacu}, A. and {Butkevich}, A.~G. and {Buzzi}, R. and {Caffau}, E. and {Cancelliere}, R. and {Cantat-Gaudin}, T. and {Carballo}, R. and {Carlucci}, T. and {Carnerero}, M.~I. and {Carrasco}, J.~M. and {Casamiquela}, L. and {Castellani}, M. and {Castro-Ginard}, A. and {Chaoul}, L. and {Charlot}, P. and {Chemin}, L. and {Chiaramida}, V. and {Chiavassa}, A. and {Chornay}, N. and {Comoretto}, G. and {Contursi}, G. and {Cooper}, W.~J. and {Cornez}, T. and {Cowell}, S. and {Crifo}, F. and {Cropper}, M. and {Crosta}, M. and {Crowley}, C. and {Dafonte}, C. and {Dapergolas}, A. and {David}, M. and {David}, P. and {de Laverny}, P. and {De Luise}, F. and {De March}, R. and {De Ridder}, J. and {de Souza}, R. and {de Torres}, A. and {del Peloso}, E.~F. and {del Pozo}, E. and {Delbo}, M. and {Delgado}, A. and {Delisle}, J. -B. and {Demouchy}, C. and {Dharmawardena}, T.~E. and {Di Matteo}, P. and {Diakite}, S. and {Diener}, C. and {Distefano}, E. and {Dolding}, C. and {Edvardsson}, B. and {Enke}, H. and {Fabre}, C. and {Fabrizio}, M. and {Faigler}, S. and {Fedorets}, G. and {Fernique}, P. and {Fienga}, A. and {Figueras}, F. and {Fournier}, Y. and {Fouron}, C. and {Fragkoudi}, F. and {Gai}, M. and {Garcia-Gutierrez}, A. and {Garcia-Reinaldos}, M. and {Garc{\'\i}a-Torres}, M. and {Garofalo}, A. and {Gavel}, A. and {Gavras}, P. and {Gerlach}, E. and {Geyer}, R. and {Giacobbe}, P. and {Gilmore}, G. and {Girona}, S. and {Giuffrida}, G. and {Gomel}, R. and {Gomez}, A. and {Gonz{\'a}lez-N{\'u}{\~n}ez}, J. and {Gonz{\'a}lez-Santamar{\'\i}a}, I. and {Gonz{\'a}lez-Vidal}, J.~J. and {Granvik}, M. and {Guillout}, P. and {Guiraud}, J. and {Guti{\'e}rrez-S{\'a}nchez}, R. and {Guy}, L.~P. and {Hatzidimitriou}, D. and {Hauser}, M. and {Haywood}, M. and {Helmer}, A. and {Helmi}, A. and {Sarmiento}, M.~H. and {Hidalgo}, S.~L. and {Hilger}, T. and {H{\l}adczuk}, N. and {Hobbs}, D. and {Holland}, G. and {Huckle}, H.~E. and {Jardine}, K. and {Jasniewicz}, G. and {Jean-Antoine Piccolo}, A. and {Jim{\'e}nez-Arranz}, {\'O}. and {Jorissen}, A. and {Juaristi Campillo}, J. and {Julbe}, F. and {Karbevska}, L. and {Kervella}, P. and {Khanna}, S. and {Kontizas}, M. and {Kordopatis}, G. and {Korn}, A.~J. and {K{\'o}sp{\'a}l}, {\'A}. and {Kostrzewa-Rutkowska}, Z. and {Kruszy{\'n}ska}, K. and {Kun}, M. and {Laizeau}, P. and {Lambert}, S. and {Lanza}, A.~F. and {Lasne}, Y. and {Le Campion}, J. -F. and {Lebreton}, Y. and {Lebzelter}, T. and {Leccia}, S. and {Leclerc}, N. and {Lecoeur-Taibi}, I. and {Liao}, S. and {Licata}, E.~L. and {Lindstr{\o}m}, H.~E.~P. and {Lister}, T.~A. and {Livanou}, E. and {Lobel}, A. and {Lorca}, A. and {Loup}, C. and {Madrero Pardo}, P. and {Magdaleno Romeo}, A. and {Managau}, S. and {Mann}, R.~G. and {Manteiga}, M. and {Marchant}, J.~M. and {Marconi}, M. and {Marcos}, J. and {Marcos Santos}, M.~M.~S. and {Mar{\'\i}n Pina}, D. and {Marinoni}, S. and {Marocco}, F. and {Marshall}, D.~J. and {Martin Polo}, L. and {Mart{\'\i}n-Fleitas}, J.~M. and {Marton}, G. and {Mary}, N. and {Masip}, A. and {Massari}, D. and {Mastrobuono-Battisti}, A. and {Mazeh}, T. and {McMillan}, P.~J. and {Messina}, S. and {Michalik}, D. and {Millar}, N.~R. and {Mints}, A. and {Molina}, D. and {Molinaro}, R. and {Moln{\'a}r}, L. and {Monari}, G. and {Mongui{\'o}}, M. and {Montegriffo}, P. and {Montero}, A. and {Mor}, R. and {Mora}, A. and {Morbidelli}, R. and {Morel}, T. and {Morris}, D. and {Muraveva}, T. and {Murphy}, C.~P. and {Musella}, I. and {Nagy}, Z. and {Noval}, L. and {Oca{\~n}a}, F. and {Ogden}, A. and {Ordenovic}, C. and {Osinde}, J.~O. and {Pagani}, C. and {Pagano}, I. and {Palaversa}, L. and {Palicio}, P.~A. and {Pallas-Quintela}, L. and {Panahi}, A. and {Payne-Wardenaar}, S. and {Pe{\~n}alosa Esteller}, X. and {Penttil{\"a}}, A. and {Pichon}, B. and {Piersimoni}, A.~M. and {Pineau}, F. -X. and {Plachy}, E. and {Plum}, G. and {Poggio}, E. and {Pr{\v{s}}a}, A. and {Pulone}, L. and {Racero}, E. and {Ragaini}, S. and {Rainer}, M. and {Raiteri}, C.~M. and {Rambaux}, N. and {Ramos}, P. and {Ramos-Lerate}, M. and {Re Fiorentin}, P. and {Regibo}, S. and {Richards}, P.~J. and {Rios Diaz}, C. and {Ripepi}, V. and {Riva}, A. and {Rix}, H. -W. and {Rixon}, G. and {Robichon}, N. and {Robin}, A.~C. and {Robin}, C. and {Roelens}, M. and {Rogues}, H.~R.~O. and {Rohrbasser}, L. and {Romero-G{\'o}mez}, M. and {Rowell}, N. and {Royer}, F. and {Ruz Mieres}, D. and {Rybicki}, K.~A. and {Sadowski}, G. and {S{\'a}ez N{\'u}{\~n}ez}, A. and {Sagrist{\`a} Sell{\'e}s}, A. and {Sahlmann}, J. and {Salguero}, E. and {Samaras}, N. and {Sanchez Gimenez}, V. and {Sanna}, N. and {Santove{\~n}a}, R. and {Sarasso}, M. and {Schultheis}, M. and {Sciacca}, E. and {Segol}, M. and {Segovia}, J.~C. and {S{\'e}gransan}, D. and {Semeux}, D. and {Shahaf}, S. and {Siddiqui}, H.~I. and {Siebert}, A. and {Siltala}, L. and {Silvelo}, A. and {Slezak}, E. and {Slezak}, I. and {Smart}, R.~L. and {Snaith}, O.~N. and {Solano}, E. and {Solitro}, F. and {Souami}, D. and {Souchay}, J. and {Spagna}, A. and {Spina}, L. and {Spoto}, F. and {Steele}, I.~A. and {Steidelm{\"u}ller}, H. and {Stephenson}, C.~A. and {S{\"u}veges}, M. and {Surdej}, J. and {Szabados}, L. and {Szegedi-Elek}, E. and {Taris}, F. and {Taylor}, M.~B. and {Teixeira}, R. and {Tolomei}, L. and {Tonello}, N. and {Torra}, F. and {Torra}, J. and {Torralba Elipe}, G. and {Trabucchi}, M. and {Tsounis}, A.~T. and {Turon}, C. and {Ulla}, A. and {Unger}, N. and {Vaillant}, M.~V. and {van Dillen}, E. and {van Reeven}, W. and {Vanel}, O. and {Vecchiato}, A. and {Viala}, Y. and {Vicente}, D. and {Voutsinas}, S. and {Weiler}, M. and {Wevers}, T. and {Wyrzykowski}, {\L}. and {Yoldas}, A. and {Yvard}, P. and {Zhao}, H. and {Zorec}, J. and {Zucker}, S. and {Zwitter}, T.},
        title = "{Gaia Data Release 3. Summary of the content and survey properties}",
      journal = {\aap},
         year = 2023,
        month = jun,
       volume = {674},
          eid = {A1},
        pages = {A1},
          doi = {10.1051/0004-6361/202243940},
archivePrefix = {arXiv},
       eprint = {2208.00211},
 primaryClass = {astro-ph.GA},
       adsurl = {https://ui.adsabs.harvard.edu/abs/2023A&A...674A...1G}
}

@article{ricker2015transiting,
  title={Transiting exoplanet survey satellite},
  author={Ricker, George R and Winn, Joshua N and Vanderspek, Roland and Latham, David W and Bakos, G{\'a}sp{\'a}r {\'A} and Bean, Jacob L and Berta-Thompson, Zachory K and Brown, Timothy M and Buchhave, Lars and Butler, Nathaniel R and others},
  journal={Journal of Astronomical Telescopes, Instruments, and Systems},
  volume={1},
  number={1},
  pages={014003--014003},
  year={2015},
  publisher={Society of Photo-Optical Instrumentation Engineers}
}

@article{dias2002new,
  title={New catalogue of optically visible open clusters and candidates},
  author={Dias, WS and Alessi, BS and Moitinho, A and L{\'e}pine, JRD},
  journal={Astronomy \& Astrophysics},
  volume={389},
  number={3},
  pages={871--873},
  year={2002},
  publisher={EDP Sciences}
}

@article{dias2021updated,
  title={Updated parameters of 1743 open clusters based on Gaia DR2},
  author={Dias, Wilton S and Monteiro, H{\'e}ktor and Moitinho, Aandr{\'e} and L{\'e}pine, J{\'a}cques RD and Carraro, Giovanni and Paunzen, Ernst and Alessi, Bruno and Villela, L{\'a}zaro},
  journal={Monthly Notices of the Royal Astronomical Society},
  volume={504},
  number={1},
  pages={356--371},
  year={2021},
  publisher={Oxford University Press}
}

@article{kharchenko2013vizier,
  title={VizieR Online Data Catalog: Milky Way global survey of star clusters. II.(Kharchenko+, 2013)},
  author={Kharchenko, NV and Piskunov, AE and Roeser, S and Schilbach, E and Scholz, R-D},
  journal={VizieR Online Data Catalog},
  volume={355},
  pages={J--A+},
  year={2013}
}

@article{maciejewski2007ccd,
  title={CCD BV survey of 42 open clusters},
  author={Maciejewski, G and Niedzielski, A},
  journal={Astronomy \& Astrophysics},
  volume={467},
  number={3},
  pages={1065--1074},
  year={2007},
  publisher={EDP Sciences}
}

@article{netopil2006photometric,
  title={Photometric survey of marginally investigated open clusters-I. Basel 11b, King 14, Czernik 43},
  author={Netopil, Martin and Maitzen, Hans Michael and Paunzen, Ernst and Claret, Antonio},
  journal={Astronomy \& Astrophysics},
  volume={454},
  number={1},
  pages={179--184},
  year={2006},
  publisher={EDP Sciences}
}

@article{subramaniam2005ngc,
  title={NGC 146: a young open cluster with a Herbig Be star and intermediate mass pre-main sequence stars},
  author={Subramaniam, Annapurni and Sahu, DK and Sagar, R and Vijitha, P},
  journal={Astronomy \& Astrophysics},
  volume={440},
  number={2},
  pages={511--522},
  year={2005},
  publisher={EDP Sciences}
}

@article{chen2020zwicky,
  title={The Zwicky Transient Facility Catalog of Periodic Variable Stars. ApJS 249, 18. doi: 10.3847/1538-4365/ab9cae},
  author={Chen, X and Wang, S and Deng, L and de Grijs, R and Yang, M and Tian, H},
  journal={arXiv preprint arXiv:2005.08662},
  year={2020}
}

@article{wang2019optical,
  title={The optical to mid-infrared extinction law based on the APOGEE, Gaia DR2, Pan-STARRS1, SDSS, APASS, 2MASS, and WISE surveys},
  author={Wang, Shu and Chen, Xiaodian},
  journal={The Astrophysical Journal},
  volume={877},
  number={2},
  pages={116},
  year={2019},
  publisher={American Astronomical Society}
}

@article{pecaut2013intrinsic,
  title={Intrinsic colors, temperatures, and bolometric corrections of pre-main-sequence stars},
  author={Pecaut, Mark J and Mamajek, Eric E},
  journal={The Astrophysical Journal Supplement Series},
  volume={208},
  number={1},
  pages={9},
  year={2013},
  publisher={IOP Publishing}
}

@inproceedings{watson2006international,
  title={The international variable star index (VSX)},
  author={Watson, Christopher L and Henden, Arne A and Price, Aaron},
  booktitle={The Society for Astronomical Sciences 25th Annual Symposium on Telescope Science. Held May 23-25, 2006, at Big Bear, CA. Published by the Society for Astronomical Sciences., p. 47},
  volume={25},
  pages={47},
  year={2006}
}

@article{gautschy1993non,
  title={On non-radial oscillations of B-type stars},
  author={Gautschy, Alfred and Saio, Hideyuki},
  journal={Monthly Notices of the Royal Astronomical Society},
  volume={262},
  number={1},
  pages={213--219},
  year={1993},
  publisher={The Royal Astronomical Society}
}

@article{de2007observational,
  title={Observational Asteroseismology of slowly pulsating B stars},
  author={De Cat, Peter},
  journal={Communications in Astroseismology, Vol. 150, p. 167},
  volume={150},
  pages={167},
  year={2007}
}

@article{papics2014kic,
  title={KIC 10526294: a slowly rotating B star with rotationally split, quasi-equally spaced gravity modes},
  author={P{\'a}pics, PI and Moravveji, Ehsan and Aerts, Conny and Tkachenko, Andrew and Triana, SA and Bloemen, Steven and Southworth, J},
  journal={Astronomy \& Astrophysics},
  volume={570},
  pages={A8},
  year={2014},
  publisher={EDP Sciences}
}

@ARTICLE{wright2010wise,
       author = {{Wright}, Edward L. and {Eisenhardt}, Peter R.~M. and {Mainzer}, Amy K. and {Ressler}, Michael E. and {Cutri}, Roc M. and {Jarrett}, Thomas and {Kirkpatrick}, J. Davy and {Padgett}, Deborah and {McMillan}, Robert S. and {Skrutskie}, Michael and {Stanford}, S.~A. and {Cohen}, Martin and {Walker}, Russell G. and {Mather}, John C. and {Leisawitz}, David and {Gautier}, III, Thomas N. and {McLean}, Ian and {Benford}, Dominic and {Lonsdale}, Carol J. and {Blain}, Andrew and {Mendez}, Bryan and {Irace}, William R. and {Duval}, Valerie and {Liu}, Fengchuan and {Royer}, Don and {Heinrichsen}, Ingolf and {Howard}, Joan and {Shannon}, Mark and {Kendall}, Martha and {Walsh}, Amy L. and {Larsen}, Mark and {Cardon}, Joel G. and {Schick}, Scott and {Schwalm}, Mark and {Abid}, Mohamed and {Fabinsky}, Beth and {Naes}, Larry and {Tsai}, Chao-Wei},
        title = "{The Wide-field Infrared Survey Explorer (WISE): Mission Description and Initial On-orbit Performance}",
      journal = {The Astronomical Journal},
         year = 2010,
        month = dec,
       volume = {140},
       number = {6},
        pages = {1868-1881},
          doi = {10.1088/0004-6256/140/6/1868},
archivePrefix = {arXiv},
       eprint = {1008.0031},
 primaryClass = {astro-ph.IM},
       adsurl = {https://ui.adsabs.harvard.edu/abs/2010AJ....140.1868W}
}

@article{brown2021gaia,
  title={Gaia Early Data Release 3-Summary of the contents and survey properties},
  author={Brown, Anthony GA and Vallenari, Antonella and Prusti, T and De Bruijne, JHJ and Babusiaux, C and Biermann, M and Creevey, OL and Evans, DW and Eyer, L and Hutton, A and others},
  journal={Astronomy \& Astrophysics},
  volume={649},
  pages={A1},
  year={{2021}},
  publisher={EDP sciences}
}

@article{Bovy2015,
  author  =  {{Bovy}, Jo},
  title  =  "{galpy: A python Library for Galactic Dynamics}",
  journal  =  {The Astrophysical Journal Supplement},
  year  =  2015,
  month  =  feb,
  volume  =  {216},
  number  =  {2},
  eid  =  {29},
  pages  =  {29},
  doi  =  {10.1088/0067-0049/216/2/29},
  archivePrefix  =  {arXiv},
  eprint  =  {1412.3451},
  primaryClass  =  {astro-ph.GA},
  adsurl  =  {https://ui.adsabs.harvard.edu/abs/2015ApJS..216...29B}
  }

@article{Bovy2012,
       author = {Bovy, Jo and Tremaine, Scott},
        title = {On the Local Dark Matter Density},
      journal = {The Astrophysical Journal},
         year = 2012,
        month = sep,
       volume = {756},
       number = {1},
          eid = {89},
        pages = {89},
          doi = {10.1088/0004-637X/756/1/89}}

@ARTICLE{Haroon2025,
       author = {{Haroon}, A.~A. and {Elsanhoury}, W.~H. and {Elkholy}, E.~A. and {Saad}, A.~S. and {{\c{C}}{\i}nar}, D.~C.},
        title = "{Study of open star clusters using the gaia DR3: I-poorly studied king 2 and king 5}",
      journal = {Physica Scripta},
         year = 2025,
        month = may,
       volume = {100},
       number = {5},
          eid = {055006},
        pages = {055006},
          doi = {10.1088/1402-4896/adbf71},
       adsurl = {https://ui.adsabs.harvard.edu/abs/2025PhyS..100e5006H}
}

@ARTICLE{Cinar2024,
       author = {{{\c{C}}{\i}nar}, Deniz Cennet and {Ta{\c{s}}demir}, Seval and {Koc}, Seliz and {Iyer}, Srishti},
        title = "{SED Analysis of the Old Open Cluster NGC 188}",
      journal = {Physics and Astronomy Reports},
         year = 2024,
        month = jun,
       volume = {2},
       number = {1},
        pages = {1-17},
          doi = {10.26650/PAR.2024.00002},
archivePrefix = {arXiv},
       eprint = {2404.13115},
 primaryClass = {astro-ph.GA},
       adsurl = {https://ui.adsabs.harvard.edu/abs/2024PARep...2....1C}
}

@ARTICLE{Yontan2022,
       author = {{Yontan}, T. and {{\c{C}}akmak}, H. and {Bilir}, S. and {Banks}, T. and {Ra{\'u}l}, M. and {Canbay}, R. and {Ko{\c{c}}}, S. and {Ta{\c{s}}demir}, S. and {Er{\c{c}}ay}, H. and {Tan{\i}k Ozt{\"u}rk}, B. and {{\c{C}}{\i}nar}, D.~C.},
        title = "{A Study of the NGC 1193 and NGC 1798 Open Clusters Using CCD UBV Photometric and Gaia EDR3 Data}",
      journal = {\rmxaa},
         year = 2022,
        month = oct,
       volume = {58},
        pages = {333-353},
          doi = {10.22201/ia.01851101p.2022.58.02.14},
archivePrefix = {arXiv},
       eprint = {2207.06407},
 primaryClass = {astro-ph.GA},
       adsurl = {https://ui.adsabs.harvard.edu/abs/2022RMxAA..58..333Y}
}

@ARTICLE{Bostanci2015,
       author = {{Bostanc{\i}}, Z.~F. and {Ak}, T. and {Yontan}, T. and {Bilir}, S. and {G{\"u}ver}, T. and {Ak}, S. and {{\c{C}}ak{\i}rl{\i}}, {\"O}. and {{\"O}zdarcan}, O. and {Paunzen}, E. and {De Cat}, P. and {Fu}, J.~N. and {Zhang}, Y. and {Hou}, Y. and {Li}, G. and {Wang}, Y. and {Zhang}, W. and {Shi}, J. and {Wu}, Y.},
        title = "{A comprehensive study of the open cluster NGC 6866}",
      journal = {\mnras},
         year = 2015,
        month = oct,
       volume = {453},
       number = {1},
        pages = {1095-1107},
          doi = {10.1093/mnras/stv1665},
archivePrefix = {arXiv},
       eprint = {1507.05968},
 primaryClass = {astro-ph.GA},
       adsurl = {https://ui.adsabs.harvard.edu/abs/2015MNRAS.453.1095B}
}

@ARTICLE{Tasdemir2025,
       author = {{Ta{\c{s}}demir}, Seval and {{\c{C}}{\i}nar}, Deniz Cennet and {Canbay}, Remziye and {Tastan}, Serkan and {Elsanhoury}, Waleed H. and {Haroon}, Aly A.},
        title = "{Comprehensive Analysis of Middle-Aged Open Cluster NGC 6793 in Vulpecula via Gaia DR3 Data}",
      journal = {Physics and Astronomy Reports},
         year = 2025,
        month = jun,
       volume = {3},
        pages = {1-12},
          doi = {10.48550/arXiv.2503.19015},
archivePrefix = {arXiv},
       eprint = {2503.19015},
 primaryClass = {astro-ph.GA},
       adsurl = {https://ui.adsabs.harvard.edu/abs/2025PARep...3....1T}
}

@ARTICLE{Elsanhoury2025,
       author = {{Elsanhoury}, W.~H. and {Haroon}, A.~A. and {Elkholy}, E.~A. and {{\c{C}}{\i}nar}, D.~C.},
        title = "{Deeply comprehensive astrometric, photometric, and kinematic studies of the three OCSN open clusters with Gaia DR3}",
      journal = {Journal of Astrophysics and Astronomy},
         year = 2025,
        month = mar,
       volume = {46},
       number = {1},
          eid = {21},
        pages = {21},
          doi = {10.1007/s12036-025-10044-0},
archivePrefix = {arXiv},
       eprint = {2412.07871},
 primaryClass = {astro-ph.SR},
       adsurl = {https://ui.adsabs.harvard.edu/abs/2025JApA...26...21E}
}

@ARTICLE{Yousef2025b,
       author = {{Alzhrani}, A.~Y. and {Haroon}, A.~A. and {Elsanhoury}, W.~H. and {{\c{C}}{\i}nar}, D.~C.},
        title = "{Enhancing SED-based astrometric, photometric, and kinematic studies of SAI 72 and SAI 75 using Gaia DR3}",
      journal = {Journal of Astrophysics and Astronomy},
         year = 2025,
        month = aug,
       volume = {46},
       number = {2},
          eid = {50},
        pages = {50},
          doi = {10.1007/s12036-025-10076-6},
archivePrefix = {arXiv},
       eprint = {2504.15341},
 primaryClass = {astro-ph.GA},
       adsurl = {https://ui.adsabs.harvard.edu/abs/2025JApA...46...50A}
}

@ARTICLE{Cinar2025,
       author = {{{\c{C}}{\i}nar}, Deniz Cennet and {Bilir}, Sel{\c{c}}uk and {{\c{S}}ahin}, Timur and {Plevne}, Olcay},
        title = "{Tracing the Galactic Origins of Selected Four G-type Stars in the Solar Neighborhood}",
      journal = {\aj},
         year = 2025,
        month = jul,
       volume = {170},
       number = {1},
          eid = {13},
        pages = {13},
          doi = {10.3847/1538-3881/add343},
archivePrefix = {arXiv},
       eprint = {2504.17844},
 primaryClass = {astro-ph.SR},
       adsurl = {https://ui.adsabs.harvard.edu/abs/2025AJ....170...13C}
}

@article{salpeter1955luminosity,
  title={The luminosity function and stellar evolution.},
  author={Salpeter, Edwin E},
  journal={The Astrophysical Journal},
  volume={121},
  pages={161},
  year={{1955}}
}

@article{balaguer1998determination,
  title={Determination of proper motions and membership of the open clusters NGC 1817 and NGC 1807},
  author={Balaguer-N{\'u}nez, L and Tian, KP and Zhao, JL},
  journal={Astronomy and Astrophysics Supplement Series},
  volume={133},
  number={3},
  pages={387--394},
  year={{1998}},
  publisher={EDP Sciences}
}

@article{girard1989relative,
  title={Relative proper motions and the stellar velocity dispersion of the open cluster M67},
  author={Girard, Terrence M and Grundy, William M and L{\'o}pez, Carlos E and van Altena, William F},
  journal={The Astronomical Journal},
  volume={98},
  pages={227--243},
  year={{1989}}
}

@article{Cantat-Gaudin-Anders_2020,
       author = {Cantat-Gaudin, T. and Anders, F.},
        title = {Clusters and mirages: cataloguing stellar aggregates in the {M}ilky {W}ay},
      journal = {Astronomy \& Astrophysics},
         year = 2020,
        month = jan,
       volume = {633},
          eid = {A99},
        pages = {A99},
          doi = {10.1051/0004-6361/201936691}}

@article{prusti2016gaia,
  title={The Gaia mission},
  author={Prusti, T and de Bruijne, JHJ and Brown, AGA and Vallenari, A and Babusiaux, C and Bailer-Jones, CAL and Bastian, U and Biermann, M and Evans, DW and Eyer, L},
  journal={Astronomy \& Astrophysics},
  volume={595},
  pages={A1},
  year={{2016}},
  publisher={EDP Sciences}
}

@article{corsaro2017spin,
  title={Spin alignment of stars in old open clusters},
  author={Corsaro, Enrico and Lee, Yueh-Ning and Garc{\'\i}a, Rafael A and Hennebelle, Patrick and Mathur, Savita and Beck, Paul G and Mathis, Stephane and Stello, Dennis and Bouvier, J{\'e}r{\^o}me},
  journal={Nature Astronomy},
  volume={1},
  number={4},
  pages={0064},
  year={{2017}},
  publisher={Nature Publishing Group UK London}
}

@article{leroy2018forming,
  title={Forming super star clusters in the central starburst of NGC 253},
  author={Leroy, Adam K and Bolatto, Alberto D and Ostriker, Eve C and Walter, Fabian and Gorski, Mark and Ginsburg, Adam and Krieger, Nico and Levy, Rebecca C and Meier, David S and Mills, Elisabeth and others},
  journal={The Astrophysical Journal},
  volume={869},
  number={2},
  pages={126},
  year={{2018}},
  publisher={IOP Publishing}
}

@article{bisht2020investigation,
  title={An investigation of poorly studied open cluster NGC 4337 using multicolor photometric and Gaia DR2 astrometric data},
  author={Bisht, D and Elsanhoury, WH and Zhu, Qingfeng and Sariya, Devesh P and Yadav, RKS and Rangwal, Geeta and Durgapal, Alok and Jiang, Guey},
  journal={The Astronomical Journal},
  volume={160},
  number={3},
  pages={119},
  year={{2020}},
  publisher={IOP Publishing}
}

@article{marigo2017new,
  title={A new generation of PARSEC-COLIBRI stellar isochrones including the TP-AGB phase},
  author={Marigo, Paola and Girardi, L{\'e}o and Bressan, Alessandro and Rosenfield, Philip and Aringer, Bernhard and Chen, Yang and Dussin, Marco and Nanni, Ambra and Pastorelli, Giada and Rodrigues, Tha{\'\i}se S and others},
  journal={The Astrophysical Journal},
  volume={835},
  number={1},
  pages={77},
  year={{2017}},
  publisher={IOP Publishing}
}

@article{cantat2018gaia,
  title={A Gaia DR2 view of the open cluster population in the Milky Way},
  author={Cantat-Gaudin, Tristan and Jordi, C and Vallenari, Antonella and Bragaglia, Angela and Balaguer-N{\'u}{\~n}ez, L and Soubiran, C and Bossini, D and Moitinho, A and Castro-Ginard, A and Krone-Martins, A and others},
  journal={Astronomy \& Astrophysics},
  volume={618},
  pages={A93},
  year={{2018}},
  publisher={EDP Sciences}
}

@article{sariya2021astrometric,
  title={Astrometric and photometric investigation of three old age open clusters in the Gaia era: Berkeley 32, Berkeley 98, and King 23},
  author={Sariya, Devesh P and Jiang, Guey and Bisht, D and Yadav, RKS and Rangwal, G},
  journal={The Astronomical Journal},
  volume={161},
  number={3},
  pages={102},
  year={2021},
  publisher={IOP Publishing}
}

@article{angelo2022investigating,
  title={Investigating Galactic binary cluster candidates with Gaia EDR3},
  author={Angelo, MS and Santos Jr, JFC and Maia, FFS and Corradi, WJB},
  journal={Monthly Notices of the Royal Astronomical Society},
  volume={510},
  number={4},
  pages={5695--5724},
  year={2022},
  publisher={Oxford University Press}
}

@article{king1962structure,
  title={The structure of star clusters. I. an empirical density law},
  author={King, Ivan},
  journal={Astronomical Journal, Vol. 67, p. 471 (1962)},
  volume={67},
  pages={471},
  year={{1962}}
}

@article{stubbs2010precise,
  title={Precise throughput determination of the PanSTARRS telescope and the gigapixel imager using a calibrated silicon photodiode and a tunable laser: initial results},
  author={Stubbs, Christopher W and Doherty, Peter and Cramer, Claire and Narayan, Gautham and Brown, Yorke J and Lykke, Keith R and Woodward, John T and Tonry, John L},
  journal={The Astrophysical Journal Supplement Series},
  volume={191},
  number={2},
  pages={376},
  year={2010},
  publisher={IOP Publishing}
}

@article{hodapp2004design,
  title={Design of the Pan-STARRS telescopes},
  author={Hodapp, KW and Kaiser, N and Aussel, H and Burgett, W and Chambers, KC and Chun, M and Dombeck, T and Douglas, A and Hafner, D and Heasley, J and others},
  journal={Astronomische Nachrichten: Astronomical Notes},
  volume={325},
  number={6-8},
  pages={636--642},
  year={2004},
  publisher={Wiley Online Library}
}

@article{de2010evolution,
  title={The Evolution of Primordial Binary Open Star Clusters: Mergers, Shredded Secondaries, and Separated Twins},
  author={de La Fuente Marcos, R and de La Fuente Marcos, C},
  journal={The Astrophysical Journal},
  volume={719},
  number={1},
  pages={104},
  year={2010},
  publisher={IOP Publishing}
}

@BOOK{2008gady.book.....B,
       author = {{Binney}, James and {Tremaine}, Scott},
        title = "{Galactic Dynamics: Second Edition}",
         year = 2008,
       adsurl = {https://ui.adsabs.harvard.edu/abs/2008gady.book.....B}
}

@article{panwar2024low,
  title={Low-mass stellar and substellar content of the young cluster Berkeley 59},
  author={Panwar, Neelam and Rishi, C and Sharma, Saurabh and Ojha, Devendra K and Samal, Manash R and Singh, HP and Yadav, Ram Kesh},
  journal={The Astronomical Journal},
  volume={168},
  number={2},
  pages={89},
  year={2024},
  publisher={IOP Publishing}
}

@article{chand2025long,
  title={Long-term investigation of an open cluster Berkeley 65},
  author={Chand, Tarak and Sharma, Saurabh and Singh, Koshvendra and Pandey, Jeewan and Verma, Aayushi and Kaur, Harmeen and Chakraborty, Manojit and Ojha, Devendra K and Singh, Ajay Kumar and others},
  journal={arXiv preprint arXiv:2505.24240},
  year={2025}
}

@article{tonry2012pan,
  title={The Pan-STARRS1 photometric system},
  author={Tonry, JL and Stubbs, Christopher W and Lykke, Keith R and Doherty, Peter and Shivvers, IS and Burgett, WS and Chambers, KC and Hodapp, KW and Kaiser, N and Kudritzki, R-P and others},
  journal={The Astrophysical Journal},
  volume={750},
  number={2},
  pages={99},
  year={2012},
  publisher={IOP Publishing}
}

@article{schlafly2012photometric,
  title={Photometric calibration of the first 1.5 years of the Pan-STARRS1 survey},
  author={Schlafly, EF and Finkbeiner, DP and Juri{\'c}, M and Magnier, EA and Burgett, WS and Chambers, KC and Grav, T and Hodapp, KW and Kaiser, N and Kudritzki, R-P and others},
  journal={The Astrophysical Journal},
  volume={756},
  number={2},
  pages={158},
  year={2012},
  publisher={IOP Publishing}
}

@article{chambers2016pan,
  title={The pan-starrs1 surveys},
  author={Chambers, Kenneth C and Magnier, EA and Metcalfe, N and Flewelling, HA and Huber, ME and Waters, CZ and Denneau, L and Draper, PW and Farrow, D and Finkbeiner, DP and others},
  journal={arXiv preprint arXiv:1612.05560},
  year={2016}
}

@article{hasan2023gaps,
  title={Gaps in the Main-Sequence of Star Cluster Hertzsprung Russell Diagrams},
  author={Hasan, Priya},
  journal={arXiv preprint arXiv:2310.17725},
  year={2023}
}

@article{bonatto2007open,
  title={Open clusters in dense fields: the importance of field-star decontamination for NGC 5715, Lyng{\aa} 4, Lyng{\aa} 9, Trumpler 23, Trumpler 26 and Czernik 37},
  author={Bonatto, Charles and Bica, Eduardo},
  journal={Monthly Notices of the Royal Astronomical Society},
  volume={377},
  number={3},
  pages={1301--1323},
  year={2007},
  publisher={Blackwell Publishing Ltd Oxford, UK}
}

@article{bate2003formation,
  title={The formation of a star cluster: predicting the properties of stars and brown dwarfs},
  author={Bate, Matthew R and Bonnell, Ian A and Bromm, Volker},
  journal={Monthly Notices of the Royal Astronomical Society},
  volume={339},
  number={3},
  pages={577--599},
  year={2003},
  publisher={Blackwell Science Ltd Oxford, UK}
}

@article{harris1994supergiant,
  title={Supergiant molecular clouds and the formation of globular cluster systems},
  author={Harris, William E and Pudritz, Ralph E},
  journal={The Astrophysical Journal, Part 1 (ISSN 0004-637X), vol. 429, no. 1, p. 177-191},
  volume={429},
  pages={177--191},
  year={1994}
}

@article{piecka2021aggregates,
  title={Aggregates of clusters in the Gaia data},
  author={Piecka, Martin and Paunzen, Ernst},
  journal={Astronomy \& Astrophysics},
  volume={649},
  pages={A54},
  year={2021},
  publisher={EDP Sciences}
}

@article{piatti2010evidence,
  title={Evidence of enhanced formation episodes in the Galactic open cluster system},
  author={Piatti, AE},
  journal={Astronomy \& Astrophysics},
  volume={513},
  pages={L13},
  year={2010},
  publisher={EDP Sciences}
}

@article{arnold2017binary,
  title={How do binary clusters form?},
  author={Arnold, Becky and Goodwin, Simon P and Griffiths, Daniel W and Parker, Richard J},
  journal={Monthly Notices of the Royal Astronomical Society},
  volume={471},
  number={2},
  pages={2498--2507},
  year={2017},
  publisher={Oxford University Press}
}

@article{mora2019collision,
  title={On collision course: The nature of the binary star cluster NGC2006/SL 538},
  author={Mora, Marcelo D and Puzia, Thomas H and Chanam{\'e}, Julio},
  journal={Astronomy \& Astrophysics},
  volume={622},
  pages={A65},
  year={2019},
  publisher={EDP Sciences}
}

@article{darma2021formation,
  title={The formation of binary star clusters in the Milky Way and Large Magellanic Cloud},
  author={Darma, R and Arifyanto, MI and Kouwenhoven, MBN},
  journal={Monthly Notices of the Royal Astronomical Society},
  volume={506},
  number={3},
  pages={4603--4620},
  year={2021},
  publisher={Oxford University Press}
}

@article{lada2003embedded,
  title={Embedded clusters in molecular clouds},
  author={Lada, Charles J and Lada, Elizabeth A},
  journal={Annual Review of Astronomy and Astrophysics},
  volume={41},
  number={1},
  pages={57--115},
  year={2003},
  publisher={Annual Reviews 4139 El Camino Way, PO Box 10139, Palo Alto, CA 94303-0139, USA}
}

@article{van1996formation,
  title={Formation of the galaxy},
  author={van den Bergh, Sidney},
  journal={Publications of the Astronomical Society of the Pacific},
  volume={108},
  number={729},
  pages={986},
  year={1996},
  publisher={IOP Publishing}
}

@article{de2009double,
  title={Double or binary: on the multiplicity of open star clusters},
  author={de La Fuente Marcos, R and de La Fuente Marcos, C},
  journal={Astronomy \& Astrophysics},
  volume={500},
  number={2},
  pages={L13--L16},
  year={2009},
  publisher={EDP Sciences}
}

@article{subramaniam1995probable,
  title={Probable binary open star clusters in the Galaxy.},
  author={Subramaniam, A and Gorti, U and Sagar, R and Bhatt, HC},
  year={1995},
  publisher={EDP Science}
}

@article{loktin1997selection,
  title={The selection of probable multiple open clusters in our Galaxy},
  author={Loktin, AV},
  journal={Astronomical and Astrophysical Transactions},
  volume={14},
  number={3},
  pages={181--193},
  year={1997},
  publisher={Taylor \& Francis}
}

@article{soubiran2019open,
  title={Open cluster kinematics with Gaia DR2 (Corrigendum)},
  author={Soubiran, C and Cantat-Gaudin, T and Romero-G{\'o}mez, M and Casamiquela, L and Jordi, C and Vallenari, A and Antoja, T and Balaguer-N{\'u}{\~n}ez, L and Bossini, D and Bragaglia, A and others},
  journal={Astronomy \& Astrophysics},
  volume={623},
  pages={C2},
  year={2019},
  publisher={EDP Sciences}
}

@article{liu2019catalog,
  title={A catalog of newly identified star clusters in Gaia DR2},
  author={Liu, Lei and Pang, Xiaoying},
  journal={The Astrophysical Journal Supplement Series},
  volume={245},
  number={2},
  pages={32},
  year={2019},
  publisher={IOP Publishing}
}

@article{cantat2020painting,
  title={Painting a portrait of the Galactic disc with its stellar clusters},
  author={Cantat-Gaudin, Tristan and Anders, Friedrich and Castro-Ginard, Alfred and Jordi, Carme and Romero-G{\'o}mez, Merc{\`e} and Soubiran, C and Casamiquela, Laia and Tarricq, Y and Moitinho, Andr{\'e} and Vallenari, Antonella and others},
  journal={Astronomy \& Astrophysics},
  volume={640},
  pages={A1},
  year={2020},
  publisher={EDP Sciences}
}

@article{casado2021list,
  title={The List of Possible Double and Multiple Open Clusters between Galactic Longitudes 240 and 270},
  author={Casado, Juan},
  journal={Astronomy Reports},
  volume={65},
  number={9},
  pages={755--775},
  year={2021},
  publisher={Springer}
}

@article{dalessandro2018unexpected,
  title={The unexpected kinematics of multiple populations in NGC 6362: do binaries play a role?},
  author={Dalessandro, Emanuele and Mucciarelli, A and Bellazzini, Michele and Sollima, A and Vesperini, Enrico and Hong, J and H{\'e}nault-Brunet, Vincent and Ferraro, FR and Ibata, R and Lanzoni, B and others},
  journal={The Astrophysical Journal},
  volume={864},
  number={1},
  pages={33},
  year={2018},
  publisher={IoP Publishing}
}

@article{kharchenko2005astrophysical,
  title={Astrophysical parameters of Galactic open clusters},
  author={Kharchenko, NV and Piskunov, AE and R{\"o}ser, S and Schilbach, E and Scholz, R-D},
  journal={Astronomy \& Astrophysics},
  volume={438},
  number={3},
  pages={1163--1173},
  year={2005},
  publisher={EDP Sciences}
}

@article{rojas2010metal,
  title={Metal-rich M-dwarf planet hosts: Metallicities with k-band spectra},
  author={Rojas-Ayala, B{\'a}rbara and Covey, Kevin R and Muirhead, Philip S and Lloyd, James P},
  journal={The Astrophysical Journal Letters},
  volume={720},
  number={1},
  pages={L113},
  year={2010},
  publisher={IOP Publishing}
}

@article{sagar1978gap,
  title={The Gap in the HR diagram of Open Clusters with Special Reference to NGC 2169, NGC 1778 and TR 1},
  author={Sagar, Ram and Joshi, UC},
  journal={Bulletin of the Astronomical Soceity of India, vol. 6, p. 37},
  volume={6},
  pages={37},
  year={1978}
}

@article{bohm1974gap,
  title={The gap in the two-color diagram of main-sequence stars},
  author={Bohm-Vitense, E and Canterna, R},
  journal={Astrophysical Journal, Vol. 194, pp. 629-636 (1974)},
  volume={194},
  pages={629--636},
  year={1974}
}

@article{rachford2000relationship,
  title={The Relationship Betweenthe B{\"o}hm-Vitense Gapand Stellar Activity inOpen Clusters},
  author={Rachford, Brian L and Canterna, R},
  journal={The Astronomical Journal},
  volume={119},
  number={3},
  pages={1296},
  year={2000},
  publisher={IOP Publishing}
}

@article{geller2010wiyn,
  title={WIYN Open Cluster Study. XXXVIII. Stellar Radial Velocities in the Young Open Cluster M35 (NGC 2168)},
  author={Geller, Aaron M and Mathieu, Robert D and Braden, Ella K and Meibom, S{\o}ren and Platais, Imants and Dolan, Christopher J},
  journal={The Astronomical Journal},
  volume={139},
  number={4},
  pages={1383},
  year={2010},
  publisher={IOP Publishing}
}

@article{sanders1971improved,
  title={An improved method for computing membership probabilities in open clusters.},
  author={Sanders, WL},
  journal={Astronomy and Astrophysics, Vol. 14, p. 226-232},
  volume={14},
  pages={226--232},
  year={1971}
}

@article{belwal2025unveiling,
  title={Unveiling dynamics and variability in open clusters: insights from a comprehensive analysis of six galactic clusters},
  author={Belwal, Kuldeep and Bisht, D and Jiang, Ing-Guey and Yadav, RKS and Raj, Ashish and Rangwal, Geeta and Dattatrey, Arvind K and Bisht, Mohit Singh and Durgapal, Alok},
  journal={Monthly Notices of the Royal Astronomical Society},
  volume={544},
  number={1},
  pages={988--1011},
  year={2025},
  publisher={Oxford University Press}
}

@article{vasilevskis1958relative,
  title={Relative proper motions of stars in the region of the open cluster NGC 6633.},
  author={Vasilevskis, S and Klemola, A and Preston, G},
  journal={Astronomical Journal, Vol. 63, p. 387-395 (1958)},
  volume={63},
  pages={387--395},
  year={1958}
}

@article{groenewegen2021parallax,
  title={The parallax zero-point offset from Gaia EDR3 data},
  author={Groenewegen, MAT},
  journal={Astronomy \& Astrophysics},
  volume={654},
  pages={A20},
  year={2021},
  publisher={EDP Sciences}
}

@article{bailer2015estimating,
  title={Estimating distances from parallaxes},
  author={Bailer-Jones, Coryn AL},
  journal={Publications of the Astronomical Society of the Pacific},
  volume={127},
  number={956},
  pages={994--1009},
  year={2015},
  publisher={The Astronomical Society of the Pacific}
}

@article{bailer2018estimating,
  title={Estimating distance from parallaxes. IV. Distances to 1.33 billion stars in Gaia data release 2},
  author={Bailer-Jones, CAL and Rybizki, J and Fouesneau, M and Mantelet, G and Andrae, R},
  journal={The Astronomical Journal},
  volume={156},
  number={2},
  pages={58},
  year={2018},
  publisher={IOP Publishing}
}

@article{lindegren2021gaia,
  title={Gaia early data release 3-the astrometric solution},
  author={Lindegren, Lennart and Klioner, SA and Hern{\'a}ndez, J and Bombrun, A and Ramos-Lerate, M and Steidelm{\"u}ller, Hea and Bastian, U and Biermann, M and de Torres, A and Gerlach, E and others},
  journal={Astronomy \& Astrophysics},
  volume={649},
  pages={A2},
  year={2021},
  publisher={EDP Sciences}
}

@ARTICLE{Song_2023,
       author = {{Song}, Fang-Fang and {Niu}, Hu-Biao and {Esamdin}, Ali and {Zhang}, Yu and {Zeng}, Xiang-Yun},
        title = "{Variable Star Detection in the Field of Open Cluster NGC 188}",
      journal = {Research in Astronomy and Astrophysics},
         year = 2023,
        month = sep,
       volume = {23},
       number = {9},
          eid = {095015},
        pages = {095015},
          doi = {10.1088/1674-4527/acd52f},
archivePrefix = {arXiv},
       eprint = {2304.12738},
 primaryClass = {astro-ph.SR},
       adsurl = {https://ui.adsabs.harvard.edu/abs/2023RAA....23i5015S}
}

@ARTICLE{Li2024,
       author = {{Li}, Zhongmu and {Zhu}, Zhanpeng},
        title = "{A catalog of newly discovered close binary open clusters in the Milky Way from Gaia DR3}",
      journal = {arXiv e-prints},
         year = 2024,
        month = may,
          eid = {arXiv:2405.02530},
        pages = {arXiv:2405.02530},
          doi = {10.48550/arXiv.2405.02530},
archivePrefix = {arXiv},
       eprint = {2405.02530},
 primaryClass = {astro-ph.GA},
       adsurl = {https://ui.adsabs.harvard.edu/abs/2024arXiv240502530L}
}

@ARTICLE{Haroon2024,
       author = {{Haroon}, A.~A. and {Elsanhoury}, W.~H. and {Saad}, A.~S. and {Elkholy}, E.~A.},
        title = "{Deep investigation of a pair of open clusters NGC 7031 and NGC 7086 utilizing Gaia DR3}",
      journal = {Contributions of the Astronomical Observatory Skalnate Pleso},
         year = 2024,
        month = aug,
       volume = {54},
       number = {3},
        pages = {22-48},
          doi = {10.31577/caosp.2024.54.3.22},
       adsurl = {https://ui.adsabs.harvard.edu/abs/2024CoSka..54c..22H}
}

@ARTICLE{Palma_2024,
       author = {{Palma}, Tali and {Coenda}, Valeria and {Baume}, Gustavo and {Feinstein}, Carlos},
        title = "{Binary and Grouped Open Clusters: A New Catalogue}",
      journal = {arXiv e-prints},
         year = 2024,
        month = dec,
          eid = {arXiv:2412.05376},
        pages = {arXiv:2412.05376},
archivePrefix = {arXiv},
       eprint = {2412.05376},
 primaryClass = {astro-ph.GA},
       adsurl = {https://ui.adsabs.harvard.edu/abs/2024arXiv241205376P}
}

@ARTICLE{TasdemirCinar2025,
       author = {{Ta{\c{s}}demir}, S. and {{\c{C}}{\i}nar}, D.~C.},
        title = "{A detailed analysis of close binary OCs}",
      journal = {Contributions of the Astronomical Observatory Skalnate Pleso},
         year = 2025,
        month = apr,
       volume = {55},
       number = {3},
        pages = {506-511},
          doi = {10.31577/caosp.2025.55.3.506},
archivePrefix = {arXiv},
       eprint = {2501.17235},
 primaryClass = {astro-ph.SR},
       adsurl = {https://ui.adsabs.harvard.edu/abs/2025CoSka..55c.506T}
}

@ARTICLE{Vines2022,
  author  =  {{Vines}, Jose I. and {Jenkins}, James S.},
  title  =  "{ARIADNE: measuring accurate and precise stellar parameters through SED fitting}",
  journal  =  {Monthly Notices of the Royal Astronomical Society},
  year  =  2022,
  month  =  jun,
  volume  =  {513},
  number  =  {2},
  pages  =  {2719-2731},
  doi  =  {10.1093/mnras/stac956},
  archivePrefix  =  {arXiv},
  eprint  =  {2204.03769},
  primaryClass  =  {astro-ph.SR},
  adsurl  =  {https://ui.adsabs.harvard.edu/abs/2022MNRAS.513.2719V}
  }

@article{bisht2020comprehensive,
       author = {{Bisht}, D. and {Zhu}, Qingfeng and {Yadav}, R.~K.~S. and {Durgapal}, Alok and {Rangwal}, Geeta},
        title = "{A comprehensive study of open clusters Czernik 14, Haffner 14, Haffner 17 and King 10 using multicolour photometry and Gaia DR2 astrometry}",
      journal = {Monthly Notices of the Royal Astronomical Society},
         year = 2020,
        month = may,
       volume = {494},
       number = {1},
        pages = {607-623},
          doi = {10.1093/mnras/staa656},
archivePrefix = {arXiv},
       eprint = {2003.02448},
 primaryClass = {astro-ph.GA},
       adsurl = {https://ui.adsabs.harvard.edu/abs/2020MNRAS.494..607B}
}

@ARTICLE{2024AJ....167..188B,
       author = {{Belwal}, Kuldeep and {Bisht}, D. and {Bisht}, Mohit Singh and {Rangwal}, Geeta and {Raj}, Ashish and {Dattatrey}, Arvind K. and {Yadav}, R.~K.~S. and {Bhatt}, B.~C.},
        title = "{Exploring NGC 2345: A Comprehensive Study of a Young Open Cluster through Photometric and Kinematic Analysis}",
      journal = {The Astronomical Journal},
         year = 2024,
        month = may,
       volume = {167},
       number = {5},
          eid = {188},
        pages = {188},
          doi = {10.3847/1538-3881/ad2fcc},
archivePrefix = {arXiv},
       eprint = {2403.04532},
 primaryClass = {astro-ph.GA},
       adsurl = {https://ui.adsabs.harvard.edu/abs/2024AJ....167..188B}
}

@ARTICLE{Hunt2024,
       author = {{Hunt}, Emily L. and {Reffert}, Sabine},
        title = "{Improving the open cluster census. III. Using cluster masses, radii, and dynamics to create a cleaned open cluster catalogue}",
      journal = {Astronomy \& Astrophysics},
         year = 2024,
        month = jun,
       volume = {686},
          eid = {A42},
        pages = {A42},
          doi = {10.1051/0004-6361/202348662},
archivePrefix = {arXiv},
       eprint = {2403.05143},
 primaryClass = {astro-ph.GA},
       adsurl = {https://ui.adsabs.harvard.edu/abs/2024A&A...686A..42H}
}

@article{sariya2023gaia,
  title={A Gaia based analysis of open cluster Berkeley 27},
  author={Sariya, Devesh P and Jiang, Guey and Bisht, D and Yadav, RKS and Rangwal, Geeta},
  journal={New Astronomy},
  volume={98},
  pages={101938},
  year={2023},
  publisher={Elsevier}
}

@article{vallenari2023gaia,
  title={Gaia data release 3-summary of the content and survey properties},
  author={Vallenari, Antonella and Brown, Anthony GA and Prusti, Timo and De Bruijne, Jos HJ and Arenou, F and Babusiaux, Carine and Biermann, Michael and Creevey, Orlagh L and Ducourant, Christine and Evans, Dafydd Wyn and others},
  journal={Astronomy \& Astrophysics},
  volume={674},
  pages={A1},
  year={2023},
  publisher={EDP sciences}
}
\bibliographystyle{aasjournalv7}



%% For example:

%% \bibitem[Aladro et al.(2015)]{Aladro15} Aladro, R., Martín, S., Riquelme, D., et al. 2015, \aas, 579, A101

%%\end{thebibliography}

\end{document}